\documentclass[12pt]{article}
\pdfoutput=1  
\usepackage{mheck}
\usepackage{amsmath}
\usepackage{amssymb,amsfonts}
\usepackage{graphicx}
\usepackage[parfill]{parskip}

\usepackage[font=small]{caption}
\usepackage{tikz}
\usetikzlibrary{calc, positioning, intersections, arrows}

\numberwithin{equation}{section}
\numberwithin{table}{section}

\newcommand{\be}{\begin{equation}}
\newcommand{\ee}{\end{equation}}
\def\beq{\begin{eqnarray}}
\def\eeq{\end{eqnarray}}
\def\ba{\begin{eqnarray}}
\def\ea{\end{eqnarray}}

\def\epi{\epsilon_1}
\def\epii{\epsilon_2}
\def\epp{\epsilon_+}
\def\epm{\epsilon_-}

\newcommand{\IZ}{\mathbb{Z}}
\newcommand{\IC}{\mathbb{C}}
\newcommand{\IP}{\mathbb{P}}

\newcommand{\IR}{\mathbb{R}}
\newcommand{\IQ}{\mathbb{Q}}

\newcommand{\IF}{\mathbb{F}}

\newcommand{\Ztop}{\mathrm{Z_{top}}}

\newcommand{\tr}{\mathrm{Tr\, }}

\newcommand{\nn}{\nonumber}

\newcommand{\cN}{{\cal N}}
\newcommand{\cM}{{\cal M}}

\newcommand{\cO}{{\cal O}}
\newcommand{\cC}{{\cal C}}

\newcommand{\cD }{{\cal D}}

\newcommand{\dualCox}{h_G^{\vee}}

\def\mb{\mathbb}

\def\md{\mathbf}
\def\mf{\mathfrak}

\def\bC{\mathbb{C}}

\def\bH{\mathbb{H}}

\def\bP{\mathbb{P}}
\def\bQ{\mathbb{Q}}
\def\bR{\mathbb{R}}
\def\bZ{\mathbb{Z}}

\def\cC{\mathcal{C}}
\def\cD{\mathcal{D}}

\def\cM{\mathcal{M}}
\def\cN{\mathcal{N}}
\def\cO{\mathcal{O}}

\def\fg{\mathfrak{g}}
\def\fh{\mathfrak{h}}
\def\fd{\mathfrak{d}}

\def\({\left(}
\def\){\right)}

\def\mass{\boldsymbol{m}}
\def\ja{\boldsymbol{z}}

\newcommand{\ri}{{\mathsf{i}}}

\newcommand{\dG}{\dim(G)}
\newcommand{\hG}{h^\vee_G}
\newcommand{\rG}{\text{rk}(G)}
\newcommand{\DG}{\Delta_G}
\newcommand{\vp}{\varphi}

\newcommand{\KB}{K_{\IF_n}}

\def\mass{\boldsymbol{m}}
\def\blambda{\boldsymbol{\lambda}}
\def\bmu{\boldsymbol{\mu}}
\def\ja{\boldsymbol{z}}
\def\bt{\boldsymbol{t}}
\def\bQ{\boldsymbol{Q}}
\def\Ztop{\widehat{Z}}

\def\aja{\widehat{\boldsymbol{z}}}
\newcommand{\inH}{\md e[\mass]}
\newcommand{\Jd}{J^{D}}
\newcommand{\Jad}{J^{\widehat D}}
\newcommand{\jout}{\Jd_{*,*}(\fg)}
\newcommand{\jtout}{\Jad_{*,*}(\fg)}
\newcommand{\jouta}{\Jd_{*,*}(\mfa)}
\newcommand{\joutd}{\Jd_{*,*}(\mfd)}
\newcommand{\jtouta}{\Jad_{*,*}(\mfa)}
\newcommand{\jtoutd}{\Jad_{*,*}(\mfd)}
\newcommand{\jweyl}{J_{*,*}(\fg)}

\newcommand{\joutas}{\Jd_{w,n}(\mfa)}
\newcommand{\joutds}{\Jd_{w,n}(\mfd)}

\newcommand{\Zgk}{Z_{G,k}}
\newcommand{\Dat}{D(\mathfrak{a}_2)}
\newcommand{\Daat}{D(\widehat{\mathfrak{a}}_2)}

\newcommand{\Dad}{D(\widehat{\mathfrak{d}}_4)}
\newcommand{\Dg}{D(\mathfrak{g})}
\newcommand{\Dag}{D(\widehat{\mathfrak{g}})}
\newcommand{\aomega}{\widehat{\omega}}
\newcommand{\mfg}{\mathfrak{g}}
\newcommand{\mfa}{\mathfrak{a}_2}
\newcommand{\mfd}{\mathfrak{d}_4}
\newcommand{\mfag}{\widehat{\mathfrak{g}}}

\newcommand{\hW}{\hat{W}}

\allowdisplaybreaks[1]

\newcommand\restrict[1]{\raisebox{-.5ex}{$|$}_{#1}}

\preprint{USTC-ICTS-??-??

LPTENS 17/34}

\begin{document}
\begin{titlepage}
{}~ \hfill\vbox{ \hbox{} }\break

\rightline{Bonn-TH-11}
\rightline{USTC-ICTS-17-15}
\rightline{LPTENS 17/34}

\vskip 1 cm

\begin{center}
		\Large \bf Topological Strings on Singular Elliptic Calabi-Yau 3-folds and Minimal 6d SCFTs
\end{center}

\vskip 0.8 cm

\centerline{Michele Del Zotto${}^a$, Jie Gu${}^{b}$, Min-xin Huang${}^{c}$, Amir-Kian Kashani-Poor${}^{b}$,} 
\centerline{ Albrecht Klemm${}^{d}$ and Guglielmo Lockhart${}^{e}$ }

\vskip 0.2in
\begin{center}{\footnotesize
		\begin{tabular}{c}
                 ${}^{\, a}${\em Simons Center for Geometry and Physics, SUNY, Stony Brook, NY, 11794-3636 USA}\\[0ex]
		${}^{\, b}${\em LPTENS, CNRS, PSL Research University, Sorbonne Universit\'{e}s, UPMC, 75005 Paris, France}\\[0ex]
		${}^{\, c}${\em ICTS, University of Science and Technology of China, Hefei, Anhui 230026, China} \\[0ex]
		${}^{\, d}${\em Bethe Center for Theoretical Physics, Physikalisches Institut, Universit\"{a}t Bonn, 53115 Bonn, Germany}\\[0ex]	
		${}^{\, e}${\em   Institute for Theoretical Physics, University of Amsterdam,  Amsterdam, The Netherlands}
		\end{tabular}
}\end{center}

\setcounter{footnote}{0}
\renewcommand{\thefootnote}{\arabic{footnote}}
\vskip 60pt
\begin{abstract} 
	{}{\bf Abstract:} We apply the modular approach to computing the topological string partition function on non-compact elliptically fibered Calabi-Yau 3-folds with higher Kodaira singularities in the fiber. The approach consists in making an ansatz for the partition function at given base degree, exact in all fiber classes to arbitrary order and to all genus, in terms of a rational function of weak Jacobi forms. Our results yield, at given base degree, the elliptic genus of the corresponding non-critical 6d string, and thus the associated BPS invariants of the 6d theory. The required elliptic indices are determined from the chiral anomaly 4-form of the 2d worldsheet theories, or the 8-form of the corresponding 6d theories, and completely fix the holomorphic anomaly equation constraining the partition function. We introduce subrings of the known rings of Weyl invariant Jacobi forms which are adapted to the additional symmetries of the partition function, making its computation feasible to low base wrapping number. In contradistinction to the case of simpler singularities, generic vanishing conditions on BPS numbers are no longer sufficient to fix the modular ansatz at arbitrary base wrapping degree. We show that to low degree, imposing exact vanishing conditions does suffice, and conjecture this to be the case generally.
	
\end{abstract}

{\let\thefootnote\relax
	\footnotetext{\tiny mdelzotto@scgp.stonybrook.edu, jie.gu@lpt.ens.fr, minxin@ustc.edu.cn, kashani@lpt.ens.fr, aklemm@th.physik.uni-bonn.de, lockhart@uva.nl}}

\end{titlepage}
\vfill \eject

\tableofcontents

\newpage

\section{Introduction}

Topological string theory on Calabi-Yau manifolds computes important terms
in the effective action of string, M-- and F--theory compactifications. It simultaneously provides a multifaceted yet computable example of a theory with both worldsheet and target space underpinning, like its parent theory, string theory proper. Three is the critical (complex) dimension in the topological setting: on Calabi-Yau threefolds, topological string amplitudes  ${\mathcal F}_g$ are particularly rich, as they receive world-sheet instanton corrections at all genera.

Tools to compute the  ${\mathcal F}_g$ include mirror symmetry, large N-- and localization techniques. Powerful constraints on these amplitudes are imposed by automorphic symmetries rooted in the monodromy group of the threefold; due to these, the ${\mathcal F}_g$ are automorphic forms. Famously, they are, at least for $g>0$, not quite holomorphic. Their anholomorphicity is captured by the holomorphic anomaly equations~\cite{BCOV,Hosono:1999qc,Huang:2010,Klemm:2012,Alim:2012,Huang:2013,Huang:2015sta}.

The ${\mathcal F}_g$ can be assembled into a generating function, the topological string partition function $Z_{top} = \exp \sum_{g=0}^{\infty} {\mathcal F}_g \lambda_s^{2g-2}$. The definition as a formal power series in $\lambda_{s}$ however does not do justice to $Z_{top}$. On certain geometries, coefficients of $Z_{top}$ in an expansion in distinguished K\"ahler classes can be computed which are automorphic forms in their own right, with analytic dependence on $\lambda_{s}$. This paper is dedicated to the study of $Z_{top}$ on such a class of geometries, elliptic fibrations underlying (minimal) six dimensional superconformal field theories (6d SCFTs).

To obtain effective theories with prescribed gauge group and matter content, it is often necessary to consider compactification on singular Calabi-Yau geometries. For example, type II compactifications on Calabi-Yau manifolds exhibiting $A_N$ singularities give rise to $N=2$ supersymmetries field theories with $SU(N)$ gauge symmetry~\cite{Katz:1996fh}. Calabi-Yau threefold singularities do not yet enjoy a complete classification (see for example~\cite{MR927963} for hypersurface and~\cite{MR1169227} 
for orbifold singularities). However, much is known, such that this framework can be used to explore
exotic quantum field theories not easily accessible via other means. 

6d SCFTs provide a prominent example in which this strategy has been pursued with success in the previous years, see e.g. \cite{Lockhart:2012vp,Haghighat:2013,Haghighat:2013tka,Hohenegger:2013ala} for some early works in this direction. To decouple gravity, a partial decompactification limit of the internal geometry is considered. As six is the maximal dimension in which the superconformal algebra can be realized~\cite{Nahm:1977tg}, these theories present an important starting point for studying lower dimensional SCFTs via compactification. An intriguing property of 6d SCFTs is that they generically contain strings in their spectrum which preserve some of the supersymmetry. The study of the worldsheet theory of these strings provides an important handle on calculating the 6d spectrum.

6d SCFTs can be constructed via F-theory compactifications on elliptically fibered Calabi-Yau threefolds. Thanks to the work of Kodaira \cite{MR0184257}, singularities of elliptic manifolds are amongst the best understood. This has led to a proposal for the classification of all 6d $(1,0)$ SCFTs~\cite{Heckman:2013pva,DelZotto:2014hpa,Heckman:2015bfa} (see also \cite{Apruzzi:2013yva,Gaiotto:2014lca,Bhardwaj:2015xxa,Bhardwaj:2015oru,Apruzzi:2017nck}), which in turn has provided novel field theoretical tools to address the topological string partition function \cite{Haghighat:2014,Kim:2014dza,Haghighat:2014vxa,Shabbir:2015oxa,Kim:2015jba,Gadde:2015tra,Hohenegger:2016yuv,Kim:2016foj,Hayashi:2017jze,Bastian:2017ing,Haghighat:2017vch}. The BPS spectrum of such theories compactified on a circle, as encapsulated in the all
genus BPS indices $I_g^\kappa\in\mathbb{Z}$~\cite{Gopakumar:1998ii,Gopakumar:1998jq} and the associated 
refined  BPS numbers\footnote{  In~\cite{CKK,NO}, the  $N^\kappa_{j_-j_+}\in \mathbb{N}$ were mathematically  
	defined by motivically refined counting of the dimensions of cohomologies on the moduli space of  stable pairs. This definition requires the Calabi-Yau to have a continuous symmetry.  Such a symmetry is generically  present in non-compact Calabi-Yau geometries; in the case of toric Calabi-Yau manifolds,
	it can be used to  calculate the $N^\kappa_{j_-j_+}$ by localization\cite{CKK}.} $N^\kappa_{j_-j_+}\in \mathbb{N}$~\cite{CKK,NO}, can be 
calculated also using methods of topological string theory.
Early work in this direction include~\cite{Klemm:1996}. This program can be pushed substantially further using the  automorphic form 
approach in combination with a holomorphic anomaly equation and boundary conditions. The holomorphic anomaly equation in question arises whenever the Calabi-Yau geometry exhibits an elliptic fibration structure. It was first introduced for elliptic surfaces in~\cite{Hosono:1999qc}, generalized to the refined topological string on elliptic surfaces in~\cite{Huang:2010,Huang:2013}, and to the non-refined case on compact elliptic Calabi-Yau threefolds in~\cite{Klemm:2012,Alim:2012}. The anholomorphicity in this setup arises already at genus zero. The relation of this holomorphic anomaly equation to the ones of \cite{BCOV} is subtle. A study was initiated in~\cite{Klemm:2012}, but further clarification is required.

Together with the automorphic symmetries imposed by the elliptic fibration, the holomorphic anomaly equation implies that in an expansion in base K\"ahler parameters, the coefficient $Z_\beta$ of $Z_{top}$ at base degree $\beta$ is a meromorphic Jacobi form, 
of vanishing weight and with index bilinear form\footnote{ This quantity was referred to as the ``index polynomial'' in \cite{Gu:2017ccq}.}  $M_\beta$ computable from intersection data of the geometry. The modular parameter of the Jacobi form is identified with the complexified K\"ahler parameter of the elliptic fiber, while the elliptic parameter is identified with the topological string coupling constant $\lambda_s$. The relation between $Z_\beta$ and the elliptic genus of the 2d quiver gauge theories describing the worldsheet theory of the 6d non-critical string~\cite{Haghighat:2014,Haghighat:2014vxa} inspires an ansatz for $Z_\beta$ as a ratio of holomorphic Jacobi forms with universal denominator~\cite{Huang:2015sta}. 
The problem of computing $Z_\beta$ on such geometries is thus reduced to determining the numerator of the ansatz, a weak Jacobi form of fixed weight and index bilinear form. This results in a 
finite dimensional problem which must be solved by imposing suitable boundary conditions.

The unrefined E--string, obtained from compactification on local $1/2$ $K3$~\cite{Morrison:1996pp,Klemm:1996}   was solved with
these methods in~\cite{Huang:2015sta}. For this theory, imposing the vanishing of the BPS indices $I_g^\kappa$ for fixed curve class $\kappa$ at sufficiently high $g$ (recall that this property is guaranteed by Castelnuovo theory), 
provides sufficient boundary conditions to solve the model.\footnote{  We refer to such boundary conditions as {\it generic} vanishing conditions, to be contrasted with {\it precise} vanishing conditions which specify the $g$ beyond  which  the invariants vanish at given class $\kappa$.}  The method was extended to the refined setting, and to arbitrary chains of M--strings terminating on an E--string, in~\cite{Gu:2017ccq}. Also here, imposing generic vanishing conditions suffices to solve the theories. The main structural modification in passing to the refined case is the necessity to accommodate the two epsilon parameters $\epsilon_\pm$ that refine 
$\lambda_s$ as elliptic parameters of the Jacobi forms $Z_\beta$.

The geometries studied in~\cite{Gu:2017ccq} are the most general based on Kodaira singularities of type $I_1$ in the fiber. In this paper, we extend the automorphic approach to higher Kodaira singularities. The main structural novelty when passing to such singularities is that their resolution gives rise to additional homology cycles in the fiber of the geometry. Their K\"ahler parameters $\mass$ contribute elliptic parameters alongside the topological string couplings $\epsilon_{\pm}$, the sole elliptic parameters in the case of the (massless) E-- and M-- strings. The parameters $\mass$ are subject to additional symmetries. The generic intersection matrix of these fiber classes within a distinguished compact elliptic surface of the geometry is given by the affine Dynkin diagram of an appropriate Lie algebra $\mathfrak{g}$. From the identification of $Z_{top}$ with the elliptic genus of the 6d string, it is immediate that the K\"ahler parameters $\mass$ must enjoy an action of the Weyl group of $\mathfrak{g}$ under which $Z_{top}$ is invariant. Happily, the question of Jacobi forms with multiple elliptic parameters endowed with a Weyl group action has been studied in the mathematics literature \cite{Wirthmuller:Jacobi,Bertola}, and the resulting ring of Weyl invariant Jacobi forms $J_{*,*}(\mathfrak{g})$ for all simple Lie algebras $\mathfrak{g}$ excluding $\mathfrak{e}_8$ constructed. For the class of theories we will discuss in this paper, the natural building blocks to implement the $\mass$ dependence of $Z_{top}$ sit in a subring $\jout$ of $\jweyl$, which we construct: the ring of Jacobi forms on which the diagram automorphisms $\Dag$ of the affine Dynkin diagram to $\mathfrak{g}$ act as a symmetry.\footnote{More precisely, the subring $\jout \subset J_{*,*}(\mathfrak{g})$ is isomorphic to the ring $\jtout$ of $\Dag$ invariant Jacobi forms.}

Six-dimensional theories generally do not  
have a Lagrangian description due to their chiral nature. Nevertheless, important information can be obtained by studying the anomaly 8-form of these theories \cite{Ohmori:2014kda,Intriligator:2014eaa,Cordova:2015fha}, related by an anomaly 
inflow argument to the anomaly 4-form of the BPS strings in their spectrum~\cite{Kim:2016foj,Shimizu:2016lbw}. It has been conjectured that the  elliptic genus of these 2d theories is simply related to their  Casimir energy, and that the latter  
can be obtained from an equivariant  integral of the anomaly 4-form~\cite{Bobev:2015kza}. This setup provides a second, and indeed currently more comprehensive path towards determining the index bilinear form $M_\beta$ of the Jacobi forms $Z_\beta$ \cite{DelZotto:2016pvm}. It is  an intriguing result of this paper that at least in the examples considered here,\footnote{  The refined theory for $\mfg=\mathfrak{a}_2$ and $\mfg=\mathfrak{d}_4$ at base degree $1$, and the unrefined theory for $\mfg=\mathfrak{a}_2$ up to base degree $3$.} this data  
together with general properties of the BPS expansion and precise vanishing conditions fix the 
 BPS spectrum completely.    

The so-called minimal 6d (1,0) SCFTs which we study in this paper arise in a family of 9 (we list all members in table \ref{non-higgsable} in the body of the paper) via F-theory compactifications on elliptic fibrations over a non-compact base surface containing an isolated rational curve $C$ with negative self-intersection within the non-compact base. The cases $- C\cdot C = 1,2$ are the E--string and M--string respectively, which, as mentioned above, are among the theories solved in \cite{Huang:2015sta,Gu:2017ccq}. In this paper, we will study the cases $- C\cdot C = 3,4$, giving rise to the gauge groups $\mfg=\mathfrak{a}_2$ and $\mfg=\mathfrak{d}_4$ respectively. The remaining cases, in particular with regard to the ring structure of $\jout$ for the respective $\mfg$, will be considered in a future publication.\footnote{We stress that the methods discussed here unveil a universal structure underlying all of these computations, providing a novel stringent consistency check for all the results that have previously appeared in the literature regarding these models \cite{Haghighat:2014vxa,Kim:2016foj,Hayashi:2017jze}.}

The rest of this paper is organized as follows: In section 2, we review topological string theory on non-compact singular elliptic Calabi-Yau threefolds and the corresponding six-dimensional superconformal field theories, with an emphasis on the minimal 6d SCFTs. In section 3, we give a detailed description of a toric realization of the Calabi-Yau threefolds corresponding to the $\mfg = \mfa$ and $\mfd$ minimal SCFTs, paying special attention to the fibration structure of the exceptional cycles in the resolved geometry. In section 4, we formulate our ansatz for the contributions $Z_\beta$ to the topological string partition function at fixed base wrapping number as a ratio of weakly-holomorphic Jacobi forms, and discuss a basis of $D(\mfag)$-invariant Jacobi forms suitable for capturing the dependence of the numerator on the gauge parameters $\mass$. In section 5, we discuss vanishing conditions on BPS numbers for the Calabi-Yau threefolds corresponding to $\mfg=\mfa$ and $\mfd$, and employ them to completely fix the contribution to the topological string partition function for various base wrapping numbers; we also discuss additional constraints that arise by taking special limits of the elliptic genera. In the appendices, we provide explicit data on the BPS invariants of the threefolds for $\mfg=\mfa,\mfd$ obtained from our ansatz and provide a detailed discussion of $D(\mfag)$-invariant Jacobi forms. We finally comment on the ring of Weyl invariant Jacobi forms for the case $\mfg = \mathfrak{e}_8$, which is not covered by the classification results of \cite{Wirthmuller:Jacobi}.

\section{Topological strings and six-dimensional theories} \label{s:top_strings_and_6d}

\subsection{Review of refined topological string invariants}
Let $X$ be a Calabi-Yau 3-fold. For each K\"ahler class $\kappa \in H_2(X,\mathbb{Z})$, we denote by $\bt_\kappa \equiv t_j(\kappa) \kappa^j$ and $\bQ^\kappa = \text{exp}(2 \pi i \boldsymbol{t}_{\kappa})$ the corresponding K\"ahler parameter, its expansion coefficients in a basis $\{\kappa_j\}_j$ of $H_2(X,\mathbb{Z})$, and its formal exponential. The expansion of the genus-$g$ worldsheet topological amplitudes ${\mathcal F}_g(\bt)$ on $X$ in terms of these formal exponentials gives rise to Gromov-Witten invariants $r^\kappa_g \in \IQ$ via
\begin{equation}
{\mathcal F}_g(\bt) = \sum_{\kappa \in H_2(X,\mathbb{Z})} r_g^\kappa \boldsymbol{Q}^\kappa \,, \quad\quad r^\kappa_g\in \IQ \,.
\end{equation}
 
The topological string free energy is the generating function of the ${\mathcal F}_g(\bt)$, \footnote{In this paper, we find it convenient to rescale the topological string coupling $\lambda_s$  by a factor of $2\pi$ as compared to the usual conventions.}
\begin{equation}\label{FTOP}
{\mathcal F}(\lambda_s,\bt)= \sum_{g\geq0} \, (2\pi\lambda_s)^{2g-2} \, {\mathcal F}_g(\bt) \ .
\end{equation}
$\lambda_s$ here is a formal parameter, the topological string coupling constant. The topological string free energy has an alternative expansion \cite{Gopakumar:1998ii,Gopakumar:1998jq} 
\begin{equation} 
\label{bpsindices}
{\mathcal F}(\lambda_s,\bt)=\sum_{m\geq 1\atop g \geq 0} \sum_{\kappa\in H_2(X,\mathbb{Z})} I^\kappa_g
 \left(2 \sin \left( \frac{2\pi \lambda_s m}{2}\right)\right)^{2 g -2} \frac{\bQ^{m \kappa}}{m}   \,,\quad \quad I^\kappa_g\in \mathbb{Z}\,,
\end{equation}
where the coefficients $I^\kappa_g$ are the integer Gopakumar-Vafa invariants. They capture the BPS degeneracies of the compactification of M-theory on $X$.

Whenever the Calabi-Yau $X$ is local, the corresponding compactification of M-theory gives rise to a five-dimensional theory with $\mathcal N = 1$ supersymmetry decoupled from gravity. Because five-dimensional $\mathcal N=1$ supersymmetric theories always have an $Sp(1)$ R-symmetry, all these theories have a well-defined curved rigid supersymmetric partition function on the $\Omega$-background \cite{Nekrasov:2002qd, Losev:2003py, Nekrasov:2003rj}, that we denote $Z(\epi,\epii,\bt)$. We stress that this true also when the corresponding local CYs are elliptic, which corresponds to the M-theory engineering of five-dimensional models that have an UV completion in six-dimensions. The corresponding free energy,
\begin{equation}
\mathcal{F}(\epi,\epii,\bt) = \text{log } Z(\epi,\epii,\bt) \ ,
\end{equation}
gives a refinement of the topological string \cite{Hollowood:2003cv}. The refined topological string free energy again gives rise to integer BPS degeneracies, now called $N^\kappa_{j_-j_+}$ and doubly indexed for each K\"ahler class $\kappa$,
\begin{equation}\label{FNEK}
\mathcal F(\epi,\epii, \bt)=\sum_{m \geq 1  \atop j_\pm \in \mathbb{N}/2 } \sum_{\kappa\in H_2(X,\mathbb{Z})} 
N^\kappa_{j_-j_+} {(-1)^{2 (j_-+j_+)} \, \chi_{j_-}(u^m) \chi_{j_+}(v^m) \over v^m + v^{-m} - u^m - u^{-m}} {\bQ^{m \kappa} \over m}  \ .
\end{equation}
We have introduced the variables
\begin{equation}\label{eq:uvdef}
u=\exp(2\pi i\epsilon_-)\ ,\qquad v=\exp(2\pi i\epsilon_+)\ ,\qquad \epsilon_\pm =\frac{1}{2}( \epi \pm \epii)\ .\\ 
\end{equation}
Moreover, for a given irreducible $SU(2)$ representation $[j]$ with highest weight $j\in \mathbb{N}/2$, we denote the corresponding character as
\be
\chi_j(x) = \sum_{k=-j}^{j} x^j \ ,
\ee
where the sum is taken in integer increments. Unlike the BPS indices $I^\kappa_g$ that are invariant under 
deformations of the complex  structure  of the Calabi-Yau manifold $M$, the BPS degeneracies $N^\kappa_{j_-j_+}$ 
can jump at complex co-dimension loci in the  complex moduli space ${\cal M}_{cs}$ of $M$ as discussed in~\cite{KKV}, see in particular footnote six for a simple example.

In the limit $\epi = - \epii = \lambda_s$, $\mathcal F(\epi,\epii, \bt)$ reduces to $\mathcal F(\lambda_s,\bt)$.
The invariants $I^\kappa_g$ can be reconstructed from the degeneracies $N^\kappa_{j_-j_+}$ via the relation
\begin{equation} \label{ItoN}
\sum_{g \in \mathbb{N}} I^\kappa_g  \cdot \left( [\,0\,] \oplus [\,0\,] \oplus [\,\tfrac12\,]\right)^{\otimes g}   = \bigoplus_{j_- \in \mathbb{N}/2} \left[ \, j_- \,\right] \sum_{j_+\in \mathbb{N}/2} (-1)^{2 j_+} (2 j_+ + 1) \, N^\kappa_{j_-j_+} \ .
\end{equation}
To see this, take the trace of $(-1)^{2\sigma_3} e^{2i \sigma_3 (2\pi\lambda_s)}$ in the representation \eqref{ItoN}, and note that
\be 
\tr_{[0]\oplus [0] \oplus[\frac{1}{2}]} (-1)^{2\sigma_3} e^{2i \sigma_3(2\pi\lambda_s)} = \left(2 \sin \frac{2\pi\lambda_s}{2}\right)^2 \,.
\ee

\subsection{6d $\Omega$-background and refined topological strings}
\label{sc:6dNek}

Whenever the local Calabi-Yau threefold $X$ is elliptically fibered over a K\"ahler base $B$, the five-dimensional M-theory background has a six-dimensional F-theory origin.\footnote{   In the case where $X$ is also K3 fibered, certain recent attempts \cite{Antoniadis:2013epe,Antoniadis:2013mna} to define the refined topological string amplitudes using the heterotic string can be understood by means of a fiberwise heterotic/F-theory duality.} The  refined topological string partition function on $X$ coincides with the $\Omega$-background partition function of the 6d theory obtained via compactification of F-theory on $X$. As $X$ is local, gravity is decoupled. The 6d $\Omega$-background \cite{Losev:2003py} is a rigid curved supersymmetric background -- the 6d theory coupled to it exhibits $(1,0)$ supersymmetry -- with topology $T^2 \times \mathbb{R}^4$, and metric
\begin{equation}
ds^2 = dz d \bar z + \sum_{\mu=1}^{4} (dx^\mu + \Omega^\mu d z + \bar \Omega^\mu d \bar z)^2 \,,
\end{equation}
where $(z,\bar z)$ are coordinates on $T^2$. The $\Omega_\mu$ satisfy
\begin{equation}
F = d \Omega = \epsilon_1 dx^1 \wedge dx^2 - \epsilon_2 dx^3 \wedge dx^4.
\end{equation}
Expressing the field strength $F$ in the spinor notation $F_{\alpha\beta}$ and $F_{\dot\alpha\dot\beta}$, we have
\begin{equation}\label{eq:epem}
\epsilon_+ = - \text{det} \, F_{\alpha\beta}  \qquad \epsilon_- = - \text{det} \, F_{\dot\alpha\dot\beta},
\end{equation}
where $\alpha,\beta=1,2$, and $\dot\alpha,\dot\beta =1,2$ are spinor indices. These are the field strengths for the spacetime $SU(2)_L \times SU(2)_R$ symmetry acting on the $\mathbb{R}^4$ factor of the $\Omega$-background. To preserve some amount of supersymmetry, the torus factor of the 6d geometry needs to be endowed with a non-trivial (flat) R-symmetry background Wilson line of the form ${A^i}_j d z + {\bar A^i}_j d \bar z$, where $i,j =1,2$ are $SU(2)_r$ R-symmetry indices and ${A^i}_j $ has eigenvalues $\pm \, \epsilon_+$.

There exist two kinds of supersymmetric systems decoupled from gravity in six spacetime dimensions, distinguished 
by their ultraviolet behavior: little string theories (LSTs) and superconformal field theories (SCFTs). 
The F-theory backgrounds corresponding to both kinds of models have recently been classified \cite{Heckman:2013pva,DelZotto:2014hpa,Heckman:2015bfa,Bhardwaj:2015oru}. A given six-dimensional theory is said to be irreducible if it cannot be decomposed in two or more mutually non-interacting subsectors. If that is the case, the corresponding F-theory model is fully characterized upon specification of a collection of mutually transversally intersecting holomorphic curves 
\begin{equation}
C_I \subset B \,,\qquad I = 1, \dots, r+f,
\end{equation}
the first $r$ of which are compact, while the remaining $f$ are not, together with the structure of their normal bundles and 
the degeneration of the elliptic fibration $\pi$ over such a collection. 
The integer $r$ is the rank of the corresponding six-dimensional theory (the tensor branch dimension), 
the integer $f$ runs over the simple factors of the non-abelian part of the flavor symmetry 
group.\footnote{   The abelian part of the flavor symmetry is captured by the Mordell-Weil group of the elliptic fibration.} The 
intersection pairing
\begin{equation}
A_{IJ} \equiv - C_I \cdot C_J \,,\qquad 1\leq I,J \leq r
\end{equation}
governs the metric along the tensor branch of the six-dimensional theory: for 6d SCFTs $A_{IJ}$ has to be positive definite, 
while for 6d LSTs $A_{IJ}$ is semi-positive definite, with a single zero eigenvalue, 
corresponding to the little string charge. 

Six-dimensional supersymmetric theories exhibit BPS strings whose F-theory origins are wrapped D3 branes on the compact curves $C_I$, $1\leq I \leq r$. The dynamics of these strings is governed by 2d $(0,4)$ worldsheet theories.
The symmetric matrix $A_{IJ}$ is interpreted as the 6d version of the Dirac pairing among 
BPS strings, and the curve classes in integer homology are interpreted as generators for the BPS string charge lattice. 
Whenever a curve $C$ is part of the discriminant of $\pi$, F-theory provides a stack of 
7-branes to wrap $C$, giving rise to a gauge symmetry of the 6d theory. The corresponding 6d gauge group is determined using the Tate algorithm \cite{Bershadsky:1996nh}. The volume of $C$ is inversely proportional to the gauge coupling squared; the BPS strings are 
instantons for the corresponding gauge group.

Up to a universal prefactor, the $\Omega$-background partition function of any six-dimensional theory localizes on contributions from such BPS strings \cite{Losev:2003py,Hollowood:2003cv}. For the theories engineered within F-theory, a given BPS string configuration with charge $\beta$ corresponds to a bound-state of wrapped D3 branes, with $\beta_I \in \mathbb{N}$ D3-branes wrapped on the curve $C_I$. We denote by
\begin{equation}\label{eq:elliptic}
Z_\beta(\tau \, | \, \epp,\epm,\mass) \equiv \text{Tr}_R \, (-1)^F q^{H_L} \bar q^{H_R} u^{2J_-} v^{ 2(J_+ + J_r) }  \,\md e[\mass] 
\end{equation}
the flavored Ramond elliptic genus for the resulting $(0,4)$ worldsheet theory. $J_\pm$ are generators of the $SU(2)_\pm$ global symmetries that rotate the four-plane transverse to the wrapped D3-branes, while $J_r$ is a  generator of the $SU(2)_r$ symmetry of the small $(0,4)$ superconformal algebra that emerges at the conformal point. The fugacities $u$ and $v$ keep track of the coupling of these theories to the $\Omega$-background --- see equations \eqref{eq:uvdef} and \eqref{eq:epem}. The remaining fugacities are denoted by $\mass$ and are associated to a global symmetry group $G$ of the 2d SCFT which arises due to the 3-7 string sectors in the F-theory engineering; the parameter $\mass$ is an element of $\fh_\bC\cong \bC^{r_G}$, the complexification of the Cartan subalgebra $\fh$ of the Lie algebra $\fg$ of rank $r_G$ associated to $G$. The $e[\mass]$ are flavor symmetry Wilson lines on the torus. Below, we will expand $\mass = \sum_i m_i \omega_i$ in a basis of fundamental weights, and define
\be 
\md e[\mass] = \prod_i \exp(2\pi \ri m_i )\,.
\ee
From the geometric perspective, the $\mass$ encode the K\"ahler classes of curves resolving the singular fibers of the fibration $\pi$.
While the $SU(2)_-\times SU(2)_+\times SU(2)_r$ factor of the global symmetry is universal, the nature of $G$ depends on the specific 6d model.  

The six-dimensional $\Omega$-background partition function has the following expression \cite{Haghighat:2013,Haghighat:2013tka,Haghighat:2014vxa}
\begin{equation}\label{eq:nekrasix}
Z(\epi,\epii,\bt) = Z_0 \left(1+ \sum_{\beta \in \Gamma} \bQ_c^\beta \widehat Z_\beta(\tau \, | \, \epp,\epm,\mass) \right),
\end{equation}
where $\Gamma$ is the BPS string charge lattice of the theory, and the $\widehat Z_\beta(\tau \, | \, \epi,\epii, \mass)$ are rescaled flavored Ramond elliptic genera (see Section \ref{sec:2danom} below). From the geometry, the string charge lattice of the theory is generated by the compact curve classes $C_I$, $1 \leq I \leq r$.

The six-dimensional $\Omega$-background partition function \eqref{eq:nekrasix} coincides with the refined topological string partition function on $X$. The structure of this expression suggests dividing the K\"ahler parameters $\bt$ of $X$ into three classes:
\begin{equation}
\bt = (\tau,\bt_{\boldsymbol{b}_c},\mass) \,.
\end{equation}
In order of appearance, these correspond to the volume of the elliptic fiber $E$, to the K\"ahler classes of the compact 2-cycles in the base of the elliptic fibration, and to the K\"ahler classes of the 2-cycles in the fiber. We also introduce the exponentiated version of these parameters,
\be
\bQ = (q, \bQ_c, Q_1, \ldots, Q_{r_G}) \,.
\ee
Note that in the elliptic genus \eqref{eq:elliptic}, the modular parameter $\tau$ is the complex structure of the $T^2$ factor of the 6d $\Omega$-background. Via the duality between F-theory and M-theory, $\mathrm{F}[X \times S^1] = \mathrm{M}[X]$, the inverse of the radius of $S^1$ is identified with the volume of the elliptic fiber, for which we therefore retain the name $\tau$. The K\"ahler classes of the base, given by $\bt_{\mathbf{b}_c}$, yield the tensions of the BPS strings. These are determined by the VEVs of the 6d scalars that parametrize the tensor branch of the 6d theory. The corresponding exponentiated variables $\bQ_c$ are formal variables in the expansion of the 6d partition function in base K\"ahler classes. Finally, the fiber classes $\mass$, fugacities for the global symmetries in \eqref{eq:elliptic}, are assigned to the exceptional cycles that resolve fiber singularities.

\subsection{Holomorphic anomaly and modular properties of elliptic genera}\label{sec:2danom}
The refined topological string partition function on elliptically fibered Calabi-Yau threefolds satisfies a holomorphic anomaly equation of the form \cite{Hosono:1999qc,Huang:2015sta,Gu:2017ccq}
\begin{equation}\label{eq:holanomaly}
\left(\partial_{E_2(\tau)} + {1 \over 12} M_\beta(\epsilon_+,\epsilon_-,\mass) \right)Z_\beta(\tau \, | \, \epp,\epm,\mass) = 0 \,,
\end{equation} 
where $M_\beta$ is the index bilinear form. In principle, it can be determined from intersection data of the geometry. In references \cite{Haghighat:2014vxa,Huang:2015sta,DelZotto:2016pvm,Gu:2017ccq}, it was argued that the index bilinear form is equal to the modular anomaly of the corresponding elliptic genus; this relation provides a more accessible path towards computing $M_\beta$, as we now explain.

Consider a 2d theory with $(0,2)$ supersymmetry. We will denote its flavored Ramond elliptic genus by $\mathbb E(\tau \, | \, \ja)$, where $\ja$ are fugacities for global symmetries. $\mathbb E(\tau \, | \, \ja)$ has a modular anomaly: under a modular transformation $ \tau\to -1/\tau $, it transforms as a weight-zero Jacobi form of modular variable $\tau$ and several elliptic variables $\ja$, 
\begin{equation}\label{eq:modularanomaly}
\mathbb E(- 1/ \tau \, | \, \ja / \tau) = e^{2 \pi i {f(\ja) \over \tau}} \mathbb E(\tau \, | \, \ja) \,.
\end{equation}
$f(\ja)$ is called the modular anomaly of  $\mathbb E(\tau \, | \,\ja)$. It is a quadratic form of the fugacities $\ja$. For models with a Lagrangian description, the modular anomaly of $\mathbb E(\tau \, | \,\ja)$ has been determined in \cite{Benini:2013xpa}. In that reference, a formula for $\mathbb E(\tau \, | \,\ja)$ is given in terms of a Jeffrey-Kirwan residue. The Lagrangian of the theory determines the integrand of the residue formula. Since the integration does not involve the modular nor the elliptic variables, the modular properties of the integrand are inherited by the integral. For the vast majority of 6d theories, however, the 2d worldsheet theory of the  BPS strings does not have a Lagrangian formulation.\footnote{  This fact is closely related to the absence of an ADHM construction for exceptional Lie groups.} Therefore, a different path towards computing the modular anomaly for this class of theories is required. We claim that this computation can proceed via the component $E(z)$ of the supersymmetric Casimir energy for any 2d $(0,2)$ theory that depends quadratically on the global symmetry fugacities. Adapting the results of \cite{Bobev:2015kza} to our setup, we find that $E(z)$ is given by\footnote{  The authors of \cite{Bobev:2015kza} write this equation as a function of $\beta$, that in our conventions is the minus the radius of the 6d circle, which is related to our $\tau$ by $\beta = - 2\pi i/\tau$. Moreover, they use rescaled fugacities $\ja^\prime = \tau \ja$ for the global symmetry Wilson lines. For this reason, the elliptic genus considered in that reference is an S-modular transform of ours.}
\begin{equation}
E(\ja) \equiv  - {1 \over 2 \pi i}  \,  \lim_{\tau \to 0} \tau^2 { \partial \over \partial \tau} \, \text{log } \mathbb E(-1/\tau \,|\, \ja / \tau).
\end{equation}
Using equation \eqref{eq:modularanomaly}, 
\begin{equation}
\lim_{\tau \to 0} \tau^2 { \partial \over \partial \tau}  \, \text{log } \left( e^{2 \pi i {f(\ja) \over \tau}}\mathbb E(\tau \, | \, \ja ) \right) =  - 2 \pi i \, f(\ja) 
\end{equation}
and therefore, 
\begin{equation}
E(\ja) = f(\ja).
\end{equation}
As $E(\ja)$ has been conjectured to be given by an equivariant integral of the anomaly polynomial \cite{Bobev:2015kza},\footnote{   To be precise, the authors of \cite{Bobev:2015kza} verify the relation between the Casimir energy and anomaly polynomial for NS boundary conditions on the non-supersymmetric side. However, this difference does not affect the quadratic component $E(\ja)$ of the Casimir energy.}
\begin{equation}\label{eq:equivariantint}
E(\ja) = \int_\text{eq} \, \mathcal A_4 \,, 
\end{equation}
this provides a computational prescription for obtaining the modular anomaly $f(p)$ also in theories which do not possess a Lagrangian description.\footnote{ An independent argument which leads to the same conclusion can be found in section 2 of \cite{Benjamin:2016fhe}.}

The anomaly polynomial of the BPS strings of 6d SCFTs is determined by anomaly inflow from the 6d anomaly polynomial \cite{Kim:2016foj,Shimizu:2016lbw}. For a BPS string with charge $\beta = \sum_{I=1}^r \beta^I [C_I]$, 
\begin{equation} \label{anom4}
\begin{aligned}
\mathcal A_4 \equiv & \, \frac{1}{2} A_{IJ} \beta^I \beta^J \Big(c_2(L)-c_2(R)\Big)\\& + \beta^J \Big(\frac{1}{4} A_{Ja} \, \text{Tr} \, \left(F^{(a)}\right)^2 \, -  \frac{1}{4} (2 - A_{JJ}) \big(p_1(T) - 2 c_2(L) - 2 c_2(R) \big) + h^\vee_{G_J} c_2(r) \Big)\\
\end{aligned}
\end{equation}
where $a=1,...,r+f$ runs also over the non-compact divisors corresponding to the flavor symmetries. The factors $c_2(L), c_2(R)$ and $c_2(r)$ indicate the second Chern classes for the $SU(2)_-\times SU(2)_+ \times SU(2)_r$ bundle reflecting the global symmetries of the 2d (0,4) SCFT governing the IR dynamics of the BPS strings. Recall that the $SU(2)_-\times SU(2)_+$ global symmetry on the worldsheet of the BPS string is interpreted geometrically as the rotation symmetry of the four-plane transverse to the string, while $SU(2)_r$ is the R-symmetry of the IR little $(0,4)$ superconformal algebra. In the $T^2 \times \mathbb{R}^4$ $\Omega$-background  the strings are wrapped around the $T^2$, hence the $\Omega$-deformed $\mathbb{R}^4$ has to be identified with the four-plane transverse to the string. This entails that the $SU(2)_L \times SU(2)_R$ corresponding to the self-dual and the anti-self dual part of the $\Omega$-background curvature in equation \eqref{eq:epem} are identified with $SU(2)_- \times SU(2)_+$, and therefore the $\Omega$-deformation parameters $\epsilon_\pm$ are identified with the fugacities of the global symmetries $SU(2)_\pm$ for the $T^2$ partition function of the BPS string. The $SU(2)_r$ symmetry of the 2d (0,4) SCFT is identified with the $SU(2)$ R-symmetry of the 6d (1,0) theory. As discussed below Equation \eqref{eq:epem}, to preserve supersymmetry, the latter is coupled to a background gauge field Wilson line on $T^2$, and this identifies the fugacity for the $SU(2)_r$-symmetry with $\epsilon_+$. With these identifications the equivariant integration in \eqref{eq:equivariantint} in the $\Omega$-background amounts to the replacement rules \cite{DelZotto:2016pvm}
\begin{equation}
\begin{aligned}
&c_2(L) \mapsto -\varepsilon_-^2 \,,\\
& c_2(R),c_2(r) \mapsto -\varepsilon_+^2\,,\\
& p_1(T) \mapsto 0 \,,\\
&\text{Tr} \, \left(F^{(a)}\right)^2 \mapsto - 2 \, (\mass^{(a)},\mass^{(a)})_{\mathfrak{g}_{(a)}}\,.
\end{aligned}
\end{equation}
The notation $(,)_\fg$ indicates the invariant bilinear form on $\fh_\bC$ normalized such that the norm square of the shortest coroot $\theta^\vee$ is 2. Substituting into \eqref{anom4}, this yields the modular anomaly of the elliptic genus $Z_\beta(\tau \, | \, \epp,\epm,\mass)$:
\begin{equation}
\begin{aligned}
f(\epp,\epm,\mass)& = \,  - \frac{\epm^2}{2}  \left((2 - A_{JJ})\beta^J + A_{IJ} \beta^I \beta^J\right) \\
& \quad - \frac{\epp^2}{2} \left( \left((2 - A_{JJ}) + 2 h^\vee_{G_J}\right) \beta^J - A_{IJ} \beta^I \beta^J \right) \\
& \quad - \beta^J A_{Ja} \frac{1}{2} (\mass_a,\mass_a)_{\mathfrak{g}_a} \,.
\end{aligned}
\end{equation}
The modular anomaly of the elliptic genus was identified with the index bilinear form $M_\beta$ featuring in the holomorphic anomaly equation \eqref{eq:holanomaly} in \cite{Haghighat:2014vxa,Huang:2015sta,DelZotto:2016pvm,Gu:2017ccq}, thus intimately linking the t'Hooft anomalies of the 6d theory and the holomorphic anomaly of the refined topological string.

We close this section by pointing out a subtlety in the relation between the topological string partition function and the elliptic genera: the chemical potential conjugate to the number of D3 branes wrapped on a specific two-cycle in the base differs from the corresponding K\"ahler parameter by a shift involving the K\"ahler parameters $\tau,\mass$ in the fiber. This can be argued for geometrically in simple examples \cite{Huang:2015sta}, and has been observed to hold generally \cite{Haghighat:2014vxa}. The relation between the elliptic genera \eqref{eq:elliptic} and the 6d $\Omega$-background partition function \eqref{eq:nekrasix} is thus the following 
\begin{equation}\label{eq:nekrasix}
Z(\epi,\epii,\bt) = Z_0 \left(1+ \sum_{\beta \in \Gamma} \bQ_c^\beta \md e[h_{\beta}(\tau,\mass)] Z_\beta(\tau \, | \, \epp,\epm,\mass) \right),
\end{equation}
where $h_{\beta}(\tau,\mass)$ is linear in the K\"ahler parameters $(\tau,\mass)$. Equivalently,  the coefficients of the expansion in equation \eqref{eq:nekrasix} can be viewed as a rescaled version of the elliptic genera $Z_\beta(\tau \, | \, \epp, \epm, \mass)$,
\begin{equation}
\widehat Z_\beta(\tau \, | \, \epp, \epm, \mass) =e[h_{\beta}(\tau,\mass)]  Z_\beta(\tau \, | \, \epp, \epm, \mass).
\end{equation}
We will discuss the precise rescaling factor for minimal SCFTs in section \ref{s:symmetries}.

\subsection{Minimal 6d SCFTs}

The 6d SCFTs and LSTs are organized hierarchically with respect to Higgs branch and tensor branch RG flows. At the very end of a chain of such flows reside the minimal 6d SCFTs. They are of rank one by definition (which situates them at the bottom of a tensor branch flow, one step before trivial theories). Most minimal SCFTs support non-Higgsable gauge groups \cite{Morrison:2012np} --- the two known exceptions are the $A_1$ (2,0) theory (the M--string) and the E--string theory.

In this section, we will describe minimal 6d SCFTs that have an F-theory realization. We consider Calabi-Yau threefolds $X$ that are elliptic fibrations over a base manifold $B$ (which at the end of our considerations we will take to be non-compact), 
\begin{equation*}
\xymatrix{ E \ar[r]&X\ar[d]^\pi\\ 
	&B\\}
\end{equation*}
Whenever the elliptic fibration $\pi$ has a global section, it is convenient to describe $X$ by means of a Weierstrass model.
Let $\mathcal O$ denote a trivial line bundle, and $K_{B}$ the canonical line bundle of the base surface $B$. The Weierstrass model is given by
\begin{equation}
Y^2 = X^3 + f \, X Z^4 + g \,Z^6 \,,
\end{equation}
where $[X:Y:Z]$ are coordinates of the ${\mathbb P}^{1,2,3}$ fibers of the bundle\footnote{\label{conv_proj}  We follow the convention that $\IP({\cal E})$ is the projectivization of the bundle $-{\cal E}$.}
\begin{equation}			
{\mathbb P}^{2,3,1}(2 K_{B} \oplus  3 K_{B} \oplus \cO)\,,
\end{equation}
and $f$ and $g$ are sections of $-4 K_{B}$ and $-6 K_{B}$ respectively. 
The discriminant of $\pi$ is
\begin{equation}
\Delta = 4 f ^3 + 27 g^2.
\end{equation}
To have a consistent F-theory model, the order of vanishing of $(f,g,\Delta)$ in codimension one must be $< (4,6,12)$. 
As a consequence, the allowed singularities in codimension one must be of Kodaira type. Higher orders of vanishing are allowed in codimension two, 
signaling the presence of tensionless strings.
			
\begin{table}
\centering
\begin{tabular}{|c|ccccccccc|}
\hline
$-C\cdot C$& 1 &2&3&4&5&6&7&8&12\phantom{$\Big|$}\\ 
\hline
fiber & $I_0$ & $I_0$  & $IV$ & $I_0^*$ & $IV^*_{ns}$ & $IV^*$ & $III^*$ & $III^*$ & $II^*$\phantom{$\Big|$}\\ 
$\mathfrak{g}_{min}$ & none & none &$\mathfrak{su}_3$&$\mathfrak{so}_8$&$\mathfrak{f}_4$&$\mathfrak{e}_6$&$\mathfrak{e}_7\oplus \tfrac{1}{2}\mathbf{56}$&$\mathfrak{e}_7$&$\mathfrak{e}_8$\phantom{$\Big|$}\\
\hline
\end{tabular}
\caption{ Minimal gauge groups for the 6d theories of rank 1. The first two entries of the table correspond respectively to the E-string and to the M-string}\label{non-higgsable}
\end{table}

By definition, minimal 6d SCFTs have rank one, i.e. the base surface $B$ contains a single holomorphic curve $C\simeq {\mathbb P}^1$ (in the enumeration of curves $C_I \subset B$ discussed in section \ref{sc:6dNek}, this is the $r=1$ case). The restriction on the order of vanishing of $(f,g,\Delta)$ entails that
\begin{equation}
1\leq - C \cdot C \leq 12.
\end{equation}
The case $- C\cdot C = 1$ corresponds to the E-string theory, the only minimal 6d SCFT with a non-trivial flavor symmetry ($f\neq 0$ in the enumeration of curves $C_I \subset B$, $f=0$ for all other cases). The curves $C$ with $ 8 < - C \cdot C < 12$ have $12 + C \cdot C$  points where $(f,g,\Delta)$ vanish at order $(4,6,12)$. These must be blown-up to describe the full tensor branch of the model. These models are hence not minimal. Whenever $-C\cdot C \geq 3$, the fibration is forced to degenerate along $C$, and the corresponding models must have non-higgsable gauge groups (see table \ref{non-higgsable}).
		
Aside from the base surface $B$, the fibration $X$ also contains an elliptic surface with base the curve $\IC$. The exceptional curves $E_i$ resolving the singularities in the fiber at $k=3,4,5,6,8,12$ are rational curves which intersect within this elliptic surface according to the Cartan matrix of the Lie algebras denoted by $\mathfrak{g}_{min}$ in table \ref{non-higgsable}. The fiber contains one additional curve class $F_0$, which intersects the base curve $C$ as $C\cdot F_0 = 1$ and intersects all exceptional curves $E_i$ according to the negative Cartan matrix of the affine Lie algebra associated to $\mathfrak{g}_{min}$, with $F_0$ identified as the zeroth node. The multiplicity of the fiber components $E_i$ are given by the corresponding comarks $a_i^\vee$,\footnote{  Note that only in the case of $k=5$ is $\mathfrak{g}_{min}=\mathfrak{f}_4$ not simply laced, requiring a distinction between marks and comarks; see \cite{Morrison:2012np} for a discussion of this case --- more details can be found in \cite{Haghighat:2014vxa,DelZotto:2017pti,Esole:2017qeh}.} such that
\be \label{fiber}
F = F_0 + \sum_i a_i^\vee E_i \,,
\ee
consistent with the fact that $F\cdot F=0$: mapping $F_0 \rightarrow \alpha_0^\vee$ and $E_i \rightarrow \alpha_i^\vee$, the sum over $i$ in \eqref{fiber} maps to the highest root $\theta$ by definition of comarks, hence $F$ maps to the imaginary root $\delta$; by $(\delta,\delta)=0$, this indeed implies that $F$ has vanishing self-intersection.
		
\subsection{Symmetries of the elliptic genus of strings in minimal SCFTs} \label{s:symmetries}
Recall that by conjugation invariance of the trace underlying the definition of the elliptic genus, we lose no information by evaluating it on an element $\inH$ of the Cartan $H$ of $G$, with $\mass \in \fh_{\mb C}$. By choosing to conjugate into a Cartan, we fix some of the conjugation invariance of the trace, but maintain the freedom to conjugate by elements of the normalizer of $H$ in $G$, $\cN_G(H)$, which map $H$ to itself under conjugation. As elements of the centralizer of $H$ in $G$, $\cC_G(H)$, fix $H$ pointwise under conjugation, we can identify the symmetry group acting on the exponentiated fugacity $\inH$ of the elliptic genus with 
\be
W(G) = \cN_G(H) / \cC_G(H) \,,
\ee
the Weyl group of the flavor group $G$. This group acts on $\mass$ via reflections $\sigma_{\alpha}$ with reflecting hyperplane orthogonal to roots $\alpha \in \Phi$ of $\mathfrak{g}$, yielding the realization of the Weyl group at the level of the Lie algebra.

A choice of basis for $\fh_{\mb C}$ is given by the set of fundamental weights $\{\omega_i\}$, $i=1,\ldots,r$, of $\mathfrak{g}$, the duals to the simple coroots $\alpha_i^\vee$,
\be
(\omega_i, \alpha_j^\vee) = \delta_{ij} \ .
\ee
We expand $\mass = m_i \omega_i$, i.e.
\begin{equation}\label{eq:mQi}
m_i = (\mass,\alpha_i^\vee)  \ ,
\end{equation}
and identify $Q_i = \md e[m_i]$ with the K\"ahler parameters of the exceptional curves $E_i$ which arise upon resolution of the singularities in the elliptic fiber.

In the case of minimal SCFT models, the identification of the 6d partition function with the refined topological string partition function takes the explicit form \cite{Haghighat:2014vxa,DelZotto:2016pvm}  
\begin{equation}\label{eq:free-energy}
	Z_{top} = Z_0(\tau,\epsilon_+,\epsilon_-,\mass)\cdot\(1+ \sum_{k=1}^\infty 
	Q_b^k \(\frac{q^{1/2}}{\prod_{i=1}^r Q_i^{a_i^\vee}}\)^{kh^\vee_G/3} Z_k(\tau,\epsilon_+,\epsilon_-,\mass)  \) \,.
\end{equation}
Here, $Q_b$ is the exponentiated K\"ahler modulus of the rational base curve denoted $C$ above, and $q=\md e[\tau]$, with $\tau$ the volume of the resolved fiber. Denoting the volume of the fiber components $F_0$ by $m_0$, the volume of the resolved fiber equals
\begin{equation}\label{eq:tauaff}
\tau = m_0 +\sum_{i=1}^r a_i^\vee m_i \ .
\end{equation}
Equivalently
\begin{equation}\label{eq:q-decomp}
q = Q_0\prod_{i=1}^r Q_i^{a_i^\vee} \ .
\end{equation}

Note that the contribution $Z_0(\tau,\epsilon_+,\epsilon_-,\mass)$ from zero base wrapping is unrelated to the elliptic genera of the strings; rather, it arises from KK reduction of the fields of the 6d SCFT.

By the identification with the topological string, the product
\begin{equation}\label{eq:inv-Zk}
	\Ztop_k:=\(\frac{q^{1/2}}{\prod_{i=1}^r Q_i^{a_i^\vee}}\)^{kh^\vee_G/3} Z_k(\tau,\epsilon_+,\epsilon_-,\mass)
\end{equation}
should have a formal power series expansion in the K\"ahler moduli $Q_i$ ($i=0,1,\ldots,r$); neither negative nor fractional powers of $Q_i$ may occur. The latter property implies in particular that $\widehat{Z}_k$ must be invariant under integer shifts of $m_i$. By \eqref{eq:mQi}, this corresponds to invariance under shifts of $\mass$ by elements of the weight lattice, the span of $\{\omega_i \}$. Furthermore, transformations acting on the fiber components exist which leave their intersection matrix invariant. These transformations coincide with the symmetry group $D(\widehat{\fg})$ of the Dynkin diagram of $\widehat{\fg}$ arising in the Kodaira classification of the fiber. Given the invariance of the intersection matrix, it is natural from the topological string point of view that $D(\widehat{\fg})$ be a symmetry of $\widehat{Z}_k$. This is confirmed via the following field theory considerations.

For the minimal SCFTs that we are studying, the BPS strings are instantons for the corresponding gauge groups. In particular, their worldsheet theories are sigma models on the instanton moduli spaces. The elliptic genus for such models is obtained as a partition function on $S^2 \times T^2$ of the 4d $\mathcal N=2$ $H^{(k)}_G$ theories discussed in \cite{DelZotto:2016pvm}. From this perspective, the discrete $D(\widehat\fg)$ symmetry can be understood from the realization of the $H^{(k)}_G$ theories as class $\mathcal{S}$ theories. Consider for instance the $G=D_4$ theories. These arise as class $\mathcal{S}[A_k]$ theories on a sphere with four identical punctures, each corresponding to an $SU(2)$ factor of the flavor symmetry. The symmetry $D(\,\widehat{\mathfrak{d}}_4)$ permutes these punctures, and is ultimately responsible for the enhancement of the flavor symmetry group from $SU(2)^4$ to $SO(8)=D_4$. As a second example, consider the models with $G=E_6$. These arise as theories of class $\mathcal{S}[A_k]$ on a sphere with three identical punctures, each corresponding to an $SU(3)$ factor of the flavor symmetry. The symmetry $D(\,\widehat{\mathfrak{e}}_6)$ permutes the $SU(3)$ punctures and is responsible for the enhancement of the flavor symmetry from $SU(3)^3$ to $E_6$. Note incidentally that all the results obtained in this paper for the elliptic genera of BPS strings have this 4d field theoretical counterpart: they yield partition functions for the 4d $\mathcal N=2$ theories $H^{(k)}_G$ on $S^2 \times T^2$.

A different field theoretical argument for reaching the same conclusions arises from the reduction of the 6d (1,0) minimal SCFTs on $S^1$, discussed in \cite{DelZotto:2015rca,DelZotto:2017pti}. The corresponding 5d gauge theories have UV completions in 6d, and are obtained by coupling together three copies of the $\hat D_{{\mathbb C}^3/{\mathbb Z}_3}$ 5d SCFT (resp. four copies of the $\hat D_2$ theory, three copies of the $\hat D_3$ theory,  and two copies of the $\hat D_4$ 5d theory plus one copy of the $\hat D_2$ model) for the $SU(3)$ (resp. $SO(8)$, $E_6$, and $E_7$) cases. It is clear that the corresponding 5d $\Omega$-background partition functions enjoy a discrete symmetry arising from permuting the various identical 5d superconformal systems. Not surprisingly, this is precisely $D(\,\widehat{\mathfrak{a}}_2)$ (resp. $D(\,\widehat{\mathfrak{d}}_4)$, $D(\,\widehat{\mathfrak{e}}_6)$, and $D(\,\widehat{\mathfrak{e}}_7)$).

\section{The toric geometry of elliptic Calabi-Yau threefolds with base $\IF_n$} \label{s:toric_geometry}
In this section, we will discuss in detail the toric realization of elliptically fibered Calabi-Yau manifolds over Hirzebruch surfaces $\IF_n$. This class of geometries was identified as relevant for F-theory compactifications in the very first papers on the subject \cite{Vafa:1996xn,Morrison:1996na,Morrison:1996pp} and has been well-studied since. Our motivation for nevertheless providing an in-depth discussion of the two cases of interest in this paper, $n=3$ and $n=4$, is two-fold. We will determine within the formalism of toric geometry how the exceptional cycles resolving singularities of the elliptic fiber are fibered over the base curve $\IP^1$ of the surface $\IF_n$. As recently discussed in \cite{DelZotto:2017pti}, this data plays an important role in the discussion of the 5d theories obtained by circle compactification of the 6d theories discussed in section \ref{s:top_strings_and_6d}. The 5d/6d relation will feature in our discussion in section \ref{s:additional}. Also, applying the formalism of mirror symmetry to such geometries requires blowing up beyond the point required to simply resolve all singularities, and to pass to a flopped phase in which the fibration structure is no longer manifest. This analysis was first performed in \cite{Haghighat:2014vxa}. We clarify this construction here.

\subsection{Elliptic fibrations over Hirzebruch surfaces}
	We want to construct an elliptic fibration over the Hirzebruch surface $\IF_n$ as a hypersurface in a weighted projective bundle over $\IF_n$. In order for the hypersurface to be Calabi-Yau and the generic fiber over $\IF_n$ an elliptic curve, we choose the weighted projective bundle to be (see footnote \ref{conv_proj} for our conventions)
	\be
	\IP^{2,3,1}(2 \KB \oplus 3 \KB \oplus \cO) \,.  \label{wpbundle}
	\ee
	Here, $\KB$ denotes the canonical line bundle of $\IF_n$, and $\IP^{2,3,1}$ indicates the weighted projectivization of the bundle with the indicated weights. We will explicitly see how this choice of geometry leads to an elliptic fibration with trivial canonical class in the toric realization of this geometry.
	
\begin{figure}[h]
		\centering
		\def\svgwidth{7cm}
		\includegraphics[width=0.4\linewidth]{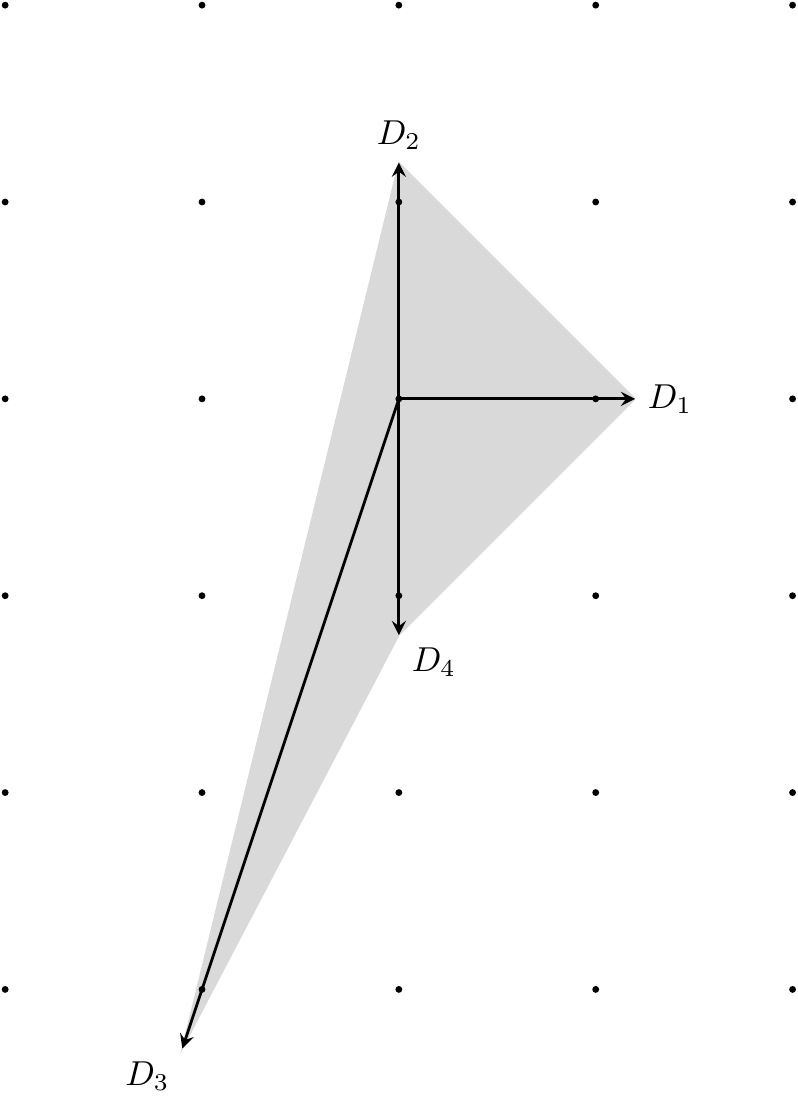}
		\caption{Toric fan for $\IF_3$.}
	\label{toricF3}
	
	\end{figure}

	The toric fan of $\IF_n$, for the example $n=3$, is depicted in figure~\ref{toricF3}. The toric fan for general $n$ is obtained by shifting the endpoint of the ray $D_3$ in figure~\ref{toricF3} from the point $(-1,-3)$ to $(-1,-n)$.

	 Eventually, we will decompactify the geometry by replacing the base surface $\IF_n$ by the total space of the line bundle $\cO_{\IP^1}(-n)$. In terms of the toric data, this decompactification will proceed by omitting the ray corresponding to $D_2$. 
	 
	 The rays of the fan underlying the total space of the vector bundle
	\be
	(-2\KB) \oplus (-3\KB) \oplus \cO  \rightarrow \IF_n \label{bundle_pre_p}
	\ee
	are given in table \ref{fig:before_proj}.
	\begin{table}[h!]
		\centering
	\begin{tabular}{c|*{5}{>{$}c<{$}}}
		$\rho_x$	&	1	&	0	&	0	&	0	&	0 \\
		$\rho_y$	&	0	&	1	&	0	&	0 	&	0 \\
		$\rho_z$	&	0	&	0	&	1	&	0	&	0 \\
		$\rho_s$	&	-2	&	-3	&	0	&	0	&	-1 \\
		$\rho_t$ &   -2  &   -3  &   0   &   0   &   1 \\
		$\rho_u$	&	-2	&	-3	&	0	&	-1	&	-n \\
		$\rho_v$	&	-2	&	-3	&	0	&	1	&	0 \\
	\end{tabular}
	\caption{Toric data for $((-2\KB) \oplus (-3\KB) \oplus \cO) \rightarrow \IF_n$.\label{fig:before_proj}}
\end{table}

	The projectivization is achieved by modding out via $2\rho_x+3\rho_y+\rho_z =0$. This yields the collection of cones given in table \ref{tdwpbundle}.
	
	\begin{table}[h]
		\centering
	\begin{tabular}{c|*{4}{>{$}c<{$}}|*{3}{>{$}c<{$}}}
			&		&		&		&		&(\IC^*)_1	&	(\IC^*)_2	&	(\IC^*)_3\\
		\hline
		$\rho_x$	&	1	&	0	&	0	&	0	&	2	&	0	&	0\\
		$\rho_y$	&	0	&	1	&	0	&	0	&	3	&	0	&	0\\
		$\rho_z$	&	-2	&	-3	&	0	&	0	&	1	&	n-2	&	-2\\
		$\rho_s$	&	-2	&	-3	&	0	&	-1	&	0	&	-n	&	1\\
		$\rho_t$ 	&   -2  &   -3  &   0   &   1 	&	0	&	0	&	1\\
		$\rho_u$	&	-2	&	-3	&	-1	&	-n	&	0	&	1	&	0\\
		$\rho_v$	&	-2	&	-3	&	1	&	0	&	0	&	1	&	0\\
	\end{tabular}
	\caption{Toric data for $\IP^{2,3,1}(2\KB \oplus 3\KB \oplus \cO) \rightarrow \IF_n$.} \label{tdwpbundle}
	\end{table}
	
	The maximal cones of this geometry, constituting the fan $\Sigma_n$, are generated by all possible combinations of two generators from the set $\{ \rho_x,\rho_y,\rho_z\}$, one from the set $\{\rho_u,\rho_v\}$, and one from the set $\{\rho_s,\rho_t\}$. In the following, we will drop the subscript ${}_n$ in the notation whenever no confusion can arise due to this omission.
	 
	Table \ref{tdwpbundle} also lists the three $\IC^*$ actions on the total coordinate ring of this geometry, allowing us to represent it as the quotient
	\be
	X_\Sigma = ( \IC^{\Sigma(1)}- Z_\Sigma ) / {(\IC^*)_1 \times (\IC^*)_2 \times (\IC^*)_3} \,.
	\ee
	The notation $\Sigma(1)$ indicates the set of all 1-cones of the fan $\Sigma$. We have already indexed the 1-cones $\rho_i$ with the corresponding variable among $(x,y,z,s,t,u,v) \in \IC^{\Sigma(1)}$. The $(\IC^*)_1$ quotient e.g. identifies $\{(x,y,z,s,t,u,v) \sim (\lambda^2 x,\lambda^3 y,\lambda z,s,t,u,v) | \lambda \in \IC^*\}$.

	$Z_\Sigma$ is the exceptional locus: the simultaneous vanishing locus of the variables $\{x_i\}$ is excluded from the geometry if no cone $\sigma \in \Sigma$ exists that contains all of the corresponding rays $\{\rho_{x_i}\}$. Given our construction of the fan, $Z_\Sigma$ can easily be determined to be 
	\be
	Z_\Sigma = \{ x=y=z=0, u=v=0, s=t=0 \} \,.
	\ee
	
	By the adjunction formula, the canonical class of a hypersurface $Y$ defined by a section $s \in H^0(X,D)$ of a holomorphic line bundle $D$ on $X$ is given by\footnote{ We will not distinguish notationally between divisors and the corresponding line bundles.}
	\be
	K_Y \cong (K_X \otimes D) \restrict{Y}  \,.
	\ee
	The anti-canonical hypersurface of any compact complex variety, defined via a section $s \in H^0(X,-K_X)$ of the anti-canonical line bundle $-K_X$, thus has trivial canonical class and describes a Calabi-Yau manifold. From the toric data of $X_\Sigma$ given in table \ref{tdwpbundle}, we can read off that
	\be
	K_{X_{\Sigma}} = - (D_x + D_y + \underbrace{D_z + D_s + D_t + D_u + D_v}_{=:D})  \,,
	\ee 
	where $D = -\KB + D_z$. Furthermore,
	\be
	D_x = 2 D \,, \quad D_y = 3 D \,,
	\ee
	hence
	\be
	K_{X_{\Sigma}} = - 6D \,.
	\ee
	A generic section of the anti-canonical bundle $-K_{X_{\Sigma}}$ will therefore be of second order in $y$ and third order in $x$, thus describing an elliptic curve in these variables. A basis of $H^0(X_\Sigma,-K_{X_{\Sigma}})$ is given by monomials with the same $(\IC^*)_i$ charges, $i=1,2,3$, as $y^2$ and $x^3$.
	
	The hypersurface defined as the zero locus of a section $s \in H^0(X,-K_X)$ will generically be singular. The type of singularity depends on the base $\IF_n$ of the fibration. In the following, we will study the cases $n=3$ and $n=4$.
	
	\subsection{The $A_2$ geometry: a resolution of the elliptic fibration over $\IF_3$}
	We first specialize to the case $n=3$. The generic section $s_{-K} \in H^0(X_{\Sigma_3},-K)$ of the anti-canonical bundle $-K$ of the weighted projective bundle \eqref{wpbundle} at $n=3$ is given by
	\begin{multline} \label{section_K}
	s_{-K} = \\
	\alpha y^2 +  y z^3 s \sum_{n=1}^6 s^{n-1} t^{6-n} f_{3(n-1)}(u,v) + y x z s \sum_{n=1}^2 s^{n-1} t^{2-n} g_{3n-1}(u,v)  +\\ \beta x^3 + x^2 z^2 s \sum_{n=1}^4 s^{n-1} t^{4-n} h_{3n-2}(u,v) +  x z^4 s^2 \sum_{n=2}^8 s^{n-2} t^{8-n} j_{3n-4}(u,v)  + \\z^6 s^2 \sum_{n=2}^{12} s^{n-2} t^{12-n} k_{3n-6}(u,v) \,.
	\end{multline}
	Here, we have denoted by $f_n(u,v), g_n(u,v), h_n(u,v), j_n(u,v),k_n(u,v)$ homogeneous polynomials in $u$, $v$ of degree $n$.
	
	The hypersurface $s_{-K} = 0$ is singular along the curve $x=y=s=0$. To resolve this singularity, we blow up the ambient space. We introduce a new generator $b$ for the total coordinate ring, such that the blow-up divisor lies at $b=0$, and require a $\IC^*$ action $\lambda(b,x,y,s) \sim (\lambda^{-1}b,\lambda x, \lambda y, \lambda s)$, leaving all other generators invariant. This uniquely fixes the ray $\rho_b$ associated to the coordinate $b$ to $\rho_b = (-1,-2,0,-1)$. The corresponding toric data is collected in table \ref{tdblowup}.
	
	\begin{table}[h] 
		\centering
		\begin{tabular}{c|*{4}{>{$}c<{$}}|*{4}{>{$}c<{$}}}
			&		&		&		&		&(\IC^*)_1	&	(\IC^*)_2	&	(\IC^*)_3	& (\IC^*)_4\\
			\hline
			$\rho_x$	&	1	&	0	&	0	&	0	&	2	&	0	&	0	&	1\\
			$\rho_y$	&	0	&	1	&	0	&	0	&	3	&	0	&	0	&	1\\
			$\rho_z$	&	-2	&	-3	&	0	&	0	&	1	&	1	&	-2	&	0\\
			$\rho_s$	&	-2	&	-3	&	0	&	-1	&	0	&	-3	&	1	&	1\\
			$\rho_t$ 	&   -2  &   -3  &   0   &   1 	&	0	&	0	&	1	&	0\\
			$\rho_u$	&	-2	&	-3	&	-1	&	-3	&	0	&	1	&	0	&	0\\
			$\rho_v$	&	-2	&	-3	&	1	&	0	&	0	&	1	&	0	&	0\\
			$\rho_b$	&	-1	&	-2	&	0	&	-1	&	0	&	0	&	0	&	-1\\
		\end{tabular}
	\caption{Toric data for $\IP^{2,3,1}(2K \oplus 3K \oplus \cO) \rightarrow \IF_3$ blown up once.}  \label{tdblowup}
\end{table}

	The maximal cones for this geometry, constituting the fan $\Sigma_B$, are given by replacing the rays $\{\rho_x,\rho_y,\rho_s\}$ in the maximal cones of the fan pre-blow up by all combinations of two rays from this set, together with $\rho_b$. The blown up geometry $X_B$ can be represented as the quotient
	\be
	X_B = (\IC^{\Sigma_B(1)}-Z_B)/G \,,
	\ee
	where we have set $G = (\IC^*)_1 \times	(\IC^*)_2 \times (\IC^*)_3 \times (\IC^*)_4$, with the $\IC^*$ actions $(\IC^*)_i$ specified in table \ref{tdblowup}. The forbidden locus $Z_B$ is now
	\be
	Z_B = \{x=y=z=0,u=v=0,s=t=0,x=y=s=0,z=b=0,t=b=0 \}  
	\ee
	and contains the previously singular locus $\{x=y=s=0\}$.

	The generic section of the anti-canonical bundle $-K_B$ of $X_B$ is
	\begin{multline} \label{section_Kb}
	s_{-K_{B}} = \\
	\alpha y^2 +  y z^3 s \sum_{n=1}^6 (b s)^{n-1} t^{6-n} f_{3(n-1)}(u,v) + b x y z s \sum_{n=1}^2 (b s)^{n-1} t^{2-n} g_{3n-1}(u,v)  +\\ \beta b x^3 + b x^2 z^2 s \sum_{n=1}^4 (b s)^{n-1} t^{4-n} h_{3n-2}(u,v) +  b x z^4 s^2 \sum_{n=2}^8 (b s)^{n-2} t^{8-n} j_{3n-4}(u,v)  + \\z^6 s^2 \sum_{n=2}^{12} (b s)^{n-2} t^{12-n} k_{3n-6}(u,v) \,.
	\end{multline}
	The corresponding zero section is smooth. The exceptional divisor lies at $b=0$. Its intersection with the anti-canonical hypersurface is given by the equation
	\be
	s_{-K_B} |_{b=0} = \alpha\, y^2 + f_0 \,y z^3 s t^5  + k_0 \,z^6 s^2 t^{10}  = 0 \,.
	\ee
	To study this equation, we consider it patchwise on the hypersurface $b=0$ in $X_B$. Recall that each 4-cone $\sigma$ determines a patch $U_\sigma$ in which all but the variables corresponding to the generators of $\sigma$ are set to 1. The patches intersecting the hypersurface $b=0$ non-trivially are hence associated to 4-cones which contain $\rho_b$. $\Sigma_B$ contains six such 4-cones,
	\begin{align}
	\sigma_{vx} &= \langle \rho_u, \rho_s, \rho_y, \rho_b \rangle \,, \quad \sigma_{ux} = \langle \rho_v, \rho_s, \rho_y, \rho_b \rangle \,,\nn\\
	\sigma_{vy} &= \langle \rho_u, \rho_s, \rho_x, \rho_b \rangle \,, \quad \sigma_{uy} = \langle \rho_v, \rho_s, \rho_x, \rho_b \rangle \,,\nn\\
	\sigma_{vs} &= \langle \rho_u, \rho_x, \rho_y, \rho_b \rangle \,, \quad \sigma_{us} = \langle \rho_v, \rho_x, \rho_y, \rho_b \rangle \,.
	\end{align}
	Note that none of these 4-cones contains either $t$ or $z$. We can hence set $t=z=1$ in all of the corresponding patches. The subscripts on the 4-cones indicate the additional coordinates, aside from $z$ and $t$, which are set to one in the corresponding patch.

	The intersection of the exceptional divisor $b=0$ with the anti-canonical hypersurface in all of these patches  is given by the two curves
	\be  \label{exceptional_locus}
	y = \frac{-f_0  \pm \sqrt{f_0^2 - 4 \alpha k_0}}{2} z^3 t^5 s \quad |_{z=t=1}
	\ee
	fibered over a base $\IP^1$ coordinatized by the homogeneous variables $u$ and $v$. We can describe these surfaces as quotients of the $\IC^4$ parametrized by the coordinates $(u,v,x,y)$, with $s$ determined by \eqref{exceptional_locus}, $b=0$, and $t=z=1$. To this end, we need to determine the $\IC^*$ actions that leave the condition $t=z=1$ fixed. These are listed in table \ref{chargesa2x}, in which we have isolated the action on $(u,v,x,y)$ in the right most columns.
	\begin{table}[h]
		\centering
		\begin{tabular}{c|*{8}{>{$}c<{$}} | *{4}{>{$}c<{$}}}
									& x	&	y &	z & s &	t	& 	u &	 v 	&	b	&	u &	v &	 x & 	y		\\				\hline
			$-Q_1 + Q_2 + 2 Q_4$ 	&	0	&	-1	&	0	&	-1	&	0	&	1	&	1	&	-2	&	1	&	1	&	0	&	-1	\\
			$Q_4$					&	1	&	1	&	0	&	1	&	0	&	0	&	0	&	-1	&	0	&	0	&	1	&	1		
		\end{tabular}
		\caption{$(\IC^*)$ actions on exceptional locus $b=0$.} \label{chargesa2x}
	\end{table}
	We thus obtain a description of each of the two surfaces as the quotient
	\be \label{f1_as_quotient}
	\left(\IC^4 - \{u=v=0, x=y=0\} \right) / \{(u,v,x,y) \sim (\lambda u, \lambda v, x, \lambda^{-1}y), (u,v,x,y) \sim (u,v, \mu x, \mu y) \}  \,,
	\ee
	allowing us to identify each with the Hirzebruch surface $\IF_1$.
	
	A third Hirzebruch surface $\IF_1$ is given by the intersection of the hypersurface $s=0$ with the zero locus of \eqref{section_Kb}, given by the equation
	\be
	s_{-K_B} |_{s=0} = \alpha\, y^2 + \beta \,b x^3  = 0 \,.
	\ee
	As no 4-cone of the fan underlying $X_B$ contains both $\rho_s$ and $\rho_t$, we may set $t=1$.
	Note that nowhere on this zero locus may $x=0$, as this would imply $y=0$, but $x=y=s=0$ lies in the forbidden locus $Z_B$. We may hence solve this equation for $b$,
	\be \label{solve_for_b}
	b = - \frac{\alpha}{\beta} \frac{y^2}{x^3} \,.
	\ee
	In patches with $z=1$, we now proceed as above to conclude that the intersection with $s=0$ is given by \eqref{f1_as_quotient}, however without the curve lying at $x=0$. This missing piece of the $\IF_1$ geometry lies in patches with $z=0$. In these, we can set $b=1$ and either $x=1$ or $y=1$. The first choice fixes the action of $(\IC^*)_1$ on $(x,y,z)$ up to the action $y \mapsto -y$, the second up to $x \mapsto \zeta_3 x$, with $\zeta_3$ a third root of unity. In either case, the equation \eqref{solve_for_b} has a unique solution, and the unconstrained variables $(u,v)$ provide the missing curve of the $\IF_1$ geometry.

	The mirror symmetry formalism developed in \cite{Batyrev:1994hm,Hosono:1993qy,Hosono:1994ax} (see \cite{Hosono:1994av} for a review) applies to Calabi-Yau hypersurfaces (or complete intersections) in toric ambient spaces whose anti-canonical bundle is semi-ample. We will call such varieties semi-Fano. In particular, the anti-canonical bundle must be NEF: the intersection product of all curve classes with the anti-canonical divisor must be non-negative. This is not the case for the geometry $X_{B}$ that we have constructed. Its Mori cone is generated by the rays displayed in table \ref{mori_cone_bp}.

	\begin{table}[h]
		\centering
		\begin{tabular}{c|*{4}{>{$}c<{$}}}
						& C_1	&	C_2	&	C_3	& C_4	\\
			\hline
			$D_x$	&	-1	&	0	&	0	&	1\\
			$D_y$	&	0	&	-1	&	0	&	1\\
			$D_z$	&	1	&	0	&	-2	&	0\\
			$D_s$	&	-3	&	-1	&	1	&	1\\
			$D_t$ 	&	0	&	0	&	1	&	0\\
			$D_u$	&	0	&	1	&	0	&	0\\
			$D_v$	&	0	&	1	&	0	&	0\\
			$D_b$	&	3	&	-2	&	0	&	-1\\  
			$-K$	&	0	&	-2	&	0	&	2\\
		\end{tabular}
		\caption{The generators of the Mori cone of $\IP^{2,3,1}(2K \oplus 3K \oplus \cO) \rightarrow \IF_3$ blown up once. } \label{mori_cone_bp}
	\end{table}
	Recall that the entries of the generators equal the intersection product $C_i \cdot D_\rho$ for $C_i$ a set of curves which represent a basis of the second homology of $X_B$. The occurrence of a negative entry in the bottom row of the table, which gives the intersection numbers of the anti-canonical divisor $-K_B$ with the $C_i$, implies that $X_B$ is not semi-fano. We can correct for this by performing a sequence of flops in the geometry. The enumerative invariants of the Calabi-Yau hypersurface will be reshuffled by these flops, but we will in any case report these with regard to the basis of curves on which $D(\mfag)$ acts as a symmetry. These numbers are invariant under flops (as the basis is reshuffled upon each flop).
	
	The fan underlying the flopped geometry retains the 1-cones $\Sigma_B(1)$, yet differs in how these combine to higher dimensional cones. The following standard construction yields fans with $\Sigma_{flopped}(1) = \Sigma_B(1)$ such that the corresponding toric variety is semi-fano. Consider a fan $\Sigma \subset N_{\IR}$ with 1-cones $\rho \in \Sigma(1)$ represented by lattice vectors $u_\rho \in N$ in a lattice $N$. For a torus invariant divisor $D = \sum_{\rho \in \Sigma(1)} a_\rho D_\rho$ of $X_\Sigma$, we consider the polyhedron in the dual lattice $M$ defined by
	\be
	P_D = \{m \in M_{\IR} | \langle m, u_{\rho} \rangle \ge -a_{\rho} \quad \forall \rho \in \Sigma(1) \} \,, \label{PD}
	\ee
	with $\langle \cdot,\cdot \rangle$ denoting the natural pairing between lattice and dual lattice. We will assume that $\{u_\rho|\rho \in \Sigma(1)\}$ span $N_{\IR}$, such that the polyhedron (an intersection of half-spaces) is a polytope (the convex hull of a finite number of points). This polytope defines a normal fan $\Sigma_{P_D}$. As $\Sigma_{P_D}(1) = \Sigma(1)$, $D$ also defines a divisor on $X_{\Sigma_{P_D}}$, guaranteed to be ample. However, $X_{\Sigma_{P_D}}$ is generically singular. Refining $\Sigma_{P_D}$ smoothens the geometry. In the process, $D$ may become semi-ample. The process of refining the fan can be conveniently recast when the divisor of interest is the anti-canonical bundle, $D = -K = \sum_{\rho \in \Sigma(1)}D_{\rho}$. In this case, the polytope \eqref{PD} is reflexive. We can hence consider its polar dual $(P_D)^{*}$, which is also reflexive. As such, it has a unique interior point $p$. The face fan of 
	$(P_D)^{*}$, defined as the fan whose top dimensional cones are generated by the rays connecting $p$ to the vertices of the faces of $(P_D)^{*}$, coincides with the normal fan of $P_D$. Refinements of $\Sigma_{P_D}$ are obtained as star triangulations of $(P_D)^{*}$ with regard to $p$.
	
	We now apply this formalism to the geometry $X_B$. The rays $\{\rho_x, \rho_y, \rho_u, \rho_t, \rho_v\} \subset \Sigma_B(1)$ listed in table \ref{tdblowup} yield a polygon $\Delta^* \subset N$ as their convex hull. We identify $\Delta^*$ with the polygon $(P_{-K})^*$ introduced above. $\Delta^*$ contains the following lattice points of $N$:
	\begin{itemize}
		\item Vertices of $\Delta^*$: endpoints of $\rho_x, \rho_y, \rho_u, \rho_t, \rho_v$.
		\item Points lying on codim 1 faces of $\Delta^*$: $(-1,-1,0,0), (-1,-2,0,0), (0,-1,0,0)$.
		\item Points lying on codim 2 faces of $\Delta^*$: endpoints of $\rho_z, \rho_s, \rho_b$.
	\end{itemize}
	The points lying on codimension 1 faces of $\Delta^*$ yield the divisors that resolve the singularities that arise when taking the weighted (vs. the ordinary) projectivization of the bundle \eqref{bundle_pre_p}. They do not intersect the hypersurface. Including all points lying on codimension 2 faces as generators of the fan  resolves the singularities of the hypersurface. Indeed, alongside the rays $\rho_z, \rho_s$, this set includes $\rho_b$, the ray corresponding to the blowup divisor that we introduced by hand above.

	There is a final point which needs to be addressed. Note that the two copies \eqref{exceptional_locus} of the Hirzebruch surface lie on the intersection of the canonical hypersurface with the divisor $D_{\rho_b}$. We can thus not vary their K\"ahler volumes independently by varying the volume of the divisor. To remedy this, we blow up the intersection point of the two surfaces at $b=y=0$. This is achieved by introducing an addition ray $\rho_{b'}$ into our geometry, corresponding to a new coordinate $b'$ in the total coordinate ring of the geometry, accompanied by a $\IC^*$ action $\lambda (b',y,b) = (\lambda^{-1} b', \lambda y, \lambda b)$. This fixes $\rho_b = (-1,-1,0,-1)$. The toric data describing this twice blown up geometry is collected in table \ref{td_twice}.
	\begin{table}[h]
		\centering
		\begin{tabular}{c|*{4}{>{$}c<{$}}|*{5}{>{$}c<{$}}}
			&		&		&		&		& (\IC^*)_1	&	(\IC^*)_2	&	(\IC^*)_3	& (\IC^*)_4 & (\IC^*)_5\\
			\hline
			$\rho_x$	&	1	&	0	&	0	&	0	&	2	&	0	&	0	&	1	&	0\\
			$\rho_y$	&	0	&	1	&	0	&	0	&	3	&	0	&	0	&	1	&	1\\
			$\rho_z$	&	-2	&	-3	&	0	&	0	&	1	&	1	&	-2	&	0	&	0\\
			$\rho_s$	&	-2	&	-3	&	0	&	-1	&	0	&	-3	&	1	&	1	&	0\\
			$\rho_t$ 	&   -2  &   -3  &   0   &   1 	&	0	&	0	&	1	&	0	&	0\\
			$\rho_u$	&	-2	&	-3	&	-1	&	-3	&	0	&	1	&	0	&	0	&	0\\
			$\rho_v$	&	-2	&	-3	&	1	&	0	&	0	&	1	&	0	&	0	&	0\\
			$\rho_b$	&	-1	&	-2	&	0	&	-1	&	0	&	0	&	0	&	-1	&	0\\
			$\rho_b'$	&	-1	&	-1	&	0	&	-1	&	0	&	0	&	0	&	0	&	-1\\
		\end{tabular}
		\caption{Toric data for $\IP^{2,3,1}(2K \oplus 3K \oplus \cO) \rightarrow \IF_3$ blown up twice. } \label{td_twice}
	\end{table}
	We call the resulting geometry $X_{B'}$. The generic section of the anti-canonical bundle $-K_{B'}$ of $X_{B'}$ has leading term
	\be
	s_{-K_{B'}} = \alpha_0\, y^2 b' + \ldots
	\ee
	in $y$. Given the action of $(\IC^*)_5$ that can be read off of table \ref{td_twice}, sections of $-K_{B}$ as given in \eqref{section_Kb} behave as follows with regard to lifts to those of $-K_{B'}$:
	\begin{itemize}
		\item Sections with at least one power of $y$ map to 
		\be
		s = y f(x,y,z,s,u,v,b) \rightarrow y f(x,b' y,z,s,u,v,b' b) \,.
		\ee
		\item Sections with at least one power of $b$ map to 
		\be
		s = b g(x,y,z,s,u,v,b) \rightarrow b g(x,b' y,z,s,u,v,b' b) \,.
		\ee
		\item Sections independent of both $y$ and $b$ do not lift.	
	\end{itemize}
	The intersection of the anti-canonical hypersurface with the divisor $D_{b}$ at $b=0$ is thus given by
	\be
	s_{-K_{B'}} |_{b=0} = \alpha\, b' y^2 + f_0 \,y z^3 s t^5  = 0 \,.
	\ee
	As $b=y=0$ now lies in the forbidden locus, this locus is parametrized by $(u,v,x,y)$ at $t=z=1$, and
	\be
	b'y = \frac{f_0}{\alpha_0} z^3 t^5 s \,.
	\ee
	Invoking $(\IC^*)_5$ to fix either $b'$ or $y$ on the LHS of the above equation, the remaining four $\IC^*$ actions determine the geometry of the intersection of the hypersurface with $D_b$ to be a single copy of $\IF_1$.
	
	By the above, at $b'=0$, $s_{-K_{B'}} |_{b'=0} = y f + b g$, with both $f$ and $g$ independent of $(y,b)$. The intersection of $D_{\rho_{b'}}$ with $s_{-K_{B'}} = 0$ yields the second $\IF_1$ by the same type of reasoning as above.

	\subsection{The $D_4$ geometry: a resolution of the elliptic fibration over $\IF_4$}
	We now discuss the case $n=4$ in somewhat less detail. The generic section $s_{-K} \in H^0(X_{\Sigma_4},-K)$ of the anti-canonical bundle $-K$ of the weighted projective bundle \eqref{wpbundle} at $n=4$ again has the form
	\be \label{section_K4}
	s_{-K} = \alpha y^2 + \beta x^3 + s f(x,y,z,s,t,u,v) \,,
	\ee
	thus exhibiting a singularity at $x=y=s=0$. Blowing up this locus as above yields the geometry $X_B$ with generic anti-canonical section 
	\be \label{section_K4B}
	s_{-K_B} = \alpha \, y^2 + \beta \, b \Big[ \prod_{i=1}^3 (x - \alpha_i z^2 s t^3 ) + y \,g(x,z,s,t,u,v,b) + b \,h(x,z,s,t,u,v,b) \Big] \,.
	\ee
	The corresponding zero locus exhibits three singular points, lying at the intersection of $s_{-K_B} = 0$ with the locus $\{y=b=0, x = \alpha_i z^2 s t^3\}$, for $i=1,2,3$.
	In contradistinction to the case $n=3$ studied in the previous subsection, one blow up hence does not suffice here to remove the singularity of the anti-canonical hypersurface. Blowing up the locus $\{y=b=0\}$ yields
	\be \label{section_K4BB}
	s_{-K_{B'}} = \alpha \, b' y^2 + \beta \, b \Big[ \prod_{i=1}^3 (x - \alpha_i z^2 s t^3 ) + b' \big[y \,g(x,z,s,t,u,v,b' b) + b \,h(x,z,s,t,u,v,b'b) \big]\Big] \,.
	\ee
	This geometry is smooth. Four surfaces lie on the exceptional locus $b'=0$.
	\be \label{skbbx}
	s_{-K_B} \restrict{b'=0} = \beta \, b \Big[ \prod_{i=1}^3 (x - \alpha_i z^2 s t^3 ) \Big] \,.
	\ee
	At $b\neq 0$, the zero locus of this section determines three surfaces parametrized by $(u,v,y,b)$ at $x= \alpha_i z^2 s t^3$, with the  $\IC^*$ actions listed in table \ref{chargesbzero}.
	\begin{table}[h]
		\centering
		\begin{tabular}{c|*{9}{>{$}c<{$}} | *{4}{>{$}c<{$}}}
							& x	&	y &	z & s &	t	& 	u &	v	&	b	&	b'	& u 	&	v &	y	& 	b		\\				\hline
	$-2 Q_1 + Q_2 + 4 Q_4$ 	&	0	&	2	&	0	&	0	&	0	&	1	&	1	&	0	&	-4	&	1	&	1	&	2	&	0	\\
	$Q_4$					&	1	&	1	&	0	&	1	&	0	&	0	&	0	&	-1	&	0	&		&		&		&	\\
	$Q_5$					&	0	&	1	&	0	&	0	&	0	&	0	&	0	&	1	&	-1	&	0	&	0	&	1	&	1
	\end{tabular}
	\caption{$(\IC^*)$ actions on exceptional locus $b'=0$ at $b \neq 0$.} \label{chargesbzero}
	\end{table}

	The action $Q_4$ together with the hypersurface equation \eqref{skbbx} fix $x$ and $s$, leaving the two $(\IC^*)$ actions in the last column of table \ref{chargesbzero}. Together with the locus $(u,v,1,0)$ at $x=s=0$ common to all three surfaces, this determines their geometry to be that of $\IF_2$. Away from $x=s=0$ at $b=0$ lies the surface parametrized by $(u,v,x,s)$, modded out by the $\IC^*$-actions listed in table \ref{chargesyzero}.
		\begin{table}[h]
		\centering
		\begin{tabular}{c|*{9}{>{$}c<{$}} | *{4}{>{$}c<{$}}}
			& x	&	y &	z & s &	t	& 	u &	v	&	b	&	b'	& u 	&	v &	x	& 	s	\\				\hline
			$-2 Q_1 + Q_2 + 4 Q_4 -2 Q_5$	&	0	&	0	&	0	&	0	&	0	&	1	&	1	&	-2	&	-2	&	1	&	1	&	0	&	0	\\
			$Q_4 - Q_5$						&	1	&	0	&	0	&	1	&	0	&	0	&	0	&	-2	&	1	&	0	&	0	&	1	&	1	\\
			$Q_5$							&	0	&	1	&	0	&	0	&	0	&	0	&	0	&	1	&	-1	&		&		&		&	
		\end{tabular}
		\caption{$(\IC^*)$ actions on exceptional locus $b'=0$ at $b=0$.} \label{chargesyzero}
		\end{table}
	We now use the $Q_5$ action to fix $y$. The resulting $\IC^*$ actions reveal the geometry to be the Hirzebruch surface $\IF_0$.	A fourth copy of the Hirzebruch surface $\IF_2$ lies at $s=0$, for which
	\be \label{skbbs}
	s_{-K_B} \restrict{s=0} = \alpha b'\, y^2	+ \beta \,b x^3 \,.
	\ee
	This can be worked out along similar lines as in the case $n=3$.

\section{Expressing the partition function in terms of $\Dag$ invariant Jacobi forms}

\subsection{Ansatz for the elliptic genus}

The elliptic genus for $k$ instanton strings for $G$ has the following form \cite{DelZotto:2016pvm}
	\begin{equation}\label{eq:ansatz-Zk}
	\begin{aligned}
		& Z_{G,k}(\tau,\epsilon_+,\epsilon_-,m) = \\
		&\frac{\cN_{G,k}(\tau,\epsilon_+,\epsilon_-,m)}{\eta^{4k\hG}\prod\limits_{i=1}^k \prod\limits_{s = \pm}\( \varphi_{-1,1/2}(i(\epsilon_++s\, \epsilon_-)) \prod\limits_{\ell=0}^{i-1}\prod\limits_{\alpha\in \Delta_+}\varphi_{-1,1/2}((i+1)\epsilon_+ +(i-1-2\ell)\epsilon_- +s\, m_\alpha) \)} \ . 
	\end{aligned}
	\end{equation}
	Here $\epsilon_+,\epsilon_-$ are the Nekrasov parameters, the fugacities of $SU(2)_R$ and $SU(2)_L$. The mass parameter $\mass$ is the fugacity of $G$, taking values in $\fh_\bC\cong \bC^r$, the complexification of the Cartan subalgebra $\fh$ of the Lie algebra $\fg$ associated to $G$ of rank $r$.
	In the denominator, we use the notation\footnote{\label{dualcartan}In the following, we will use the same symbol $\mass$ for an element of $\fh_\IC$ and the corresponding dual element in $\fh^*_\IC$. This identification relies on a choice of bilinear form on $\fh$.}
	\begin{equation}  \label{malpha}
		m_\alpha = (\mass, \alpha^\vee)_\fg
	\end{equation}
	for any positive root $\alpha$. To fix conventions, we normalize the invariant bilinear form $(,)_\fg$ on $\fh^*_\bC$ such that the norm square of the shortest coroot $\theta^\vee$ is 2.

	We discussed the modular transformation properties of $\Zgk$ in section \ref{sec:2danom} above. By working out the modular properties of the denominator of $\Zgk$ as written in \ref{eq:ansatz-Zk}, we obtain the transformation properties of the numerator $\cN_{G,k}$ \cite{Huang:2015sta, DelZotto:2016pvm,Gu:2017ccq}. It is a holomorphic function on $\bH \times \bC^2\times \fh_\bC$, satisfying 
	\begin{equation}
		\cN_{G,k}(-1/\tau, \epsilon_+/\tau, \epsilon_-/\tau,\mass/\tau) = \tau^{w(G,k)}\md e[\frac{d_{G,k}(\epsilon_+,\epsilon_-,\mass)}{\tau}] \cN_{G,k}(\tau, \epsilon_+,\epsilon_-,\mass) \ .
	\end{equation}
	The weight and index of $\cN_{G,k}$ follow by subtracting the weight and index of the denominator of \eqref{eq:ansatz-Zk} from those of the elliptic genus, which is of vanishing total weight, and index bilinear form $i_{G,k}$ given by \cite{DelZotto:2016pvm}
	\begin{equation}
	i_{G,k}(\epsilon_+,\epsilon_-,m) = i_+(G,k) \epsilon_+^2 + i_-(G,k) \epsilon_-^2 + i_f(G,k) \frac{(\mass,\mass)_\fg}{2} \ ,
	\end{equation}
	with coefficients
	\ba \label{eq:indices}
	i_+ &=& (1+ \frac{\dualCox}{6})k^2 - \frac{5}{6} \dualCox k \,, \\
	i_- &=& -(1+ \frac{\dualCox}{6})k^2 + \frac{1}{6} \dualCox k \,, \nn\\
	i_f &=& - ( 2 + \frac{\dualCox}{3}) k \nn \,.
	\ea
	The modular weight $w(G,k)$ of the numerator then follows as
	\begin{equation}\label{eq:weight}
		w(G,k) = 2k (\hG - 1) - \frac{1}{2}\DG \, k(k+1)\ ,
	\end{equation}
	where $\hG$ is the dual Coxeter number, and $\DG$ denotes $\dG - \rG$. The index bilinear form of the numerator is
	\begin{equation}\label{eq:anom_poly}
		d_{G,k}(\epsilon_+,\epsilon_-,m) = d_+(G,k) \epsilon_+^2 + d_-(G,k) \epsilon_-^2 + d_f(G,k) \frac{(\mass,\mass)_\fg}{2} \ ,
	\end{equation}
	with coefficients
	\begin{align}
		& d_+(G,k) = \frac{k}{24}\(4(2k^2+9k+1) + 4 \hG(k-5) + \DG (k+1)(k+2)(3k+5)\) \ ,\label{eq:ind-p}\\
		& d_-(G,k) = \frac{k(k-1)}{24}\(4(2k-1) - 4 \hG + \DG(k+1)(k+2)\) , \label{eq:ind-m}\\
		& d_f(G,k) = \dualCox \,k^2 +  (\frac{\dualCox}{3}-1)2k \ . \label{eq:ind-f}
	\end{align}
	The modular weight $w(G,k)$ is an integer for any simple Lie group, while the indices $d_\pm(G,k), d_f(G,k)$ are positive integers for all the Lie groups listed in table \ref{non-higgsable}.
	
	We discussed the symmetries of $\Zgk$ under transformations of the elliptic parameters in section \ref{s:symmetries} above. In particular, it is invariant under Weyl transformations of the mass parameter $\mass$. As the denominator in \ref{eq:ansatz-Zk} is manifestly invariant under such transformations, $\cN_{G,k}$ must be as well.
	Based on the properties just discussed, it is natural \cite{Huang:2015sta, DelZotto:2016pvm,Gu:2017ccq} to propose an ansatz for the numerator $\cN_{G,k}(\tau,\epsilon_+,\epsilon_-,\mass)$ as a sum of products of Weyl invariant Jacobi forms for $SU(2)_R$, $SU(2)_L$, and $G$ with indices $d_+(G,k)$, $d_-(G,k)$, and $d_f(G,k)$ respectively with $w(G,k)$ the total modular weight of each summand. We will refine this proposal in section \ref{s:dginv}.
		
	We review the notion of Weyl invariant Jacobi forms in appendix \ref{a:WeylJac}.
		In this paper, we will be interested in the cases $A_2$ and $D_4$, i.e. $SU(3)$ and $SO(8)$. The ring $J_{*,*}(A_2)$ is generated by
	\begin{equation}
		\varphi_0 \in J_{0,1}(A_2) \ ,\quad  \varphi_2 \in J_{-2,1}(A_2) \ ,\quad \varphi_3 \in J_{-3,1}(A_2)  \ ,
	\end{equation}
	the ring $J_{*,*}(D_4)$ by
	\begin{equation}
		\varphi_0 \in J_{0,1}(D_4) \ ,\quad\varphi_2 \in J_{-2,1}(D_4)  \ ,\quad\varphi_4, \psi_4 \in J_{-4,1}(D_4) \ ,\quad \varphi_6 \in J_{-6,2}(D_4) \ .
	\end{equation}
	Explicit expressions for these generators, constructed by Bertola \cite{Bertola:Jacobi}, are given in appendix~\ref{sc:Bertola-basis}.
	
	With these ingredients, the ansatz for the numerator of the $k$ string elliptic genus takes the form
	\begin{equation}\label{eq:N-ansatz}
		\cN_{G,k}(\tau,\epsilon_+,\epsilon_-,\mass) = \sum_{i} c_i \, g_{0}^{(i)}(\tau)g_{+}^{(i)}(\tau,\epsilon_+)g_{-}^{(i)}(\tau,\epsilon_-)g_{f}^{(i)}(\tau,\mass)\ .
	\end{equation}
	$g_{0}^{(i)}$ are products of $E_4(\tau), E_6(\tau)$, $g_\pm^{(i)}$ are products of $A(\epsilon_\pm), B(\epsilon_\pm)$ with total indices $d_\pm(G,k)$, $g_f^{(i)}$ are products of $W(\fg)$ invariant Jacobi forms with total index $d_f(G,k)$, such that the total modular weight of the monomial is $w(G,k)$. The numbers of all possible such monomials for $G=A_2, D_4$ are given in Tabs.~\ref{tb:A2-Bertola}, \ref{tb:D4-Bertola} (the single string case allows for a simplification resulting in a reduction in the number of monomials required, as described presently). The unrefined case is obtained by setting $\epsilon_+ = 0$, which results in replacing $g_{+}^{(j)}(\tau,\epsilon_+)$ by a constant in \eqref{eq:N-ansatz}. As demonstrated by these tables, the number of possible terms in the numerator grows rapidly with respect to the number of strings. Indeed, the modular indices  $d_\pm(G,k), d_f(G,k)$ grow like $\cO(k^3)$, and the number of terms of \eqref{eq:N-ansatz} is roughly a restricted partition number of $d_\pm(G,k), d_f(G,k)$, hence grows exponentially fast.  The coefficients $c_j$ in the ansatz can be determined by imposing vanishing conditions on BPS numbers, as we will discuss in section \ref{s:computing}.
	
	\begin{table}
		\centering
		\begin{tabular}{*{7}{>{$}c<{$}}}
			\hline
			k & w & d_+ & d_- & d_f & \text{unrefined} & \text{refined}\\
			\hline
			1 & -2 & 12 & 0 & 3 & 4 & 126  \\
			2 & -10 & 72 & 6 & 12 & 583  & 192859 \\
			3 & -24 & 230 & 32 & 27 & 33154 & 38331108 \\
			\hline
		\end{tabular}
		\caption{Numbers of possible terms in the numerator of the elliptic genus of $k$ strings for $G=A_2$ in terms of  elements of the ring $J_{*,*}(\mfa)$ of Weyl invariant Jacobi forms both in the unrefined and refined case. The modular weight $w$ and indices $d_\pm, d_f$ of the numerator are also listed.}\label{tb:A2-Bertola}
	\end{table}
	
	\begin{table}
		\centering
		\begin{tabular}{*{7}{>{$}c<{$}}}
			\hline
			k & w & d_+ & d_- & d_f & \text{unrefined} & \text{refined}\\
			\hline
			1 & -14 & 46 & 0 & 8 & 287 & 66598 \\
			2 & -52 & 267 & 23 & 28 & 1871672 & 3241308332 \\
			\hline
		\end{tabular}
		\caption{Numbers of possible terms in the numerator of the elliptic genus of $k$ strings for $G=D_4$ in terms of elements of the ring $J_{*,*}(\mfd)$ of Weyl invariant Jacobi forms both in the unrefined and refined case. The modular weight $w$ and indices $d_\pm, d_f$ of the numerator are also listed.}\label{tb:D4-Bertola}
	\end{table}

	In the case of a single string, a simplification was observed in \cite{DelZotto:2016pvm} in the massless limit of the refined elliptic genus: the elliptic parameter $\epsilon_+$ can be replaced by $2 \epsilon_+$. Given the ansatz \eqref{eq:N-ansatz}, this simplification carries over to the general massive case. Replacing $\epsilon_+$ by $2 \epsilon_+$ divides the corresponding index by $4$, reducing the number of monomials in the generators of the corresponding ring of required weight and index. The new index is given by
	\begin{equation}
	\frac{d_+(G,1)}{4} = \frac{1}{2}(1+\DG-\hG/3) \ ,
	\end{equation}
	is hence integral or half-integral, depending on $G$. As $w(G,1)$ is even, we can incorporate the latter case by recourse to a lemma of Gritsenko \cite{Gritsenko:1999fk}, which demonstrates that any Jacobi form of even weight and half-integral index $m+1/2$ can be written as the product of 
	\begin{equation}
	\varphi_{0,3/2}(\tau,z) = \frac{\theta_1(\tau,2z)}{\theta_1(\tau,z)}\ ,\quad\quad  z\in \bC
	\end{equation}
	and a Jacobi form of weight $2k$ and integral index $m-1$. We can thus modify the ansatz for the numerator of the elliptic genus of 1 string for $G=A_2, D_4$ to be
	\begin{equation}\label{eq:N-ansatz-k1}
	\cN_{G,1}(\tau,\epsilon_+,\epsilon_-,m) = \varphi_{0,3/2}(\tau,2\epsilon_+)\(\sum_{i} c_i g_{0}^{(i)}(\tau) \tilde{g}_{+}^{(i)}(\tau,2\epsilon_+) g_{f}^{(i)}(\tau,m)\)\ ,
	\end{equation}
	with the index of $\tilde{g}_{+}^{(j)}(\tau,2\epsilon_+)$, given by $d_+(G,1)/4-3/2$, guaranteed to be a positive integer. We juxtapose the number of monomials needed upon passing to elliptic parameter $2 \epsilon_+$ with the generic choice $\epsilon_+$ in Tab.~\ref{tb:2e+}. 
	\begin{table}
		\centering
		\begin{tabular}{*{3}{>{$}c<{$}}}
			\hline
			\text{gauge group} & \epsilon_+ &  2\epsilon_+ \\
			\hline
			A_2 & 126 & 21 \\
			D_4 & 66598 & 5854 \\
			\hline
		\end{tabular}
		\caption{Numbers of possible terms in the numerators of the elliptic genus for $1$ string in terms of  Weyl invariant Jacobi forms with $SU(2)_R$ elliptic parameter $\epsilon_+$ vs. $2\epsilon_+$.}\label{tb:2e+}
	\end{table}

	At higher string numbers, this choice is not possible. Already in the case of 2 or 3 strings for $G=A_2$ and even for the case of a single string case for $G=D_4$, the numbers of terms are so large that fixing the free coefficients is computationally extremely onerous. In the remaining parts of this section, we will argue for expanding the numerator $\cN_{G,k}$ in a subring of $\jweyl$ with elements invariant under the diagram automorphisms of the affine Dynkin diagram of $\mfg$. This will sufficiently reduce the number of expansion coefficients, compare tables~\ref{tb:A2-affine} and \ref{tb:D4-affine} to the tables~\ref{tb:A2-Bertola} and \ref{tb:D4-Bertola}, to make the computation feasible for low string numbers.

	\subsection{Properties of the denominator of the ansatz}
	\label{sc:num}

	In section \ref{s:symmetries}, we argued that the elliptic genus should be invariant under a shift of the fugacities $\mass$ by elements $\blambda$ of the weight lattice, as well as under the action of the symmetry group $D(\widehat{\fg})$ of the affine Dynkin diagram on the K\"ahler parameters of the fiber components. In this section, we will study the behavior of the denominator of our ansatz, 
	\ba
		\cD_{G,k} = \eta^{4kh^\vee_G}\prod_{i=1}^k \prod_{s = \pm}\(\varphi_{-1,1/2}(i(\epsilon_++s\, \epsilon_-))\prod_{\ell=0}^{i-1}\prod_{\alpha\in \Delta_+}\varphi_{-1,1/2}((i+1)\epsilon_+ +(i-1-2\ell)\epsilon_- +s \, m_\alpha)\) \ ,\nn\\ \label{eq:den-Zk}
	\ea
	under these transformations.
	
	Regarding the shift invariance, as any root $\alpha$ is a linear combination of simple roots $\alpha_i$ with integer coefficients, $\alpha = \sum_i c_i \alpha_i$, shifting $\mass \rightarrow \mass + {\boldsymbol \lambda}$ shifts $m_\alpha$ by an integer, as
	\be
	m_\alpha \rightarrow (\mass + \blambda , \alpha^\vee) = m_\alpha + \sum_i c_i (\blambda,\alpha_i^\vee) \,. 
	\ee
	By invariance of $\varphi_{-1,1/2}$ under integer shifts of its elliptic parameter, $\mass \rightarrow \mass + {\boldsymbol \lambda}$ is thus an invariance of $\cD_{G,k}$.

	For the invariance under $D(\widehat{\fg})$, we will consider the two cases $G=A_2, D_4$ in turn. 
	\begin{figure}
		\centering
		\includegraphics[width=0.14\linewidth]{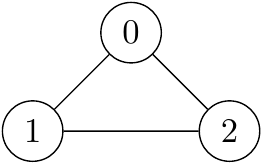}\hspace{6ex}
		\includegraphics[width=0.12\linewidth]{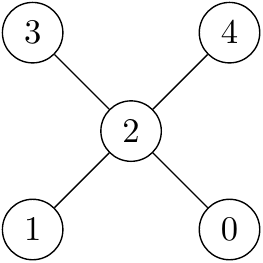}
		\caption{Affine Dynkin diagrams for $\widehat{A}_2$ and $\widehat{D}_4$. \label{fig:dynkin}}
	\end{figure}
	
	From its Dynkin diagram depicted in figure \ref{fig:dynkin}, we can read off $D(\widehat{\mathfrak{a}}_2)$ to be the permutation group on three letters, $S_3$. The set of positive roots of $\mfa$ is
	\begin{equation}\label{eq:A2-rootsplus}
		\Delta_+(\mathfrak{a}_2) = \{ \alpha_1,\alpha_2, \alpha_1+\alpha_2\} \ .
	\end{equation}
	From the definitions of $\varphi_{-1,1/2}(\tau,z)$ and $\eta(\tau)$ given in the appendix
	and the decomposition of $q$ in \eqref{eq:q-decomp}, we conclude that the denominator has the form
	\begin{equation}\label{eq:A2-den-power}
		\cD_{A_2,k} = q^{k/2} (Q_1 Q_2)^{-k(k+1)}  P_{A_2,k}(Q_0,Q_1,Q_2) \ ,
	\end{equation}
	where $P_{A_2,k}(Q_0,Q_1,Q_2)$ is a power series in $Q_0,Q_1,Q_2$ with leading term a non-vanishing constant. The factor $q^{k/2}$ comes from $\eta^{4kh^\vee_G}$ in $\cD_{G,k}$, where $h^\vee_{A_2} = 3$, and the negative powers of $Q_1 Q_2$ from the coefficient $(y^{1/2}-y^{-1/2})$ in the product form of $\varphi_{-1,1/2}$,
	\be
	\varphi_{-1,1/2}(\tau,z)= i\frac{\theta_1(\tau,z)}{\eta(\tau)^3} = - (y^{\frac{1}{2}} - y^{-\frac{1}{2}}) \prod_{n=1}^\infty \frac{(1 + q^n y) (1+ \frac{q^n}{y})}{(1-q^n)^2} \,.
	\ee
	We can argue that $P_{A_2,k}(Q_0,Q_1,Q_2)$ is invariant under the symmetry group $\Daat = S_3$ as follows.
	$S_3$ is generated by the two transpositions $(01)$ and $(12)$. As $\Delta_+$ is invariant under $(12)$, it suffices to show the invariance of $P_{A_2,k}(Q_0,Q_1,Q_2)$ under $(01)$. By $\tau = m_0 + m_1 + m_2$,
	\be
	(01) : (\tau,m_1,m_2) \mapsto (\tau, \tau-m_1 -m_2, m_2) \ ,
	\ee
	and hence 
	\begin{equation}
	(01) : \prod_{\alpha \in \Delta_+}\theta_1(c + m_\alpha)\theta_1(c - m_\alpha) \mapsto \md e[-2\tau+2m_1+4m_2] \prod_{\alpha\in \Delta_+} \theta_1(c + m_\alpha)\theta_1(c - m_\alpha)
	\end{equation}
	for any constant $c$. The coefficient $\md e[-2\tau+2m_1+4m_2] = \frac{Q_1^4 Q_2^2}{q^2}$ combines with 
	\be
	(01): (Q_1 Q_2)^2 \quad \mapsto \quad \frac{q^2}{Q_1^2}
	\ee
	to indeed guarantee the invariance of $P_{A_2,k}(Q_0,Q_1,Q_2)$ under the transposition $(01)$ and thus under all of $\Daat$.
	
	Turning now to $\mfg = \mfd$, we read off the group $D(\widehat{\mathfrak{d}}_4)$ from the affine Dynkin diagram depicted in figure \ref{fig:dynkin} to be $S_4$, the permutation group acting on the four nodes labeled by $0,1,3,4$. The set of positive roots of $\mathfrak{d}_4$ is given by
	\begin{equation}
		\Delta_+(\mathfrak{d}_4) = \left\{\begin{gathered}
			\alpha_1, \alpha_2, \alpha_3, \alpha_4, \\
			\alpha_2+\alpha_1, \alpha_2+\alpha_3, \alpha_2+\alpha_4, \\
			\alpha_2+\alpha_1+\alpha_3, \alpha_2+\alpha_1+\alpha_4, \alpha_2+\alpha_3+\alpha_4, \\
			\alpha_1+\alpha_2+\alpha_3+\alpha_4, \alpha_1+2\alpha_2+\alpha_3+\alpha_4
		 \end{gathered}\right\} \ .
	\end{equation}
	Evaluating the sum over positive roots in \eqref{eq:den-Zk}, we find that $\cD_{D_4,k}$ has the form
	\begin{equation}\label{eq:D4-den-power}
		\cD_{D_4,k} = q^kQ_2^{-5k(k+1)}(Q_1Q_3Q_4)^{-3k(k+1)}P_{D_4,k}(Q_0,Q_1,Q_2,Q_3,Q_4)
	\end{equation}
	with $P_{D_4,k}(Q_0,Q_1,Q_2,Q_3,Q_4)$ a power series in $Q_0,Q_1,Q_2,Q_3,Q_4$ with leading term a non-vanishing constant. We have used $h^\vee_{D_4} = 6$ to obtain the first factor $q^k$.
	
	We will now argue that $P_{D_4,k}(Q_0,Q_1,Q_2,Q_3,Q_4)$ is invariant under the action of $\Dad$. The symmetry group $D(\mathfrak{d}_4) = S_3$ acting on the nodes $1,3,4$ of the Dynkin diagram maps 
	$\Delta_+(\mathfrak{d}_4)$ to itself, is hence a symmetry of $P_{D_4,k}(Q_0,Q_1,Q_2,Q_3,Q_4)$. It thus remains to show invariance under the transposition $(01)$, which together with $D(\mathfrak{d}_4)$ generates $D(\widehat{\mathfrak{d}}_4)$. Recalling that $q = Q_0 Q_1 Q_2^2 Q_3 Q_4$, we see that
	\begin{equation}\label{eq:gen-V4}
	(01): (\tau,m_1,m_2,m_3,m_4) \mapsto (\tau,\tau-m_1-2m_2-m_3-m_4, m_2,m_3,m_4) \,,
	\end{equation}
	resulting in
	\begin{align}
	(01) :  \prod_{\alpha\in \Delta_+} \theta_1(c + m_\alpha) & \theta_1(c - m_\alpha) \\ 
	&\longmapsto \quad \md e[-6(\tau-2m_1-2m_2 -m_3-m_4)] \prod_{\alpha\in \Delta_+}\theta_1(c + m_\alpha)  \theta_1(c - m_\alpha) \,. \nn
	\end{align}
	Combining the coefficient $\md e[-6(\tau-2m_1-2m_2 -m_3-m_4)]=\frac{(Q_1^2 Q_2^2 Q_3 Q_4)^6}{q^6}$ on the RHS with 
	\be
	(01): (Q_1 Q_3 Q_4)^6 \mapsto \frac{q^6}{Q_1^6 Q_2^{12}}
	\ee
	allows us to conclude that $P_{D_4,k}(Q_0,Q_1,Q_2,Q_3,Q_4)$ is invariant under the transposition $(01)$, whence under the full symmetry group $D(\widehat{\mathfrak{d}}_4)$.

	We have thus shown that the products
	\begin{equation}
		q^{-k/2}(Q_1Q_2)^{k(k+1)} \cD_{A_2,k}\quad \text{and}\quad q^{-k}Q_2^{5k(k+1)}(Q_1Q_3Q_4)^{3k(k+1)}\cD_{D_4,k}
	\end{equation}
	can be expanded as power series in $Q_i$, $i=0,1,2$ and $i=0,1,\ldots,4$ respectively (in particular, neither fractional nor negative powers occur), and that they are invariant under the actions of the symmetry groups of the corresponding affine Dynkin diagrams. As we have argued, the elliptic genus $Z_k$ must exhibit both of these properties. We conclude that the numerators $\cN_{G,}$ of $Z_k$ as defined in \eqref{eq:ansatz-Zk}, appropriately normalized as
	\begin{equation}\label{eq:inv-N-A2}
		(Q_1Q_2)^{k^2} \cN_{A_2,k}(\tau,\epsilon_+, \epsilon_-,m)
	\end{equation}
	for $G= A_2$, and by
	\begin{equation}\label{eq:inv-N-D4}
		Q_2^{5k^2+k}(Q_1Q_3Q_4)^{3k^2+k} \cN_{D_4,k}(\tau,\epsilon_+, \epsilon_-,m)
	\end{equation}
	for $G= D_4$ must exhibit these two properties as well. The generators of the ring $J_{*,*}(\fg)$ for $\mfg=\mfa,\mfd$ as constructed abstractly by Wirthm\"{u}ller \cite{Wirthmuller:Jacobi} and concretely by Bertola \cite{Bertola:Jacobi} do not satisfy either of these properties, as one can check explicitly with the formulae given in appendix~\ref{sc:Bertola-basis}. As we will see in the next section, it is possible to find a subring of $J_{*,*}(\fg)$ whose generators do satisfy these properties. Using the generators of the subring, the number of possible terms in our ansatz \eqref{eq:N-ansatz} for the numerator $\cN_{G,k}(\tau,\epsilon_+,\epsilon_-,m)$ can be greatly reduced.

	\subsection{$D(\mfag)$ invariant Jacobi forms} \label{s:dginv}
	
	In this section, we will construct a subring $\jout$ of the ring of Weyl invariant Jacobi forms. We will argue that the numerator $\cN_{G,k}$ of the elliptic genus $Z_{G,k}$ lies within this subring, thus reducing the number of coefficients required to determine it.

	We argued in the previous subsection that based on considerations from the topological string, the numerator $\cN_{G,k}$ should have a Laurent series in the variables $Q_i$ ($i=0, 1,\ldots,r$) defined in \eqref{eq:mQi}, \eqref{eq:q-decomp}, be invariant under the Weyl group $W(\mfg)$, and transform in a simple way (it should become invariant upon rescaling with a power of a product of the $Q_i$) under the symmetry group $\Dag$ of the affine Dynkin diagram of $\mfag$. We will show that it is possible to impose these conditions already at the level of the ring generators from which we construct the expansion forms $g_f^{(j)}$ in \eqref{eq:N-ansatz}. In fact, we will see that it suffices to impose, aside from the Laurent series condition, invariance under the group of diagram automorphisms $\Dg$ of the finite Lie algebra on elements of $\jweyl$; simple transformation properties (invariance upon rescaling with a weight dependent integer power of a product of the $Q_i$) under the symmetry group $\Dag$ of the affine Dynkin diagram then follow. Conversely, we will show that imposing $\Dag$ invariance upon rescaling on elements of $J_{*,*}(\fg)$ entail that they have a Laurent series expansion in the variables $Q_i$ (a general Weyl invariant Jacobi form expanded in $Q_i$ will exhibit fractional powers).
	
	We denote the subring of $\jweyl$ whose elements have Laurent series expansions and are $\Dg$ invariant by $\jout$, and the isomorphic ring of $\Dag$ invariant Jacobi forms by $\jtout$.
	
	We can construct elements of $\jout$ at given weight $w$ and index $n$ by considering the general element of $J_{w,n}(\fg)$, a linear combination of the products of generators of $J_{*,*}(\fg)$ and the Eisenstein series $E_4, E_6$, and imposing the Laurent condition and symmetry under $\Dg$ to fix the coefficients.
	
	Following this procedure, we have found three algebraically independent elements of $\jouta$,
	\begin{equation}\label{eq:gens-affine-A2}
	\phi_0 \in \jouta_{0,3}\ ,\quad \phi_2 \in \jouta_{-2,3} \ ,\quad \phi_6 \in \jouta_{-6,6} \ ,
	\end{equation}
	and five algebraically independent elements of $\joutd$,
	\begin{equation}\label{eq:gens-affine-D4}
	\phi_0 \in \joutd_{0,2} \ ,\quad \phi_2\in \joutd_{-2,2}\ ,\quad \phi_6\in \joutd_{-6,4} \ ,
	\ee
	\be
	\phi_8 \in \joutd_{-8,4} \ ,\quad \phi_{12} \in \joutd_{-12,6} \ .\nn
	\end{equation}
	Their explicit expressions in terms of the generators of the respective Weyl invariant Jacobi forms are given in appendix~\ref{sc:affine-basis}. 
	
	 We conjecture that \eqref{eq:gens-affine-A2} and \eqref{eq:gens-affine-D4} generate the subrings $\jouta$ and $\joutd$ of $\jweyl$ respectively over $\IC[E_4,E_6]$. Independently of this conjecture, we will show in section \ref{s:computing} that the numerator $\cN_{G,k}$ of the elliptic genera for $G=A_2, D_4$ is an element of the subring generated by these elements.
	
	To study the transformation properties of elements of $\jout$ under $D(\widehat{\fg})$, we first note that having a Laurent series expansion in the variables $Q_i$ implies quasi-periodicity with regard to translations by elements of the lattice $P \oplus \tau P$, with $P$ the weight lattice of $\fg$, rather than merely invariance under the sublattice $Q^\vee \oplus \tau Q^\vee$, i.e.
		\begin{equation}\label{eq:qp-P}
		\phi_{w,n}(\tau, \mass+\blambda \tau+\bmu) = \md e\left[-n\(\frac{(\blambda,\blambda)_\fg}{2}\tau+(\blambda,\mass)_\fg\)\right] \phi_{w,n}(\tau,\mass) \ , \quad \blambda,\bmu \in P \ .
		\end{equation} 
	 Periodicity under $\mass \rightarrow \mass + \blambda$, $\blambda \in P$, follows immediately from $(\mass + \blambda, \alpha^{\vee}_i) = m_i +  \lambda_i$, where $\blambda = \sum_i \lambda_i \omega_i$, $\lambda_i \in \IZ$. The transformation behavior under shifts by $\blambda \tau$, $\blambda \in P$ is obtained by applying the modular transformation $S$ on both sides of the equality
	 \begin{equation}
	 \phi_{w,n}\(-\frac{1}{\tau}, \frac{\mass}{\tau} +\bmu\) = \phi_{w,n}\(-\frac{1}{\tau}, \frac{\mass}{\tau}\) \ ,\quad\quad \bmu \in P \ .
	 \end{equation}
	 $\Dag$ for both $\mfg= \mfa$ and $\mfg = \mfd$ is generated by $\Dg$ in conjunction with the transposition $(01)$ exchanging $\aomega_0$ and $\aomega_1$. As elements of the ring $\jout$ are invariant under $\Dg$ by definition, it remains to study their behavior under this transposition. We do this for $\mfg= \mfa$ and $\mfg = \mfd$ in turn. 
	
	For $\mfa$, we have
	\ba
	(01) : m_0 \aomega_0 + m_1 \aomega_1 + m_2 \aomega_2 \mapsto m_0 \aomega_1 + m_1 \aomega_0 + m_2 \aomega_2 \nn = \tau \aomega_0 + (\tau - m_1 - m_2) \omega_1 + m_2 \omega_2 \,,
	\ea
	whence, for $\phi \in \joutas$,
	\begin{align}
	(01) : \phi(\tau& \aomega_0 + m_1 \omega_1 + m_2 \omega_2) \quad \mapsto \quad \phi(\tau \aomega_0 + (\tau - m_1 - m_2) \omega_1 + m_2 \omega_2) \\
	&= \md e\left[-n\(\frac{(\omega_1,\omega_1)_\fg}{2}\tau-(\omega_1,  (m_1 + m_2) \omega_1 - m_2 \omega_2)_\fg\)\right]  \phi(\tau \aomega_0 - (m_1 + m_2) \omega_1 + m_2 \omega_2)\nn \\
	&=\md e\left[-\frac{n}{3}(\tau - 2m_1 -m_2)\right] \phi(\tau \aomega_0 +m_1 \omega_1 + m_2 \omega_2) \nn\,.
	\end{align}
	We have here used the quasi-periodicity under shifts in $P \oplus \tau P$ for the first equality, and
	\be
	- (m_1 + m_2) \omega_1 + m_2 \omega_2 = s_{\alpha_1}s_{\alpha_2} \circ (12) \left(m_1 \omega_1 + m_2 \omega_2 \right)
	\ee
	together with Weyl invariance and $\Dat$ invariance of the elements of $\jouta$ for the second. If we extend the action of $\Daat$ onto a prefactor $(Q_1 Q_2)^{\frac{n}{3}}$ in the natural way,
		\be
		(01):\quad (Q_1 Q_2)^{\frac{n}{3}} \mapsto (Q_0 Q_2)^{\frac{n}{3}} = \md e[\frac{n}{3}(\tau - m_1)] \,,
		\ee
	we see that 
	\be \label{tildea}
	\tilde{\phi}_{w,n}=(Q_1 Q_2)^\frac{n}{3} \phi_{w,n}
	\ee 
	is invariant under $\Daat$ for $\phi_{w,n} \in \joutas$. The tilded variants of the generating set introduced in \eqref{eq:gens-affine-A2}, $\{\tilde{\phi_0},\tilde{\phi_2},\tilde{\phi_6}\}$, hence generate $\Daat$ invariant Jacobi forms over the ring $\IC[E_4,E_6]$. Our conjecture $\jouta = \langle \phi_0, \phi_2, \phi_6 \rangle_{\IC[E_4,E_6]}$ would imply that the ring thus generated coincide with $\jtouta$.
	
	Turning now to the case $\mfg = \mfd$, we note that 
	\be
	(01) : m_0 \aomega_0 + \sum_{i=1}^4 m_i \aomega_i \quad \mapsto  \quad
	\tau \aomega_0 + (\tau - m_1 - 2m_2-m_3-m_4) \omega_1 + \sum_{i=2}^4 m_i \omega_i \,,
	\ee
	and therefore
	\begin{align}
	(01) : \phi(\tau& \aomega_0 + \sum_{i=1}^4 m_i \omega_i) \quad \mapsto \quad \phi(\tau \aomega_0 + (\tau - m_1 - 2m_2-m_3-m_4) \omega_1 + \sum_{i=2}^4 m_i \omega_i ) \nn \\
	&= \md e\left[-n\(\frac{(\omega_1,\omega_1)_\fg}{2}\tau- (\omega_1, ( m_1 + 2m_2+m_3+m_4) \omega_1 - \sum_{i=2}^4 m_i \omega_i )_\fg\)\right]  \nn\\
	&\hspace{6cm}\phi( \tau \aomega_0 -( m_1 + 2m_2+m_3+m_4) \omega_1 + \sum_{i=2}^4 m_i \omega_i )\nn \\ 
	&=\md e\left[-\tfrac{n}{2}(\tau-2m_1-2m_2-m_3-m_4)\right] \phi(\tau \aomega_0 + \sum_{i=1}^4 m_i \omega_i) \nn	\,.
	\end{align}
	Again, we have used quasi-periodicity for the first equality, and 
	\begin{equation}
		(s_{\alpha_1} s_{\alpha_3} s_{\alpha_4})\circ s_{\alpha_2}\circ (s_{\alpha_1} s_{\alpha_3} s_{\alpha_4})\circ s_{\alpha_2}\circ s_{\alpha_1}\circ s_{\alpha_2}\circ (s_{\alpha_3}s_{\alpha_4}) \ ,
	\end{equation}
	followed by an exchange of $m_3,m_4$ for the second. As under a natural extension of the $\Dad$ action to the product $(Q_1 Q_2^2 Q_3 Q_4)^{\frac{n}{2}}$, we have
	\be
	(01):\quad (Q_1 Q_2^2 Q_3 Q_4)^{\frac{n}{2}} \mapsto (Q_0 Q_2^2 Q_3 Q_4)^{\frac{n}{2}} = \md e[\frac{n}{2}(\tau - m_1)] \,,
	\ee
	we find that the forms 
	\be \label{tilded}
	\tilde{\phi}_{w,n}=(Q_1 Q_2^2 Q_3 Q_4)^{\frac{n}{2}} \phi_{w,n}
	\ee
	are invariant under the $\Dad$ action for $\phi_{w,n} \in \joutds$. Introducing the tilded variants to the generating set introduced in equation \eqref{eq:gens-affine-D4}, we conclude that $\langle \tilde{\phi_0},\tilde{\phi_2},\tilde{\phi_6},\tilde{\phi_8},\tilde{\phi_{12}} \rangle_{\IC[E_4,E_6]}$ is a ring of $\Dad$ invariant Jacobi forms, and our conjecture regarding the completeness of the generating set \eqref{eq:gens-affine-D4} would imply that the ring thus generated coincide with $\jtoutd$.
	
	As the indices of the generating set $\{\phi_0,\phi_2,\phi_6\}$ introduced in \eqref{eq:gens-affine-A2} for $\mfg=\mfa$ are divisible by 3, and those of the generating set $\{\phi_0,\phi_2,\phi_6,\phi_8,\phi_{12}\}$ introduced in \eqref{eq:gens-affine-D4} for $\mfg=\mfd$ are divisible by 2, the prefactors relating $\phi_{w,n}$ to $\tilde{\phi}_{w,n}$ introduced in \eqref{tildea} and \eqref{tilded} do not introduce any fractional powers in the $Q_i$ expansion of the tilded forms. More is true: from explicit expansion of the generators \eqref{eq:gens-affine-A2} and \eqref{eq:gens-affine-D4}, we find in fact that the forms $\tilde{\phi}_{n,w}$ have Taylor series expansions in the variables $Q_i$, $i=0,1, \ldots, r$, both in the case $\mfg = \mfa$ and $\mfg = \mfd$, i.e. no negative powers of these variables arise.
	
	Note that $\Dag$ invariance of the forms $\tilde{\phi}_{n,w}$ for $\mathfrak{g}=\mathfrak{a}_2,\mathfrak{d}_4$, together with periodicity in $\tau$, already implies quasi-periodicity under the lattice $P \oplus \tau P$. For shifts of the argument by $\omega_i$ for which $\omega_i$ is swapped with $\omega_0$ under an element $(0i) \in \Dag$, the shift is mapped to a shift of $\tau$ by 1 under the action of this transposition. In more detail, 
	\begin{align}
	\tau \aomega_0 +& (m_i+1 ) \omega_i + \sum_{k\neq i} m_k \omega_k = m_0 \aomega_0 + \sum_k m_k (\aomega_k - a_k^\vee \aomega_0) + \aomega_i - a_i^\vee \aomega_0  \\
	&\rightarrow \quad m_0 \aomega_0 + \sum_k m_k \aomega_k + \aomega_i \quad \rightarrow \quad m_0 \aomega_i + m_i \aomega_0 + \sum_{k\neq i}  m_k \aomega_k + \aomega_0 \\
	&\rightarrow \quad m_0 \aomega_i + m_i \aomega_0 + \sum_{k\neq i}  m_k \aomega_k \quad \rightarrow \quad m_0 \aomega_0 + \sum_k m_k \aomega_k = \tau \aomega_0 + \sum_k m_k \omega_k \,,
	\end{align}
	where each arrow corresponds to an invariance of $\tilde{\phi}_{n,w}$: the first and third under translations by $\aomega_0$, the second and fourth under the transposition $(0i)  \in \Dag$. In the case of $\mathfrak g = \mfd$, shifts by the central node $\omega_2$, which is not connected to $\aomega_0$ by a transposition in $\Dad$, can be mapped to a shift in $\omega_1$ by invoking periodicity under $\alpha_1^\vee = 2 \omega_1 - \omega_2 \in Q^\vee$.
	
	Expanding the numerator of our ansatz for the elliptic genus in terms of elements of the subring $\jout$ of $\jweyl$ cuts down the number of coefficients that need to be determined by a polynomial factor in the base wrapping $k$. We display these numbers for $k=1, \ldots, 4$ in tables~\ref{tb:A2-affine} and \ref{tb:D4-affine}; the counterparts in terms of the basis $\jweyl$ are displayed in tables~\ref{tb:A2-Bertola} and \ref{tb:D4-Bertola}. 	
	\begin{table}
		\centering
		\begin{tabular}{*{7}{>{$}c<{$}}}
		\hline
			k & w & d_+ & d_- & d_f & \text{unrefined} & \text{refined}\\
			\hline
			1 & -2 & 12 & 0 & 3 & 1 & 5 \\
			2 & -10 & 72 & 6 & 12 & 38 & 30362 \\
			3 & -24 & 230 & 32 & 27 & 2299 & 4904253 \\
			4 & -44 & 554 & 98 & 48 & 57378 & 237021553 \\		
			\hline	
		\end{tabular}
		\caption{Numbers of possible terms in the numerator of the $k$-string for $\mfg = \mfa$ in terms of $\Daat$ invariant Jacobi forms.}\label{tb:A2-affine}
	\end{table}

	\begin{table}
		\centering
		\begin{tabular}{c *{7}{>{$}c<{$}}}
		\hline
			k & w & d_+ & d_- & d_f & \text{unrefined} & \text{refined}\\
			\hline
			1 & -14 & 46 & 0 & 8 & 2 & 149 \\
			2 & -52 & 267 & 23 & 28 & 10719 & 58896996 \\
			\hline
		\end{tabular}
		\caption{Numbers of possible terms in the numerator of the $k$-string for $\mfg = \mfd$ in terms of $\Dad$ invariant Jacobi forms.}\label{tb:D4-affine}
	\end{table}			

	\section{Computing the partition function} \label{s:computing}
	
	\subsection{The $\bQ$ expansion}
	We have argued above that the topological string partition function on the $G=A_2,D_4$ geometries should take the form
	\begin{align} \label{eq:ztopG}
	& \Ztop_{G,k}(\tau,\epsilon_+,\epsilon_-,\mass) = \(\frac{q^{1/2}}{\prod_{i=1}^r Q_i^{a_i^\vee}}\)^{kh^\vee_G/3} \times\\
	&\frac{\cN_{G,k}(\tau,\epsilon_+,\epsilon_-,\mass)}{\eta^{4k\hG}\prod\limits_{i=1}^k \prod\limits_{s = \pm}\( \varphi_{-1,1/2}(i(\epsilon_++s\, \epsilon_-)) \prod\limits_{\ell=0}^{i-1}\prod\limits_{\alpha\in \Delta_+}\varphi_{-1,1/2}((i+1)\epsilon_+ +(i-1-2\ell)\epsilon_- +s \, m_\alpha) \)} \ , \nn
	\end{align}
	with
	\be
	\cN_{G,k}(\tau,\epsilon_+,\epsilon_-,\mass) \in J_{*,d_+}(\epsilon_+) \otimes J_{*,d_-}(\epsilon_-) \otimes \Jd_{*,d_f}(\mfg) \,,
	\ee
	of total weight $w(G,k)$ and indices $d_+(G,k)$, $d_-(G,k)$, $d_f(G,k)$, as given in \eqref{eq:weight} -- \eqref{eq:ind-f}, in particular
	\be
	d_f(A_2,k) = 3k^2 \,, \quad d_f(D_4,k) = 2k(3k+1) \,.
	\ee
	The factor
	\be
	\(\frac{q^{1/2}}{\prod_{i=1}^r Q_i^{a_i^\vee}}\)^{kh^\vee_G/3} \label{eq:ztozhat}
	\ee
	relating $Z_{G,k}$ to $\Ztop_{G,k}$ essentially maps 
	\be
	\cN_{G,k}(\tau,\epsilon_+,\epsilon_-,\mass)  \quad \mapsto \quad \widehat{\cN}_{G,k}(\tau,\epsilon_+,\epsilon_-,\mass) \in J_{*,d_+}(\epsilon_+) \otimes J_{*,d_-}(\epsilon_-) \otimes \Jad_{*,d_f}(\mfg) \,, \label{eq:ntonhat}
	\ee
	via the isomorphism between the rings $\Jd_{w,n}(\mfg)$ and $\Jad_{w,n}(\mfg)$ given by 
	\be
	\begin{aligned}
		&\Jd_{w,n}(\mfa) & \longrightarrow \quad &\Jad_{w,n}(\mfa) 	  \\ 
		&\phi_{w,n} & \longmapsto \quad&(Q_1 Q_2)^\frac{n}{3} \phi_{w,n} =\tilde{\phi}_{w,n} 
	\end{aligned}
	\ee
	and
	\be
	\begin{aligned}
		&\Jd_{w,n}(\mfd) & \longrightarrow \quad &\Jad_{w,n}(\mfd) \\
		&\phi_{w,n} & \longmapsto \quad &(Q_1 Q_2^2 Q_3 Q_4)^{\frac{n}{2}} \phi_{w,n} = \tilde{\phi}_{w,n} \nn
	\end{aligned}
	\ee
	respectively.
	
	To see this, recall from section \ref{sc:num} that the denominator of the elliptic genus $Z_{G,k}$, given by the second line in \eqref{eq:ztopG}, has the form
	\begin{equation}\label{eq:A2-den-power}
	\cD_{A_2,k} = q^{k/2} (Q_1 Q_2)^{-k(k+1)}  P_{A_2,k}(Q_0,Q_1,Q_2) \ ,
	\end{equation}
	\begin{equation}\label{eq:D4-den-power}
	\cD_{D_4,k} = q^kQ_2^{-5k(k+1)}(Q_1Q_3Q_4)^{-3k(k+1)}P_{D_4,k}(Q_0,Q_1,Q_2,Q_3,Q_4) \ ,
	\end{equation}
	respectively, where $P_{G,k}$ are power series in $Q_i$ with non-vanishing constant term. The monomials in $q$ and $Q_i$ multiplying $P_{G,k}$ combine with the factor \eqref{eq:ztozhat} to yield the monomials $(Q_1 Q_2)^{k^2}$, $(Q_1 Q_3 Q_4)^{k(3k+1)} Q_2^{k(5k+1)}$ for $G=A_2$, $G=D_4$ respectively. Multiplying $\cN_{G,k}(\tau,\epsilon_+,\epsilon_-,\mass)$, these factors implement the map \eqref{eq:ntonhat}, yielding
	\be
	\Ztop_{A_2,k}(\tau,\epsilon_+,\epsilon_-,\mass) = \frac{\widehat{\cN}_{A_2,k}(\tau,\epsilon_+,\epsilon_-,\mass)}{P_{A_2,k}(\bQ)} \ , 
	\ee
	\be \label{eq:ztopd4}
	\Ztop_{D_4,k}(\tau,\epsilon_+,\epsilon_-,\mass) = \frac{1}{Q_2^{k(k+1)}}\frac{\widehat{\cN}_{D_4,k}(\tau,\epsilon_+,\epsilon_-,\mass)}{P_{D_4,k}(\bQ)} \ .
	\ee
	Both $\widehat{\cN}_{G,k}(\tau,\epsilon_+,\epsilon_-,\mass)$ and $P_{G,k}$ have power series expansions in the $Q_i$, and $P_{G,k}$ has non-vanishing constant coefficient. The only source of negative powers of $Q_i$ is hence the coefficient in $Q_2$ on the RHS of \eqref{eq:ztopd4}. The vanishing of negative powers of $Q_2$ in the power series expansion of $\Ztop_{D_4,k}$ gives rise to constraints on the expansion coefficients of $\widehat{\cN}_{D_4,k}(\tau,\epsilon_+,\epsilon_-,\mass)$.
	
	Note that the $\mass$ dependent factors in the denominator of \eqref{eq:ztopG} partially resum the expansion of the topological string partition function in the fiber directions coordinatized by $Q_i$, introducing additional poles in the partition function in the parameters $m_\alpha, \epsilon_\pm$ that are not visible in the large radius expansion. This phenomenon in the decompactification limit $q \rightarrow 0$ of the fiber was already observed in \cite{Iqbal:2003ix,Iqbal:2003zz}, and played an important role there for the existence of the 4d limit.
	
	Note also that unlike the case of the E-- and M--string (and more generally chains of M--strings ending on an E--string), even once the partition function is put in Gopakumar-Vafa form, it is not possible to take the $Q_i \rightarrow 0$ limit. This would introduce an unweighted sum over all Weyl orbits of the fiber parameters at given degree in $Q_b$, $q$, and $\epsilon_\pm$. In contradistinction to the case of the E-- and M--string, these sums are infinite. This behavior can be traced back to the contribution of the fiber  classes to the vanishing condition: it is positive in the gauged cases and negative for the E-- and M--string, rendering the aforementioned sum effectively finite in the latter case, infinite in the former.

	\subsection{Non-sufficiency of generic vanishing conditions for fixing numerator of ansatz}
	In the case of the E-- and the M--string and related models studied in \cite{Gu:2017ccq}, all coefficients $c_j$ that arise in the expansion \eqref{eq:N-ansatz} of the numerator $\cN_{G,k}$ of the ansatz \eqref{eq:ztopG} can be fixed\footnote{  At base degree 1, several non-vanishing invariants must also be provided to fix the ansatz completely.} by imposing the vanishing of the BPS degeneracies $N^\kappa_{j_-j_+}$ at given K\"ahler class $\kappa$ at sufficiently high $j_{-},j_{+}$. We speak of a generic vanishing condition when the bound beyond which the BPS invariants vanish need not be specified. For the models studied in this paper, imposing a generic vanishing condition does not suffice to fix all coefficients: we can identify a subfamily of the ansatz which satisfies the generic vanishing condition for arbitrary coefficients (this type of subfamily was called {\it the restricted ansatz} in \cite{Huang:2015sta}). To see this, consider the expansion of $\varphi_{-1,\frac{1}{2}}(\tau, z)$,
	\be \label{eq:exp-phi}
	\varphi_{-1,\frac{1}{2}}(\tau,z) = - \sqrt{x(z)} \prod_{n=1}^\infty \frac{1+ x(z)\, q^n -2q^n + q^{2n}}{(1-q^n)^2} \,,
	\ee
	where $x(z) = (2 \sin \frac{z}{2})^2$. The vanishing condition requires the powers in $x(\epsilon_+)$ and $x(\epsilon_-)$ in a $q$ and $\boldsymbol{Q}$ expansion of \eqref{eq:ztopG} to be bounded at fixed order in $q$ and $\boldsymbol{Q}$. Given that the $x(z)$ dependence within the product in \eqref{eq:exp-phi} is accompanied by a factor of $q^n$ it is the coefficient $\sqrt{x(z)}$ that determines whether arbitrary powers of $x(z)$ can be generated at finite order in K\"ahler parameters. We distinguish between three types of contributions of $\varphi_{-1,\frac{1}{2}}$ to the denominator in \eqref{eq:ztopG}:
	\begin{enumerate}
		\item $\varphi_{-1,\frac{1}{2}}(\epsilon_1),\varphi_{-1,\frac{1}{2}}(\epsilon_2)$: 	these terms contribute the $\sqrt{x(\epi)}\sqrt{x(\epii)}$ factor in the denominator of \eqref{FNEK}, hence do not induce a violation of the vanishing condition. 
		\item $\varphi_{-1,\frac{1}{2}}(k\epsilon_1),\varphi_{-1,\frac{1}{2}}(k\epsilon_2)$, $k>1$: these are the terms that generically lead to a violation of the vanishing condition. To see this, note that
		\be
			x(nz) = 2 \left(1- T_n(1-\frac{x(z)}{2}) \right) \,,
		\ee
		with $T_n(x)$ denoting the Chebyshev polynomials.\footnote{  These polynomials are uniquely determined by the equation $T_n(\cos \theta) = \cos n \theta$. They satisfy the recursion relation $T_0(x)=1$,  $T_1(x)=x$, $T_{n+1}(x) = 2x T_n(x) - T_{n-1}(x)$.} Except for the case $n=1$, $\frac{1}{\sqrt{x(nz)}}$ hence leads to an infinite series in $x(z)$ (with leading term $x(z)^{-1/2}$ for $n$ odd) multiplying a power series in $\boldsymbol{Q}$.
		\item $\varphi_{-1,\frac{1}{2}}(m_\alpha + j \epsilon_{+} + k \epsilon_{-})$: in the $\boldsymbol{Q}$ expansion of this term, powers of $x(\epsilon_{\pm})$ are accompanied by factors of $\boldsymbol{Q}$. These terms hence do not lead to a violation of the vanishing condition.
	\end{enumerate} 	
	If one can eliminate all terms of type 2 upon diligent choice of the numerator $\cN_{G,k}(\tau,\epsilon_+,\epsilon_-,m)$, then the remaining expression satisfies the vanishing condition. If this choice leaves coefficients $c_i$ unfixed, then clearly the vanishing conditions are not sufficient to fix all $c_i$.
	
	Terms of type 2 can be cancelled as the product $\varphi_{-1,\frac{1}{2}}(\tau,k \epsilon_1) 	\varphi_{-1,\frac{1}{2}}(\tau,k \epsilon_2)$ can for any $k$ be re-expressed as a polynomial in Jacobi forms with elliptic parameters $\epsilon_{\pm}$. The question of sufficiency of imposing generic vanishing conditions hence hinges on the possibility of doing this while respecting the indices \eqref{eq:indices}. In the case of the E- and M-string, the constraint arises from $i_-$: eliminating all terms of type 2 would require the numerator to have negative index $d_-$ such that the total index of \eqref{eq:ztopG} be $i_-$, which is not possible. On the other hand, for all gauged cases, the index remains positive upon removing all type 2 contributions, thanks to contributions of type 3. The resulting expressions can therefore be completed by elements of $\jout$ to have the desired indices, and any completion will satisfy the generic vanishing condition.
	
In the following, we will find that imposing precise vanishing conditions is sufficient for fixing all unknown coefficients in the ansatz \eqref{eq:ztopG} for all cases that we consider: the refined $A_2$ and $D_4$ model at base degree one, and the unrefined $A_2$ model up to base degree 3.

	\subsection{Fixing the ansatz by imposing precise vanishing conditions}
	As argued in the previous subsection, imposing generic vanishing conditions on the BPS numbers is not sufficient to fix the unknown coefficients $c_i$ in the ansatz \eqref{eq:N-ansatz} for the numerator $\cN_{G,k}$ of $\Ztop_{G,k}$. Our strategy is thus to divine precise vanishing conditions from BPS data available by other means (e.g. as computed in \cite{Haghighat:2014vxa}, \cite{Kim:2016foj}). For the cases that are computationally feasible (refined $A_2$ and $D_4$ string at base degree 1, unrefined $A_2$ string up to base degree 3), this proves sufficient to fix the ansatz completely. 
	
	Labeling the degrees $d_i$ in K\"ahler classes in accordance with the labeling $Q_i$ introduced in section \ref{sc:num}, we obtain
	
	\paragraph{Refined vanishing conditions at base degree 1, $G=A_2$}
	\begin{eqnarray} \label{ref_a2}
	2j_{-}^{\max} = \min(d_0, d_{1}, d_{2}),~~~~ 2j_{+}^{\max} = 2\max (d_0, d_{1}, d_{2}). 
	\end{eqnarray}
	We find that by computing the BPS invariants up to  $\max (d_0, d_{1}, d_{2})=1$ and imposing these vanishing conditions, we can fix the base degree one ansatz up to normalization. 
	
	\paragraph{Refined vanishing conditions at base degree 1, $G=D_4$} \mbox{} \newline
	For
	\be 
	\quad d_0\leq d_1 \leq d_3\leq d_4 \quad \text{(enforced by invoking }\Dad \text{ invariance),} \nn
	\ee
	\be \label{ref_d4}
	2j_{-}^{\max} =\min(d_2-d_3, d_0),~~~~~ 2j_{+}^{\max}  =\max(2d_2+1, 2d_4-1) \,,
	\ee
	$(g_L)_{\max}<0$, which arises for $d_2<d_3$, is to be interpreted as the vanishing of all BPS invariants for the corresponding K\"ahler class. As noted below equation \eqref{eq:ztopd4}, the $\bQ$ expansion of the generic ansatz is a Laurent series in $Q_2$ with lowest power $Q_2^{-2}$. The vanishing of the principal part of this Laurent series (the coefficients of $Q_2^{-2}$ and $Q_2^{-1}$) imposes additional constraints on the ansatz. Overall, we find that by computing the BPS invariants up to  $d_2=4$, $\max (d_0, d_1,  d_3, d_4)=3$, and imposing the vanishing conditions, we can fix the base degree one ansatz with 149 coefficients up to a normalization. 

	\paragraph{Unrefined vanishing conditions at general base degree $d_b$, $G=A_2$}
	\ba \label{a2higher}
	g_{max} &=& (d_b - 1) \max(d_0,d_{1},d_{2}) - \frac{(d_b+2)(d_b-1)}{2}+ \min(d_0,d_{1},d_{2})  +  \nn \\ 
	&& +\begin{cases} 
		\left(\frac{d_b-\Delta}{2}\right)\left(\frac{d_b-\Delta}{2}-1\right) & \mbox{if } {\cal R} \mbox{ and } \Delta \le d_b-3 \mbox{ and } d_b+\Delta \mbox{ even,} \\
		\left(\frac{d_b-\Delta-1}{2}\right)^2 & \mbox{if } {\cal R} \mbox{ and }\Delta \le d_b-3 \mbox{ and } d_b+\Delta \mbox{ odd,} \\
		0 & \mbox{else,}
	\end{cases}
	\ea
	where
	\ba
	\Delta &=& | \text{difference between the two largest elements of } \{d_0,d_{1},d_{2}\}| \,, \nn\\
	{\cal R} &=& \begin{cases} \text{true if the two smallest elements of the set } \{d_0,d_{1},d_{2}\} \text{ do not both vanish,} \\
		\text{false else.} 
		\end{cases} \nn
	\ea

	These vanishing conditions reproduce the bounds presented in \cite{Gu:2017ccq} when all but one fiber class degree is set to zero. In this limit, we can focus on the local geometry of one exceptional fiber class, which, as derived in section \ref{s:toric_geometry}, is that of the local Hirzebruch surface $\IF_1$ in the case of the three fiber classes of the $G=A_2$ geometry, and is local $\IF_2$ for all fiber classes of the resolved $G=D_4$ geometry other than the class whose degree is denoted by $d_2$ above. For this latter curve class, the local geometry is that of $\IF_0$ \cite{DelZotto:2017pti}.
	
	In local $\IF_k$, denoting the base class as $C$ and the fiber class as $F$, the BPS invariants $I_g^\kappa$ and  $N_{j_- j_+}^\kappa$ for the curve class $\kappa = d_b\, C + d \,F$ vanish beyond the bounds \cite{Gu:2017ccq} 
	\be \label{surfaceboundsL}
	g_{max} = 2j_{-}^{max} = \frac{1}{2}k d_b(1-d_b) + (d_b-1)(d-1)
	\ee
	and 
	\be
	2j_{+}^{max} = (d_b+1)d + \frac{1}{2}(2d_b - kd_b(d_b+1)) \,.
	\ee
	Setting $d_b=1$, these formulae reduce to \eqref{ref_a2} and \eqref{ref_d4} with $d$ denoting the sole non-vanishing degree. for $d = d_0$, $k=2$, and $d = d_2$, $k=0$. 
	
	Beyond the local Hirzebruch limit, we can e.g., after reorganizing \eqref{surfaceboundsL} at $k=1$ for better comparison with \eqref{a2higher},
	\be
	g_{max} = (d_b -1) d - \frac{(d_b+2)(d_b-1)}{2} \,,
	\ee
	interpret the first correction term $\min(d_0,d_{1},d_{2})$ as arising from the $g$ handles that arise when the fiber curve configuration with intersections as depicted on the LHS of figure~\ref{fig:dynkin} is wrapped $\min(d_0,d_{1},d_{2})$ times. Deriving the vanishing conditions \eqref{ref_a2}, \eqref{ref_d4}, \eqref{a2higher} within algebraic geometry or otherwise, and extending \eqref{ref_a2}, \eqref{ref_d4} to higher base degree, remains an interesting open problem.

\subsection{Additional constraints on the elliptic genus from gauge theory} \label{s:additional}

Although the main tool we have exploited in this paper has been vanishing conditions of certain BPS numbers, it is also possible to impose constraints on the modular ansatz by taking specific limits in which the topological string partition function simplifies. In this section we discuss two such limits.

\textbf{Five-dimensional limit.} The first limit is attained by taking the K\"ahler parameter of the two-cycle associated to the affine node, $m_0$, to infinity. By equation \eqref{eq:tauaff}, this corresponds to decompactifying the elliptic fiber, thus taking the 5d limit of the 6d SCFT by shrinking the radius of the 6d circle to zero. The Calabi-Yau geometry then reduces to an ALE fibration over $\bP^1$, which geometrically engineers pure 5d SYM theory with gauge group $G$. This in particular means that the topological string partition function reduces to the 5d Nekrasov partition function for this theory\cite{Kim:2016foj, DelZotto:2016pvm}. The terms corresponding to wrapping number $k$, i.e. $ \widehat{Z}_{k}$, become the $k$-instanton piece of the 5d Nekrasov partition function \cite{Nekrasov:2002qd},
\be \sum Q_b^n \widehat Z_k \longrightarrow Z_{inst} = \sum Q_b^n \int_{\cM(G,k)}1,\label{eq:5dneklim}\ee
where the right hand side is a sum of equivariant integrals over the moduli spaces of $k$ $G$-instantons.

For example, for $G=SU(3)$ the equivariant integrals evaluate to
\be \int_{\cM(G,k)}1 = e^{-3\epp k}\sum_{\vert\vec{Y}\vert=k}\frac{1}{\prod_{i,j=1}^3 n^{\vec{Y}}_{i,j}(\epp,\epm,\mass)}, \ee
obtained by specializing the results of \cite{Nakajima:2005fg} to the case $G=SU(3)$; here, $\vec Y = \{Y_1,Y_2,Y_3\}$ is a collection of three Young diagrams, and
\be n^{\vec{Y}}_{i,j}(\epp,\epm,\mass) = \prod_{s\in Y_{i}}(1-e^{-(-l_{Y_j}(s)\epi+(a_{Y_i}(s)+1)\epii+u_j-u_i)})\prod_{s\in Y_{j}}(1-e^{-((l_{Y_i}(s)+1)\epi-a_{Y_j}(s)\epii+u_j-u_i)})\ee
is a product over boxes $(i,j)$ in the Young diagrams;
\be a_Y(s)=Y_i-j;\qquad l_Y(s) = Y^t_j-i\ee
are the usual arm length and leg length of a box in a Young diagram, and $u_1,u_2,u_3$ are $SU(3)$ fugacities related to the parameters $\mass$ via equation \eqref{eq:params-A2}. For $k=1$,\footnote{  For $k= 1$, a universal formula valid for any $G$ is also known \cite{Keller:2011ek}.} 
\be \int_{\cM(SU(3),1)}1= \frac{v^3x\md (1+2v^2+2v^6+v^8-v^4(2-\chi_{Adj}(\mass)))}{(v-x)(1- v x)\prod_{s=\pm}
(1-v^2\md e[s\, m_1])(1-v^2\md e[s\, m_2])(1-v^2\md e[s(m_1+m_2)])}, \ee
where 
\be\chi_{Adj}(\mass) = \md e[-m_1]+\md e[-m_2]+\md e[-m_1-m_2]+2+\md e[m_1]+\md e[m_2]+\md e[m_1+m_2].\label{eq:su35dlim}\ee
One can verify explicitly that the rescaled one-string elliptic genus $\widehat Z_{1}$ obtained from the modular ansatz reduces precisely to equation \eqref{eq:su35dlim} in the $q\to 0$ limit. Similarly, we find the expected agreement between $\widehat Z_k$ and $\int_{\mathcal{M}(SU(3),k)}1$ for $k=2,3$ (in the unrefined limit $v\to1$ which we used to compute $\widehat{Z}_2$ and $\widehat{Z}_3$ with our modular approach).\newline

For $G=SO(8)$, the computation of the 5d Nekrasov partition function was first addressed in \cite{Nekrasov:2004vw}; here we use the expression for the $k$-instanton term of the Nekrasov partition function as presented in appendix B of \cite{Hayashi:2017jze}, and find
\be
\int_{\mathcal{M}(SO(8),k)}1=\frac{1}{2^k k!}\oint\prod_{I=1}^k (-i\, \mathrm{d}\phi_I) Z_{vec},
\ee
where
\ba
Z_{vec} &=& \frac{\prod_{1\leq I<J\leq k}\prod_{s=\pm}2\sin(\pi(\phi_I+s\,\phi_J))2\sin(\pi(\phi_I+s\,\phi_J+2\epp))}{\prod_{1\leq I<J\leq k}\prod_{s=\pm}\prod_{t=\pm}2\sin(\pi(s\,\phi_I+t\,\phi_J+\epi))2\sin(\pi(s\,\phi_I+t\,\phi_J+\epii))}\nonumber\\
&\times& \left(\frac{2\sin(2\pi\epp)}{2\sin(\pi \epi)2\sin(\pi \epii)}\right)^k\frac{\prod_{I=1}^k\prod_{s=\pm}2\sin(2\pi s\,\phi_I)2\sin(2\pi(s\,\phi_I+\epp))}{\prod_{I=1}^k\prod_{i=1}^{4}\prod_{s=\pm}\prod_{t=\pm}2\sin(\pi(s\, \phi_I+t\, a_i+\epp))} \,.\nonumber\\
\ea
The contour of the integral is specified in \cite{Nekrasov:2004vw,Hwang:2014uwa}. The $a_i$ are the coefficients of $\mass$ in the orthogonal basis for $SO(8)$; these are related to the parameters $m_i = (\mass,\alpha^\vee_i)$ employed in the rest of the paper via
\be a_1=m_1+m_2+\frac{m_3}{2}+\frac{m_4}{2},\qquad a_2=m_2+\frac{m_3}{2}+\frac{m_4}{2},\qquad a_3=\frac{m_3}{2}+\frac{m_4}{2},\qquad a_4=-\frac{m_3}{2}+\frac{m_4}{2}.
\ee

For $k=1$, the integral evaluates to a sum over four residues at $\phi_1 = -\epp+a_i$ and four residues at $\phi_1 = -\epp-a_i$, leading to
\begin{align}
\int_{\mathcal{M}(SO(8),1)}1=-\sum_{i=1}^4\sum_{s=\pm}\frac{v^4}{2\md e[2s a_i]}&\frac{(1-v^2\md e[2sa_i])(1-v^4\md e[2sa_i])}{(1-v x)(1-v/x)}\times\nonumber\\
&\frac{1}{\prod_{j\neq i}\prod_{t=\pm}(1-\md e[-sa_i+t a_j])(1-v^2\md e[s a_i+t a_j])} \,,
\end{align}
which indeed agrees with the $q\to 0$ limit of $\widehat{Z}_1$.\newline\\

\textbf{Factorization.} Next, we discuss another set of constraints on the partition functions $ Z_{G,k} $ for $k>1$, which have a natural interpretation from the point of view of 6d SCFTs on $T^2\times \bR^4$. The constraints arise when one tunes the fugacities $\epp,\epm$ to particular values for which the elliptic genus for $k$ strings factorizes into a product of elliptic genera for lower numbers of strings.\footnote{  A similar factorization was observed in \cite{Gaiotto:2012uq,Hanany:2012dm} for the Hall-Littlewood index of certain $4d $ $\cN=2$ theories. When the 4d theory has a 6d origin, the factorization observed in \cite{Gaiotto:2012uq,Hanany:2012dm} can be derived from the factorization of the elliptic genera discussed in the present section, by taking the $\tau\to i\infty$ limit.}

Specifically, recall that the denominator of the elliptic genus of $k$ strings includes a factor of the form
\be\prod\limits_{i=1}^k \prod\limits_{s = \pm}\( \varphi_{-1,1/2}(i(\epsilon_++s \, \epsilon_-))\),\ee
which vanishes when
\be \epm\to \epm^*=\pm \epp + \frac{n_1\tau+n_2}{i},\qquad i=1,\dots,k,\qquad n_1\in\bZ,\qquad n_2\in\bZ. \label{eq:eplim}\ee
The corresponding pole in the elliptic genus arises because the chemical potentials for $SU(2)_+\times SU(2)_-$ have been tuned in such a way that any collection of $i$ strings can be separated from the remaining strings at no energy cost, giving rise to an infinitely degenerate set of ground states even for nonzero chemical potentials. For example, in the simplest case $\epsilon_-\to \pm\epsilon_+$, the chemical potential for one of the two isometries $U(1)_{\epi,\epii}$ that rotate the two $\bR^2$'s in $\bR^4$ has been turned off, and individual strings are no longer confined at the origin of that plane. The key point is that once the chemical potentials have been tuned so that a collection of $i$ strings can be separated from the rest, the elliptic genus will factorize into the elliptic genus from this collection of $i$ strings times the elliptic genus of the remaining $k-i$ strings.

A given singular value $\epm^*$ may in general be realized for several different choices of $(i,n_1,n_2)$; among those, let $(i',n_1',n_2')$ be the one corresponding to the smallest value of $i$. Upon specializing $\epm\to\epm^*$, the elliptic genus will acquire a leading order pole of order $\ell=\lfloor \frac{k}{i'}\rfloor$; and the leading order term will capture the contribution of $\ell$ collections of $i'$ strings which have been moved infinitely far from each other, as well as the contribution from the remaining $i''=k\mod i$ strings (which is just equal to $1$ if $i$ divides $k$):
\be\lim_{\epm\to\epm^*}\left[(\epm-\epm^*)^\ell Z_{G,k}\right] = \frac{1}{\ell!}\left[\lim_{\epm\to\epm^*}(\epm-\epm^*)(Z_{G,i'})\right]^\ell \cdot Z_{G,i''}\bigg\vert_{\epm\to\epm^*}\label{eq:ellfactor}.\ee
Here, the factorial accounts for the fact that the $\ell$ bound states of $i'$ strings are indistinguishable from each other.

Furthermore, in the limit we are considering, the contribution from each set of $i'$ strings to the elliptic genus is the same as that of a single string which is wound multiple times around the $T^2$. Indeed, suppose for a moment that the $i'$ strings did not form bound states. Then their elliptic genus would be given by the $i'$-th Hecke transform of the elliptic genus for one string \cite{Dijkgraaf:1996xw}:
\be \mathcal Z_{G,i'}(\tau,\epp,\epm,m) = \sum_{\substack{a d =i'\\0\leq b <d}} \frac{1}{i'}Z_{G,1}\left(\frac{a\tau+b}{d},a\epp,a\epm,a m\right),\label{eq:hecke}\ee
where the term on the right hand side corresponding to a given choice of $(a,b,d)$ is the elliptic genus of a single string wrapped $a$ times along the A-cycle of the $T^2$ and $d$ times around the $\frac{a B+b A}{gcd(a,b)}$-cycle of the $T^2$.

In the present context, the strings do form bound states, and the expression \eqref{eq:hecke} is not valid; nevertheless, one can still isolate the contribution to the elliptic genus from a single multiply wrapped string in the $(a,b,d)$ sector, by taking the limit \eqref{eq:eplim} with $i'=a\, d$, $n_1=a,n_2=b$:
\be \lim_{\epm\to\epm^*} (\epm-\epm^*)Z_{G,i'}(\tau,\epp,\epm,m)  = \frac{1}{i'} \lim_{\epm\to\epm^*} (\epm-\epm^*)Z_{G,1}(\frac{a^2 \tau+a b}{i'},a\epp,a\epm,a m),\ee
while the first contributions from bound states of multiple strings appear at $\cO(\epm-\epm^*)^0$.

Combining this with equation \eqref{eq:ellfactor}, we obtain the result
\begin{align}\lim_{\epm\to\epm^*}\bigg[(\epm-\epm^*)^\ell Z_{G,k}(\tau,&\epp,\epm,m)\bigg]=\nonumber\\
& \frac{1}{\ell!}\left[\frac{1}{i'}\lim_{\epm\to\epm^*}(\epm-\epm^*)Z_{G,1}(\frac{a^2 \tau+a b}{i'},a\epp,a\epm,a m)\right]^\ell \cdot Z_{G,i''}\bigg\vert_{\epm\to\epm^*}\label{eq:ellfactor2}.\end{align}

The functions $\widehat{Z}_k$ are required to satisfy equations \eqref{eq:5dneklim} and \eqref{eq:ellfactor2}. This leads to several constraints on the coefficients appearing in the ansatz \eqref{eq:ansatz-Zk}, which may be used as consistency checks of the modular approach.

\section{Conclusions and Outlook}
In this paper, we extend the modular approach of \cite{Huang:2015sta,Gu:2017ccq} to topological string theory on elliptically fibered threefolds with higher Kodaira singularities. A central step in our analysis consists in identifying the universal pole structure of the topological string partition function at given base degree \cite{DelZotto:2017pti}. 

Compared to the smooth geometries underlying the E-- and M--strings (and chains of M--strings potentially ending on an E--string) discussed in \cite{Gu:2017ccq}, the geometries studied in this paper require introducing a new class of Jacobi forms to incorporate the dependence of the topological string partition function on the exceptional classes which arise upon resolving singularities in the fiber. We call these $D(\fg)$-invariant Jacobi forms.
	
	Unlike the case of the E-- and M--string, generic vanishing conditions are no longer sufficient in the case of higher Kodaira singularities to compute the topological string partition function $Z_\beta$ at given base degree. In all cases that we consider, imposing exact vanishing conditions does suffice.

	The methods we have developed in this paper clearly extend beyond the geometries we have considered here; they should be applied to more general 6d SCFTs, or even little string theories.
	
	The $SL(2,\IZ)$ symmetry underlying our analysis is the automorphic symmetry characteristic of elliptic fibrations. More generally, such automorphic symmetries can be identified with the monodromy action on the period matrix of the Calabi-Yau manifold. In this form, they are exceedingly difficult to compute. A more accessible approach proceeds via the Fourier-Mukai transform acting on the derived category of coherent sheaves on the geometry. With this machinery, the symmetry can be derived purely from the knowledge of the classical topological data of the Calabi-Yau. In~\cite{Cota:2017aal}, this approach was used to  find novel forms of the holomorphic anomaly equations for Calabi-Yau fourfolds. These match conjectures based on recent progress in the theory of stable pairs which apply to Calabi-Yau manifolds in any dimension~\cite{OP}. These methods apply both in the local and compact setting, but so far, they have only been applied to the case of $I_1$ fiber singularities (yielding the E-- and M--string in the corresponding F-theory compactifications). The extension of these methods to the general elliptic singularities discussed in this paper is an obvious next step in this program. The techniques of \cite{OP} could potentially lead to additional constraints on the topological string partition function sufficient to fix it even in the compact case.

	We have seen that our 6d results reproduce the BPS content of corresponding 5d theories in appropriate limits. Indeed, there exists a hierarchy of rational, trigonometric and elliptic integrable 
	systems associated to theories with eight supercharges \cite{Hollowood:2003cv}, in ascending order from 4 to 6 dimensions. The  
	elliptic  theory of the  E--string solved in~\cite{Kim:2015jba,Gu:2017ccq} incorporates via blow downs the local type 
	II theories on all del Pezzo surfaces except for $\mathbb{P}^1\times\mathbb{P}^1$.  Taking further five (trigonometric) 
	and four dimensional (rational) limits permits extracting information about the topological sector of all $E_n$ 
	theories in the corresponding dimensions \cite{Ganor:1996pc}.
	Similarly here, our analytic solutions for $Z_\beta$ permit us to recover information about analogous theories with higher dimensional Coulomb branches and more interesting gauge content.

	Let us conclude this paper with a more far reaching speculation. The same conjectures relating the anomaly polynomial of BPS strings 
	to the Casimir energy of the corresponding 2d SCFT have been formulated for 6d SCFTs as well \cite{Bobev:2015kza}, establishing a relation between the superconformal index and the partition function on $S^1\times S^5$. The 6d anomaly polynomial determines the corresponding Casimir energy via an equivariant integration similar to the 2d case. The 6d superconformal index therefore exhibits interesting transformation properties with respect to a generalization of the modular group. This raises the question whether it might be possible to constrain it by exploiting properties of a corresponding family of special functions. Finally, the 6d superconformal index is related to the non-perturbative completion of the topological string proposed in \cite{Lockhart:2012vp}, suggesting perhaps that a broad extension of the ideas discussed in this paper could ultimately lead to constraining the whole non-perturbative topological string partition function.

\section*{Acknowledgements}
We are grateful to Denis Bernard, Marco Bertola, Cesar Alberto Fierro Cota, Kazuhiro Sakai, Thorsten Schimannek, and Don Zagier for discussions.

AK would like to thank the ENS for hospitality during a CNRS visiting professorship, at which time this work was initiated. MDZ thanks the ENS for hospitality during the completion of this manuscript.
GL is grateful to the ENS and Universit\"at Bonn for hospitality at different stages of this work. AK, AKKP, GL and MH would like to acknowledge the Tsinghua Summer Workshop in Geometry and Physics 2017, where a preliminary version of this work was presented.

JG and AKKP acknowledge support from the grant ANR-13-BS05-0001. MH is supported by Natural Science Foundation of China grant number 11675167, the ``Young Thousand People" plan by the Central Organization Department in China, and CAS Center for Excellence in Particle Physics (CCEPP). This project has received funding from the European Union's Horizon 2020 research and innovation programme under the Marie Sklodowska-Curie grant agreement No 708045.

	\appendix
	
		\section{BPS invariants}
		\subsection{Unrefined BPS invariants for the $A_2$ model}		
		We have solved the $G=A_2$ model completely up to base degree 3 purely by imposing vanishing conditions. We could easily give higher fiber and genus results than we do below for base degree up to 3. The extraction of GV invariants becomes cumbersome for order $m_i$ beyond 10. For base degree 4, we have determined 250 coefficients by also imposing fiber degree 0 invariants (obtained via the vertex) as boundary conditions, giving us access to all invariants up to genus 6. For base degree 5, we have determined 52 coefficients, again with vertex boundary conditions, fixing invariants up to genus 1. (At base degrees 6 and 7, we have determined the fiber degree 0 invariants up to mass degree 10).
		\subsubsection{Base degree 1}
		\paragraph{Fiber degree 0, genus 0}
		\be
		\begin{tabular}{|c|cccccc|}
			\hline
			$m_2/m_1$ & 0 & 1 & 2 & 3 & 4 & 5 \\
			\hline
			 0 & 1 & 3 & 5 & 7 & 9 & 11 \\
			1 & 3 & 4 & 8 & 12 & 16 & 20 \\
			2 & 5 & 8 & 9 & 15 & 21 & 27 \\
			3 & 7 & 12 & 15 & 16 & 24 & 32 \\
			4 & 9 & 16 & 21 & 24 & 25 & 35 \\
			5 & 11 & 20 & 27 & 32 & 35 & 36 \\
			\hline
		\end{tabular}
		\ee
		
		\paragraph{Fiber degree 5, genus 4}
		\be
		\begin{tabular}{|c|*{6}{>{$}c<{$}}|}
			\hline
			$m_2/m_1$ & 0 & 1 & 2 & 3 & 4 & 5 \\
			\hline
			0 & 0 & 0 & 0 & 0 & 0 & 0 \\
			1 & 0 & 0 & 0 & 0 & 0 & 0 \\
			2 & 0 & 0 & 0 & 0 & 0 & 0 \\
			3 & 0 & 0 & 0 & 0 & 0 & 0 \\
			4 & 0 & 0 & 0 & 0 & 15 & 20 \\
			5 & 0 & 0 & 0 & 0 & 20 & 128 \\
			\hline
		\end{tabular}
		\ee
		
		\subsubsection{Base degree 2}
		\paragraph{Fiber degree 0, genus 0}
		\be
		\begin{tabular}{|c|*{6}{>{$}c<{$}}|}
			\hline
			$m_2/m_1$ & 0 & 1 & 2 & 3 & 4 & 5 \\
			\hline
			0 & 0 & 0 & -6 & -32 & -110 & -288 \\
			1 & 0 & 0 & -10 & -70 & -270 & -770 \\
			2 & -6 & -10 & -32 & -126 & -456 & -1330 \\
			3 & -32 & -70 & -126 & -300 & -784 & -2052 \\
			4 & -110 & -270 & -456 & -784 & -1584 & -3360 \\
			5 & -288 & -770 & -1330 & -2052 & -3360 & -6076 \\
			\hline
		\end{tabular}
		\ee
		
		\paragraph{Fiber degree 5, genus 4}
		\be
		\begin{tabular}{|c|*{6}{>{$}c<{$}}|}
			\hline
			$m_2/m_1$ & 0 & 1 & 2 & 3 & 4 & 5 \\
			\hline
			0 & 0 & 0 & 0 & 0 & 0 & 0 \\
			1 & 0 & -26 & -48 & -66 & -80 & -210 \\
			2 & 0 & -48 & -560 & -1030 & -1518 & -3996 \\
			3 & 0 & -66 & -1030 & -6170 & -12862 & -36120 \\
			4 & 0 & -80 & -1518 & -12862 & -61252 & -203380 \\
			5 & 0 & -210 & -3996 & -36120 & -203380 & -814088 \\
			\hline
		\end{tabular}
		\ee
		
		\subsubsection{Base degree 3}
		\paragraph{Fiber degree 0, genus 0}
		\be
		\begin{tabular}{|c|cccccc|}
			\hline
			$m_2/m_1$ & 0 & 1 & 2 & 3 & 4 & 5 \\
			\hline
			0 & 0 & 0 & 0 & 27 & 286 & 1651 \\
			1 & 0 & 0 & 0 & 64 & 800 & 5184 \\
			2 & 0 & 0 & 25 & 266 & 1998 & 11473 \\
			3 & 27 & 64 & 266 & 1332 & 6260 & 26880 \\
			4 & 286 & 800 & 1998 & 6260 & 21070 & 70362 \\
			5 & 1651 & 5184 & 11473 & 26880 & 70362 & 191424 \\
			\hline
		\end{tabular}
		\ee
		
		\paragraph{Fiber degree 5, genus 4}
		\be
		\begin{tabular}{|c|cccccc|}
			\hline
			$m_2/m_1$ & 0 & 1 & 2 & 3 & 4 & 5 \\
			\hline
			0 & 212 & 432 & 624 & 879 & 2568 & 17935 \\
			1 & 432 & 4156 & 8316 & 14180 & 42532 & 253744 \\
			2 & 624 & 8316 & 45503 & 110055 & 363678 & 1981134 \\
			3 & 879 & 14180 & 110055 & 526469 & 2105226 & 11006493 \\
			4 & 2568 & 42532 & 363678 & 2105226 & 9919241 & 49726249 \\
			5 & 17935 & 253744 & 1981134 & 11006493 & 49726249 & 218999610 \\
			\hline
		\end{tabular}
		\ee
		
		\subsubsection{Base degree 4}
		\paragraph{Fiber degree 0, genus 0}
		\be
		\begin{tabular}{|c|*{6}{>{$}c<{$}}|}
			\hline
			$m_2/m_1$ & 0 & 1 & 2 & 3 & 4 & 5 \\
			\hline
		0 & 0 & 0 & 0 & 0 & -192 & -3038 \\
		1 & 0 & 0 & 0 & 0 & -572 & -10374 \\
		2 & 0 & 0 & 0 & -160 & -2980 & -33192 \\
		3 & 0 & 0 & -160 & -2058 & -18270 & -129910 \\
		4 & -192 & -572 & -2980 & -18270 & -103872 & -536620 \\
		5 & -3038 & -10374 & -33192 & -129910 & -536620 & -2169828 \\
			\hline
		\end{tabular}
		\ee
		
		\paragraph{Fiber degree 5, genus 4}
		\be
		\begin{tabular}{|c|*{6}{>{$}c<{$}}|}
			\hline
			$m_2/m_1$ & 0 & 1 & 2 & 3 & 4 & 5 \\
			\hline
			0 & -1542 & -3382 & -6138 & -20316 & -131048 & -1306790 \\
			1 & -3382 & -27746 & -72132 & -261980 & -1616152 & -14987122 \\
			2 & -6138 & -72132 & -409688 & -1836286 & -11284130 & -98232706 \\
			3 & -20316 & -261980 & -1836286 & -9994938 & -60923844 & -493074498 \\
			4 & -131048 & -1616152 & -11284130 & -60923844 & -334012600 & -2323334482 \\
			5 & -1306790 & -14987122 & -98232706 & -493074498 & -2323334482 & -12681667680 \\
			\hline
		\end{tabular}
		\ee

		\subsection{Refined BPS invariants for the $D_4$ model at base degree one}		
		
		 For the $D_4$ model, fixing the coefficients in our ansatz takes considerable computing time even for the base degree one case. However, once the ansatz is fixed, extracting the refined BPS invariants by expanding it in K\"ahler classes is computationally more efficient than computing these via localization methods \cite{Haghighat:2014vxa}. Below, we list some refined BPS invariants extracted from our base degree 1 result.
		
		\paragraph{Fiber degrees $\{d_2, \{d_0,d_1,d_3,d_4 \} \} = \{0, \{{0, 0, 0, 3} \} \} $  } \be \begin{tabular} {|c|cccccc|} \hline $2j_L \backslash 2j_R$  & 0 & 1 & 2 & 3 & 4 & 5 \\  \hline  0 & 0 & 0 & 0 & 0 & 0 & 1 \\  \hline \end{tabular} \ee 	
		
		\paragraph{Fiber degrees $\{d_2, \{d_0,d_1,d_3,d_4 \} \} = \{1, \{{1, 1, 1, 3} \} \} $  } \be \begin{tabular} {|c|cccccc|} \hline $2j_L \backslash 2j_R$  & 0 & 1 & 2 & 3 & 4 & 5 \\  \hline  0 & 0 & 0 & 0 & 1 & 0 & 1 \\  \hline \end{tabular} \ee 	
		
		\paragraph{Fiber degrees $\{d_2, \{d_0,d_1,d_3,d_4 \} \} = \{2, \{{2, 2, 2, 3} \} \} $  } \be \begin{tabular} {|c|cccccc|} \hline $2j_L \backslash 2j_R$  & 0 & 1 & 2 & 3 & 4 & 5 \\  \hline  0 & 0 & 1 & 0 & 1 & 0 & 1 \\  \hline \end{tabular} \ee 
		
		\paragraph{Fiber degrees $\{d_2, \{d_0,d_1,d_3,d_4 \} \} = \{3, \{{3, 3, 3, 3} \} \} $  } \be \begin{tabular} {|c|cccccccc|} \hline $2j_L \backslash 2j_R$  & 0 & 1 & 2 & 3 & 4 & 5 & 6 & 7 \\  \hline  0 & 0 & 4 & 0 & 4 & 0 & 4 & 0 & 1 \\  \hline \end{tabular} \ee 
		
		\paragraph{Fiber degrees $\{d_2, \{d_0,d_1,d_3,d_4 \} \} = \{4, \{{3, 3, 3, 3} \} \} $  } \be \begin{tabular} {|c|cccccccccc|}\hline 
			$2j_L \backslash 2j_R$  & 0 & 1 & 2 & 3 & 4 & 5 & 6 & 7 & 8 & 9 \\  \hline  
			0 & 0 & 81 & 0 & 85 & 0 & 40 & 0 & 9 & 0 & 1 \\ 
			1 & 4 & 0 & 8 & 0 & 5 & 0 & 1 & 0 & 0 & 0 \\  \hline 
		\end{tabular} 
		\ee

	\section{Weyl invariant and  $\Dag$ invariant Jacobi forms} \label{a:WeylJac}
	We will use the following conventions throughout regarding Jacobi $\theta$ functions:
	\be 
	\Theta\left[a \atop b\right](\tau,z)=\sum_{n\in \mathbb{Z}} e^{\pi i (n + a)^2 \tau + 2 \pi i z (n+a) + 2 \pi i bn}\ ,
	\ee
	and 
	\be
	\theta_1=i\Theta\left[\frac{1}{2} \atop \frac{1}{2}\right] \,, \quad 
	\theta_2=\Theta\left[\frac{1}{2} \atop 0\right] \,,\quad \theta_3=\Theta\left[0 \atop 0\right] \,, \quad \theta_4=\Theta\left[0 \atop \frac{1}{2}  \right] \,.
	\ee

	\subsection{Review of Weyl invariant Jacobi forms}
	Let $Q^\vee, P$ be the coroot lattice and the weight lattice of a Lie algebra $\fg$ with rank $r$. $W$ is the Weyl group of $\fg$. Then following \cite{Wirthmuller:Jacobi}, a Weyl-invariant Jacobi form of weight $w$ and index $n$ $(w\in \mb Z, n\in \mb N)$ is defined to be a holomorphic function $\varphi_{w,n}: \mb H\times \fh_{\mb C} \rightarrow \mb C \;$   satisfying the following conditions:\footnote{As in the body of the paper, we identify $\fh$ and $\fh^*$ via a choice of bilinear form.}
	\begin{itemize}
		\item Modularity: for any $\(\begin{smallmatrix}
		a & b \\ c & d
		\end{smallmatrix}\) \in SL(2,\mb Z)$
		\begin{equation}
			\varphi_{w,n}\( \frac{a\tau+b}{c\tau +d}, \frac{\ja}{c\tau +d}\) = (c\tau+d)^w \md e\left[\frac{n c}{2(c\tau+d)} (\ja,\ja)_\fg\right] \varphi_{w,n}(\tau,\ja) \ .
		\end{equation}
		\item Quasi-periodicity: for any $\lambda, \mu \in Q^\vee$
		\begin{equation}
			\varphi_{w,n}(\tau, \ja+\lambda \tau+\mu) = \md e\left[-n\(\frac{(\lambda,\lambda)_\fg}{2}\tau+(\lambda,\ja)_\fg\)\right] \varphi_{w,n}(\tau,\ja) \ .
		\end{equation}
		\item Weyl symmetry: for any $w\in W$
		\begin{equation}
			\varphi_{w,n}(\tau, w\ja) = \varphi_{w,n}(\tau,\ja) \ .
		\end{equation}
		\item Fourier expansion: $\varphi_{w,n}$ can be expanded as
		\begin{equation}
			\varphi_{w,n}(\tau, \ja) = \sum_{\substack{\ell\in \mb N_0, \gamma \in P}} c(\ell,\gamma) q^\ell \zeta^\gamma \ ,
		\end{equation}
		where $\zeta^\gamma = \md e[(\ja,\gamma)_\fg]$.
	\end{itemize}
	We have written these equations in terms of the invariant bilinear form $(,)_\fg$ on $\fh_\bC$ normalized such that the norm square of the shortest coroot $\theta^\vee$ is 2. The argument $\tau\in \bH$ of $\varphi_{w,n}(\tau,\ja)$ is called the modular parameter, and $\ja\in \fh_\bC\cong \bC^r$ the elliptic parameter. Weyl invariant Jacobi forms are generalizations of weak Jacobi forms. In fact, a weak Jacobi form $\varphi_{w,n}(\tau,\ja)$ $(\ja\in \bC)$ can be thought of as a $W(\mathfrak{a}_1)$ invariant Jacobi form of the same weight and index with the elliptic parameter given by $\ja = z\alpha^\vee $, $\alpha^\vee$ the unique (upon making the standard choices) simple coroot of $\mathfrak{a}_1$. 
	
	For a given Lie algebra $\fg$, the vector space of $W(\fg)$ invariant Jacobi forms of weight $w$ and index $n$ is denoted by $J_{w,n}(\fg)$. The bigraded ring $J_{*,*}(\fg) = \oplus_{w,n} J_{w,n}(\fg)$ is a polynomial ring over the ring of $SL(2,\bZ)$ modular forms generated by the Eisenstein series $E_4, E_6$. It was shown in \cite{Wirthmuller:Jacobi} that if $\fg$ is a simple Lie algebra besides $E_8$, $J_{*,*}(\fg)$ is freely generated by
	\begin{equation}
		\varphi_0, \varphi_1, \ldots, \varphi_r
	\end{equation}
	whose weights and indices are given respectively by
	\begin{equation}\label{eq:km-wjacobi}
		(-d_i , a_i^\vee) \ .
	\end{equation}
	$d_0 = 0, a_0^\vee =1$, while for $i=1,\ldots,r$, $d_i$ are the exponents of the Casimirs of $\fg$, and $a_i^\vee$ comarks. We call the generators of the ring $J_{*,*}(\fg)$ the fundamental Jacobi forms. 
	
	The ring generators for $A_1$ are given by \cite{Eichler,Dabholkar:2012nd}
	\begin{equation}
	\begin{aligned}
		& \varphi_{-2,1}(\tau,z\alpha^\vee) =\varphi_{-2,1}(\tau,z)= -\frac{\theta_1(\tau,z)^2}{\eta(\tau)^6} \in J_{-2,1}\ , \\
		& \varphi_{0,1}(\tau,z\alpha^\vee) =\varphi_{0,1}(\tau,z)= 4\sum_{i=2}^3 \frac{\theta_i(\tau,z)^2}{\theta_i(\tau,0)^2} \in J_{0,1}\ .
	\end{aligned}
	\end{equation}
	In the above formulae, we denote them first as Weyl invariant Jacobi forms and then as weak Jacobi forms. These two functions are also conventionally denoted as $A, B$ respectively. As for higher rank Lie algebras, the explicit forms of the ring generators were constructed in Bertola's thesis \cite{Bertola:Jacobi} for $\fg = \mathfrak{a}_n, \mathfrak{b}_n, \mathfrak{g}_2, \mathfrak{c}_3, \mathfrak{d}_4$, and in \cite{Sakai:2017ihc} for $\fg = \mathfrak{e}_6, \mathfrak{e}_7$; fundamental Jacobi forms for $\mathfrak{e}_8$ were proposed in \cite{Sakai:2011xg}, and they indeed do not have the weights given in \eqref{eq:km-wjacobi}. 

	For the introduction of $\Dag$ invariant Jacobi forms, it is convenient to combine the modular argument $\tau$ and the elliptic parameter $\mass$ to an element of the affine root space. We quickly review the necessary notions from the theory of affine Lie algebras: in terms of the imaginary root $\delta$, which satisfies $(\delta,\delta)_{\mathfrak{g}} = 0$, and the highest root $\theta$, the zeroth simple root is defined as
	\be
	\alpha_0 = - \theta + \delta \,.
	\ee
	The dual basis to the set obtained by adjoining the zeroth coroot to the finite simple coroots, $\{\alpha_0^\vee,\alpha_i^\vee\}$, is given by $\{\aomega_0,\aomega_i\}$, where
	\be
	\aomega_i = \omega_i + a_i^\vee \aomega_0 \,,
	\ee
	with the shift of the finite fundamental weight $\omega_i$ by $a_i^\vee \aomega_0$ introduced to impose orthonormality with regard to $\alpha_0^\vee$. We now combine the modular and the elliptic arguments of the Weyl invariant Jacobi forms into a single quantity $\aja$ taking values in the affine weight lattice, 
	\be
	\aja = m_0\, \aomega_0 + \sum_i m_i \aomega_i = (m_0 + \sum_i a_i^\vee m_i)\,\aomega_0 + \sum_i m_i \omega_i = \tau\, \aomega_0 + \ja \,.
	\ee
	$m_0$, which we will identify with the K\"ahler modulus for the zeroth fiber component, is thus the coefficient of $\aomega_0$ in the affine basis, while $\tau$ is the coefficient of $\aomega_0$ in the finite basis with $\aomega_0$ adjoined.
	
	The finite Weyl group $W(\mathfrak g)$ leaves $\omega_0$ fixed and acts via the conventional reflections on the finite part $\ja$ of $\aja$. The affine Weyl reflection associated to the root $\alpha + n \delta$ can be written in the form
	\be
	\sigma_{\alpha + n \delta} = \sigma_\alpha \circ (t_{\alpha^\vee})^n
	\ee
	where $t_{\alpha^\vee}$ acts via
	\be
	t_{\alpha^\vee}:
	\begin{cases}
		\aomega_0 \mapsto \aomega_0 + \alpha^\vee &\mod \delta \,,\\
		\omega_i \mapsto \omega_i &\mod \delta \,.
	\end{cases}
	\ee
	The $\mod \delta$ refers to the fact that if we keep track of the level, the affine weight lattice has rank $n+2$ compared to the rank $n+1$ of the affine root lattice: it is spanned by $\{\aomega_0,\aomega_i\}$ and $\delta$. $t_{\alpha^\vee}$ acting on $\{\aomega_0,\omega_i\}$ also generates contributions in the $\delta$ direction, which will not be relevant for our considerations.
	
	Weyl invariant Jacobi forms thus transform well under affine Weyl reflections $\sigma_\alpha \circ (t_{\alpha^\vee})^n \in \hW(\mathfrak g)$: they are invariant under $\sigma_{\alpha}$, and transform quasi-periodically under $t_{\alpha^\vee}$.

	\subsection{Bertola's basis of Weyl invariant Jacobi forms}
	\label{sc:Bertola-basis}
	
	To match the notation in the main part of the text, we will call the elliptic parameter $\ja = \mass$ in the following. In the case of $A_2$, a vector $\mass$ in $\fh_\bC$ can be parametrized as
	\begin{equation}
		\mass = \sum_{j=1}^3 u_j e_j = \sum_{j=1}^2 x_j \alpha_j^\vee = \sum_{j=1}^2 m_j w_j \ .
	\end{equation}
	$\{e_j\}$ is the standard basis of $\bR^3$, in which $\fh_\bR$ can be embedded as a hyperplane. The parameters $u_j$ should satisfy
	\begin{equation}
		u_1 + u_2 + u_3 = 0 \ .
	\end{equation}
	$\alpha_j$ are the simple roots and $\omega_j$ are the fundamental weights. They can be taken to be \cite{Bourbaki}
	\begin{equation}
		\alpha_1 = e_1 - e_2 \ , \quad \alpha_2 = e_2 - e_3 \ ,
	\end{equation}
	and
	\begin{equation}
		w_1 = \frac{1}{3}(2 e_1 - e_2 - e_3) \ ,\quad 	w_2 = \frac{1}{3}( e_1 + e_2 - 2 e_3 )\ .
	\end{equation}
	Accordingly, the different parametrizations are related by
	\begin{equation}\label{eq:params-A2}
	\left\{
	\begin{aligned}
		& u_1 = x_1 &&= \tfrac{2}{3} m_1 + \tfrac{1}{3} m_2 \ , \\
		& u_2 = -x_1 + x_2 &&= -\tfrac{1}{3}m_1 +\tfrac{1}{3}m_2 \ ,\\
		& u_3 = -x_2 &&= -\tfrac{1}{3}m_1 -\tfrac{2}{3}m_2 \ .
	\end{aligned}\right.
	\end{equation}
	
	The doubly graded algebra $J_{*,*}(A_2)$ of $W(A_2)$ invariant Jacobi forms is generated by forms
	\begin{equation}
		\varphi_3 \in J_{-3,1} \ ,\quad \varphi_2 \in J_{-2,1} \ ,\quad \varphi_0 \in J_{0,1} 
	\end{equation}
	which were formally constructed in \cite{Wirthmuller:Jacobi}. The construction was made more explicit in Bertola's thesis \cite{Bertola:Jacobi}: defining
	\begin{equation}
		\fd_x = \frac{1}{2\pi}\frac{\partial}{\partial x} \ ,
	\end{equation}
	the ring generators as constructed by Bertola read
	\begin{equation}
	\begin{aligned}
		& \varphi_3 = -\ri\,\eta(\tau)^{-9} \prod_{j=1}^3 \theta_1(\tau,u_j) \Big|_{u_*\rightarrow x_*} \ ,\\
		& \varphi_2 = \(\sum_{j=1}^3 \frac{{\mf d}_{u_j}\theta_1(\tau,u_j)}{\theta_1(\tau,u_j)}\)\cdot \varphi_3\Big|_{u_*\rightarrow x_*}\ ,\\
		& \varphi_0 = \(-{\mf d}_\tau - \frac{E_2(\tau)}{4} + \frac{1}{3}({\mf d}_{x_1}^2+ {\mf d}_{x_2}^2+{\mf d}_{x_1}{\mf d}_{x_2})\)\circ \varphi_2 \ ,
	\end{aligned}
	\end{equation}
	where $u_* \rightarrow x_*$ means a change of parametrization according to \eqref{eq:params-A2}. 
	
	For the case of $D_4$, we again introduce the three parametrizations
	\begin{equation}
		\mass = \sum_{j=1}^4 u_j e_j = \sum_{j=1}^{4} x_j \alpha_j = \sum_{j=1}^4 m_j w_j 
	\end{equation}
	for a vector $\mass \in \fh_\bC$. The simple roots can be taken to be \cite{Bourbaki}
	\begin{equation}
		\alpha_1 = e_1 -e_2 \ , \quad \alpha_2 = e_2 - e_3 \ ,\quad \alpha_3 = e_3 - e_4 \ ,\quad \alpha_4 = e_3+ e_4 \ ,
	\end{equation}
	and correspondingly the fundamental weights are
	\begin{equation}
		w_1 = e_1 \ ,\; w_2 = e_1 + e_2 \ ,\; w_3 = \frac{1}{2}(e_1+e_2+e_3-e_4) \ ,\; w_4 = \frac{1}{2}(e_1 + e_2 + e_3 + e_4) \ .
	\end{equation}
	
	The doubly graded algebra $J_{*,*}(D_4)$ of $W(D_4)$ invariant Jacobi forms is generated by the forms
	\begin{equation}
		\varphi_0 \in J_{0,1} \ ,\quad \varphi_2 \in J_{-2,1} \ ,\quad \varphi_4,\psi_4\in J_{-4,1} \ ,\quad \varphi_6 \in J_{-6,2} \ ,
	\end{equation}
	which were formally constructed in \cite{Wirthmuller:Jacobi}. The explicit construction is given in \cite{Bertola:Jacobi}. Using the notation $\theta_i(z) = \theta_i(\tau,z)$, $\theta_i = \theta_i(\tau,0)$, $\eta=\eta(\tau)$, the elements of the basis read\footnote{  They are slightly different from \cite{Bertola:Jacobi} for better presentation or better expansion.}
	\begin{equation}
	\begin{aligned}
		& \varphi_0  = \eta^{-12}\(\theta_3^8\prod_{j=1}^4\theta_3(u_j) -\theta_4^8 \prod_{j=1}^4\theta_4(u_j) - \theta_2^8 \prod_{j=1}^4 \theta_2(u_j)\) \ , \\
		& \varphi_2  = \eta^{-12} \( (\theta_4^4 - \theta_2^4)\prod_{j=1}^4 \theta_3(u_j) - (\theta_2^4+\theta_3^4)\prod_{j=1}^4 \theta_4(u_j)  +(\theta_3^4 + \theta_4^4 ) \prod_{j=1}^4 \theta_2(u_j)\) \ ,\\
		& \varphi_4 = \eta^{-12} \( \prod_{j=1}^4 \theta_3(u_j) - \prod_{j=1}^4 \theta_4(u_j) - \prod_{j=1}^4 \theta_2(u_j) \) \ ,\\
		& \psi_4  = \eta^{-12}\prod_{j=1}^4 \theta_1(u_j) = \prod_{j=1}^4 \phi_{-1,\tfrac{1}{2}} (u_j) \ ,\\
		& \varphi_6 = \pi^{-2}\eta^{-24} \prod_{j=1}^4 \theta_1^2(u_j) \cdot \sum_{j=1}^4 \wp(u_j) \ .
	\end{aligned}
	\end{equation}

	\subsection{Basis of  $\Dag$ invariant Jacobi forms}
	\label{sc:affine-basis}
	
	In the case of $\widehat{A}_2$, the conjectured basis of the ring of $\Dat$ invariant Jacobi forms consists of
	\begin{equation}
		\phi_0 \in \widehat{J}_{0,3} \ ,\quad \phi_2 \in \widehat{J}_{-2,3} \ ,\quad \phi_6 \in \widehat{J}_{-6,6} \ .
	\end{equation}
	Their explicit forms in terms of generators of $J_{*,*}(\fg)$ and modular forms are
	\begin{equation}\label{eq:affine-basis-A2}
	\begin{aligned}
		\phi_0 &= 6\varphi_0^3 + \frac{E_4 \varphi_0 \varphi_2^2}{8} - \frac{E_6 \varphi_2^3}{72} -\frac{E_6 \varphi_0 \varphi_3^2}{16} + \frac{E_4^2 \varphi_2 \varphi_3^2}{192}  \ ,\\
		\phi_2 &= 24 \varphi_0^2 \varphi_2  - E_4 \varphi_0 \varphi_3^2 - \frac{E_4 \varphi_2^3}{6}  + \frac{E_6 \varphi_2 \varphi_3^2}{12} \ ,\\
		\phi_6 &= 4 \varphi_0^3 \varphi_2^3
		-27 \varphi_0^4\varphi_3^2
		+\frac{5}{8} E_4 \varphi_0^2 \varphi_2^2 \varphi_3^2 
		+\frac{E_6 \varphi_0^2 \varphi_3^4}{16} 
		-\frac{E_4 \varphi_0\varphi_2^5}{12}
		-\frac{E_6 \varphi_0 \varphi_2^3\varphi_3^2}{24}				
		-\frac{E_4^2 \varphi_0 \varphi_2 \varphi_3^4 }{96}\\
		&+\frac{E_6 \varphi_2^6}{216}
		+\frac{E_4^2 \varphi_2^4\varphi_3^2 }{2304}
		+\frac{E_4 E_6 \varphi_2^2\varphi_3^4 }{2304}
		-\frac{E_6^2\varphi_3^6}{27648}
		+\frac{E_4^3 \varphi_3^6}{27648} 	 \ .
	\end{aligned}
	\end{equation}
	They are normalized so that in the power series expansion in $Q_0,Q_1,Q_2$ with
	\begin{equation}
		q = Q_0 Q_1 Q_2
	\end{equation}
	the leading term has coefficient 1.

	These generators can also be expressed in the following way
	\begin{equation}
	\begin{aligned}
		& \phi_6 = -\,\eta^{-18} \prod_{j=1}^3 \theta_1^2(m_j) \Big|_{m_3 = -m_1 - m_2} \ ,\\
		& \phi_2 = -8\ri \, \eta^{-9} \prod_{j=1}^3 \theta_1(m_j) \(\sum_{k=1}^3 \frac{{\mf d}_{m_k}\theta_1(m_k)}{\theta_1(m_k)}\)\Big|_{m_3 = -m_1-m_2}  \ ,\\
		& \phi_0 = \frac{3}{4} \(-{\mf d}_\tau - \frac{E_2}{4} + \frac{1}{3}({\mf d}_{m_1}^2+ {\mf d}_{m_2}^2-{\mf d}_{m_1}{\mf d}_{m_2})\)\circ \phi_2 \ .
	\end{aligned}
	\end{equation}
	
	In the case of $\widehat{D}_4$, a ring of $D(\mathfrak{d}_4)$ invariant Jacobi forms is spanned by
	\begin{equation}
		\phi_0 \in \widehat{J}_{0,2} \ ,\quad \phi_2 \in \widehat{J}_{-2,2}\ ,\quad \phi_6 \in \widehat{J}_{-6,4}\ ,\quad \phi_8 \in \widehat{J}_{-8,4} \ ,\quad \phi_{12}\in \widehat{J}_{-12,6} \ .
	\end{equation}
	The explicit forms of these generators are
	\begin{equation}\label{eq:affine-basis-D4}
	\begin{aligned}
		\phi_0 &= \frac{\vp_0^2}{32} + \frac{E_4 \vp_2^2}{288} -\frac{E_4 \vp_0\vp_4}{24} -\frac{E_6 \vp_2\vp_4}{72} + \frac{E_4^2 \vp_4^2}{48}+ \frac{E_4^2\psi_4^2}{48} + \frac{E_6 \vp_6}{16}  \ , \\
		\phi_2 &= \frac{\vp_0 \vp_2}{48} -\frac{E_4 \vp_2\vp_4}{36} + \frac{E_6 \vp_4^2}{144} + \frac{E_6\psi_4^2}{48} + \frac{E_4\vp_6}{16} \ , \\
		\phi_6 &=\frac{\vp_0 \vp_2^3}{6912} - \frac{\vp_0^2 \vp_2 \vp_4}{1152} - \frac{E_4 \vp_2^3 \vp_4}{5184} + \frac{13 E_4 \vp_0 \vp_2 \vp_4^2}{6912} + \frac{E_6 \vp_2^2 \vp_4^2}{20736} - \frac{E_6 \vp_0 \vp_4^3}{3456} - \frac{5 E_4^2 \vp_2 \vp_4^3}{5184} \\
		&+ \frac{5 E_4 E_6 \vp_4^4}{20736} + \frac{\vp_0^2 \vp_6}{256}  + \frac{E_4\vp_2^2 \vp_6}{2304} - \frac{E_4 \vp_0 \vp_4 \vp_6 }{192} - \frac{E_6 \vp_2 \vp_4 \vp_6 }{576} + \frac{E_4^2 \vp_4^2 \vp_6}{384}  + \frac{E_6 \vp_6^2}{256} \\
		&  + \frac{5 E_4 \vp_0 \vp_2 \psi_4^2}{2304} - \frac{E_6 \vp_2^2 \psi_4^2}{2304} + \frac{E_6 \vp_0 \vp_4 \psi_4^2}{384}  - \frac{5 E_4^2 \vp_2 \vp_4 \psi_4^2}{1728}- \frac{5 E_4 E_6 \vp_4^2 \psi_4^2}{3456} + \frac{E_4^2 \vp_6 \psi_4^2}{384} \\
		&   + \frac{E_4 E_6 \psi_4^4}{2304} \ ,\\
		\phi_8 &=\frac{\vp_2^4}{20736} - \frac{\vp_0 \vp_2^2 \vp_4}{1728} + \frac{\vp_0^2 \vp_4^2}{576}  + \frac{5 E_4 \vp_2^2 \vp_4^2}{10368} - \frac{5 E_4 \vp_0 \vp_4^3}{1728} + \frac{25 E_4^2 \vp_4^4}{20736} + \frac{\vp_0 \vp_2 \vp_6}{384} \\
		& - \frac{E_4 \vp_2 \vp_4 \vp_6}{288}  + \frac{E_6 \vp_4^2 \vp_6}{1152} + \frac{E_4 \vp_6^2}{256}  + \frac{\vp_0^2 \psi_4^2}{192}  - \frac{E_4 \vp_2^2 \psi_4^2}{3456} - \frac{E_4 \vp_0 \vp_4 \psi_4^2}{576}  - \frac{E_6 \vp_2 \vp_4 \psi_4^2}{432}  \\
		&- \frac{E_4^2 \vp_4^2 \psi_4^2}{1152} + \frac{E_6 \vp_6 \psi_4^2}{384}  + \frac{E_4^2 \psi_4^4}{2304} \ ,\\
		\phi_{12} &= \frac{\vp_2^6}{2985984} - \frac{\vp_0 \vp_2^4 \vp_4}{165888} + \frac{\vp_0^2 \vp_2^2\vp_4^2}{27648} + \frac{5 E_4 \vp_2^4 \vp_4^2}{995328} - \frac{\vp_0^3\vp_4^3}{13824} - \frac{5 E_4 \vp_0 \vp_2^2 \vp_4^3}{82944} + \frac{5 E_4 \vp_0^2\vp_4^4}{27648} \\
		&+ \frac{25 E_4^2 \vp_2^2 \vp_4^4}{995328} - \frac{25 E_4^2 \vp_0\vp_4^5}{165888} + \frac{125 E_4^3 \vp_4^6}{2985984} + \frac{\vp_0 \vp_2^3\vp_6}{36864} - \frac{\vp_0^2 \vp_2 \vp_4 \vp_6}{6144} - \frac{E_4 \vp_2^3\vp_4 \vp_6}{27648} \\
		&+ \frac{13 E_4 \vp_0 \vp_2 \vp_4^2\vp_6}{36864} + \frac{E_6 \vp_2^2 \vp_4^2 \vp_6}{110592} - \frac{E_6 \vp_0\vp_4^3 \vp_6}{18432} - \frac{5 E_4^2 \vp_2 \vp_4^3 \vp_6}{27648} + \frac{5 E_4E_6 \vp_4^4             \vp_6}{110592} + \frac{3 \vp_0^2 \vp_6^2}{8192} \\
		&+ \frac{E_4 \vp_2^2\vp_6^2}{24576} - \frac{E_4 \vp_0 \vp_4 \vp_6^2}{2048} - \frac{E_6 \vp_2\vp_4 \vp_6^2}{6144} + \frac{E_4^2 \vp_4^2 \vp_6^2}{4096} + \frac{E_6\vp_6^3}{4096} +              \frac{\vp_0^2 \vp_2^2 \psi_4^2}{18432} - \frac{E_4 \vp_2^4\psi_4^2}{331776} \\
		&+ \frac{\vp_0^3 \vp_4 \psi_4^2}{1536} - \frac{E_4 \vp_0 \vp_2^2\vp_4 \psi_4^2}{3072} + \frac{E_6 \vp_2^3 \vp_4 \psi_4^2}{20736} - \frac{5 E_4\vp_0^2 \vp_4^2      \psi_4^2}{4608} - \frac{5 E_6 \vp_0 \vp_2 \vp_4^2\psi_4^2}{27648} \\
		& + \frac{41 E_4^2 \vp_2^2 \vp_4^2 \psi_4^2}{165888} + \frac{17 E_4^2\vp_0 \vp_4^3 \psi_4^2}{27648} + \frac{E_4 E_6 \vp_2 \vp_4^3\psi_4^2}{10368} - \frac{5 E_4^3 \vp_4^4     \psi_4^2}{36864} + \frac{E_6^2 \vp_4^4\psi_4^2}{55296} \\
		&+ \frac{5 E_4 \vp_0 \vp_2 \vp_6 \psi_4^2}{12288} - \frac{E_6\vp_2^2 \vp_6 \psi_4^2}{12288} + \frac{E_6 \vp_0 \vp_4 \vp_6\psi_4^2}{2048} - \frac{5 E_4^2 \vp_2 \vp_4 \vp_6              \psi_4^2}{9216} - \frac{5 E_4E_6 \vp_4^2 \vp_6 \psi_4^2}{18432} \\
		&+ \frac{E_4^2 \vp_6^2\psi_4^2}{4096} + \frac{E_4 \vp_0^2 \psi_4^4}{3072} - \frac{E_6 \vp_0 \vp_2\psi_4^4}{9216} + \frac{E_4^2 \vp_2^2 \psi_4^4}{110592} - \frac{5 E_4^2 \vp_0 \vp_4\psi_4^4}{18432} + \frac{E_4^3 \vp_4^2 \psi_4^4}{12288} - \frac{E_6^2 \vp_4^2\psi_4^4}{27648} \\
		&+ \frac{E_4 E_6 \vp_6 \psi_4^4}{12288} - \frac{E_4^3\psi_4^6}{110592} + \frac{E_6^2 \psi_4^6}{55296} \ .
	\end{aligned}
	\end{equation}
	They are normalized so that in the power series expansion in $Q_0,Q_1,Q_2,Q_3,Q_4$ with
	\begin{equation}
		q = Q_0 Q_1 Q_2^2 Q_3 Q_4
	\end{equation}
	the leading term has coefficient 1. We conjecture that this set generates the ring $\joutd$.

\subsection{$E_8$ Weyl invariant Jacobi forms}
The classification results of \cite{Wirthmuller:Jacobi} discussed above apply to all simply laced Lie algebras except for $\mathfrak{e}_8$. For this final case, Sakai has constructed certain holomorphic Jacobi forms \cite{Sakai:2011xg}, denoted as $A_1, A_2, A_3, A_4, A_5, B_2, B_3, B_4, B_6$, with the subscript indicating the $E_8$ elliptic index. The $A_n$'s have modular weight 4 and reduce to the Eisenstein series $E_4$ in the massless limit, while the $B_n$'s have modular weight 6 and reduce to the Eisenstein series $E_6$. This set of forms generates a ring over the space of holomorphic modular forms which we shall refer to as $R_{Sakai}$. $R_{Sakai}$ does not coincide with the full ring of $E_8$ Weyl invariant Jacobi forms \cite{Sakai:2017ihc}. For example, since the leading term in the $q$-series expansion of the $A_n$'s and $B_n$'s is 1, we can easily construct polynomials in these forms and the Eisenstein series which vanish at $q=0$. As the discriminant function $\Delta=\frac{1}{1728}( E_4^3-E_6^2)$ vanishes on the upper half plane only at $\tau=i \infty$, i.e. at $q=0$, we can then construct new holomorphic Weyl invariant Jacobi forms by dividing such polynomials with zero at $q=0$ by $\Delta$. Examples of such forms are $\frac{1}{\Delta}(A_2 E_4-A_1^2)$, $\frac{1}{\Delta}(B_2 E_4-A_2 E_6)$, $\frac{1}{\Delta}(B_2E_6-A_2E_4^2)$. These forms are holomorphic in the upper half plane including the point at infinity for general $E_8$ elliptic parameters, but they are clearly not elements of $R_{Sakai}$.

To elucidate further the set of forms obtained by dividing by the discriminant function, we introduce some notation. We call an element of $R_{Sakai}$ which vanishes at $q=0$ for general $E_8$ elliptic parameters a {\it cusp polynomial}. At given weight and index, we can introduce a basis of such cusp polynomials: we consider all pairs of generators of $R_{Sakai}$ at this weight and index endowed with appropriate coefficients to cancel the constant term in the $q$ expansion. We discard pairs that exhibit common factors in $\{A_n, B_n,E_4,E_6\}$. To reduce this set further, we define the notion of {\it connection} between two generators at given weight and index. If they have a common factor, we connect them. For example $E_4^2A_2$ and $E_4A_1^2$ are connected, while $E_4^2A_2$ and $E_6B_2$ are not connected. In this way, the finite set of generators of $R_{Sakai}$ at given index and weight are separated into disjoint components. The number of linearly independent cusp polynomials at given weight and index that cannot be generated by cusp polynomials over lower weight/index over $R_{Sakai}$ is the number of disjoint components minus one: choose one representative per disjoint component, and form all pairs between these and a given distinguished representative. For example, there are 4 generators of $R_{Sakai}$ for index 3 and weight 12, separated into 2 disjoint components as $\{ E_4^2 A_3, E_4 A_1A_2, A_1^3 \}, \{ E_6 B_3 \}$. We can choose an independent cusp polynomial $E_4^2 A_3 - E_6 B_3$. 

Via a computer search, we find a finite number, namely 43, of cusp polynomials independent in the sense just described, including $1728 \Delta = E_4^3-E_6^2$. The highest index in this set is 12. We enumerate these cusp polynomials as $p_1=E_4^3-E_6^2,  p_2= A_2 E_4-A_1^2, p_3= B_2 E_4-A_2 E_6, \cdots, p_{43} = A_4^3 - B_6^2$. We conjecture that any cusp polynomial can be written as a linear combination of these 43 cusp polynomials over the ring $R_{Sakai}$. On the other hand, by construction, none of the 43 cusp polynomials can be written as a linear combination of the others over $R_{Sakai}$.

Beyond division by the discriminant function, a more intricate cancellation of zeros in a ratio of Jacobi forms was encountered in the study of topological strings on local $1/2 K3$  \cite{Huang:2013}, corresponding to the E--string theory. It was found that the topological string amplitudes on this geometry can be written as elements of $R_{Sakai}$ divided by $E_4$. The numerators found in \cite{Huang:2013} at low degree have the following common factor with $E_8$ index 5 and modular weight 16:
\begin{eqnarray} \label{polynomialB.35}
	P =  864 A_1^3 A_2+21 E_6^2 A_5 -770 E_6 A_3 B_2+3825 A_1 B_2^2-840 E_6 A_2 B_3+60 E_6 A_1 B_4. 
\end{eqnarray}
We have checked numerically that this polynomial vanishes at the zero points $\tau = \pm \frac{1}{2} +\frac{\sqrt{3}}{2} i $ of $E_4$ for general $E_8$ elliptic parameters such that the poles introduced by dividing by $E_4$ are in fact only apparent. It would be nice to derive the location of the poles of \eqref{polynomialB.35} analytically. If the zeros of $P$ and $E_4$ do indeed coincide, then we must include $\frac{P}{E_4}$ in our list of holomorphic $E_8$ Weyl invariant Jacobi forms linearly independent over $R_{Sakai}$.

Motivated by this example, we have systematically searched numerically for other Jacobi forms independent of (\ref{polynomialB.35}) whose zeros coincide with those of $E_4$, without success. We therefore conjecture that any Jacobi form expressed as a polynomial of $A_n$'s, $B_n$'s and $E_6$ which vanishes at the zero points of $E_4$ for general $E_8$ elliptic parameters must be divisible by the polynomial (\ref{polynomialB.35}). We have also searched for Jacobi form that vanishes at the zero point $\tau=i$ of $E_6$, without success. 

We thus arrive at a conjecture for a complete set of generators of the ring of $E_8$ invariant Jacobi forms over the ring of holomorphic forms: the $A_n$'s and $B_n$'s of \cite{Sakai:2011xg}, the forms $\frac{p_i}{\Delta}$ for $i=2, \ldots, 43$, and finally $\frac{P}{E_4}$. An open challenge is to study the algebraic dependence among these generators.

	\bibliography{draft_refEM}
	
\end{document}